\documentclass{aa}  
\usepackage{lscape}
\usepackage{inputenc}
\usepackage[usenames, dvipsnames]{color}
\usepackage{booktabs, caption, makecell}
\usepackage{epstopdf}
\usepackage{gensymb}
\usepackage{threeparttable}
\usepackage{geometry}
\usepackage{hyperref}
\usepackage{capt-of}

\DeclareUnicodeCharacter{FEFF}{}

\graphicspath{ }
\usepackage{epstopdf}

\definecolor{mypink3}{cmyk}{0.26, 1.0, 0.0, 0.56}

\usepackage{color}
\usepackage{graphicx}
\usepackage{txfonts}
\begin{document} 

%{scrbook}
%\addtokomafont{disposition}{\normalfont}

   \title{Chemical and kinematic structure of extremely high-velocity molecular jets in the Serpens Main star-forming region}

   \author{{\L}ukasz Tychoniec \inst{1}, Charles L. H. Hull \inst{2,3,9}, Lars E. Kristensen\inst{4}, John J. Tobin \inst{5}, Valentin J. M. Le Gouellec \inst{6,7}, Ewine~F.~van Dishoeck \inst{1,8}
          }
   \institute{Leiden Observatory, Leiden University, PO Box 9513, 2300RA, Leiden, The Netherlands\\
              \email{tychoniec@strw.leidenuniv.nl}
         \and
           National Astronomical Observatory of Japan, NAOJ Chile Observatory,
Alonso de C{\'o}rdova 3788, Office 61B, Vitacura 763 0422, Santiago, Chile
         \and
         Joint ALMA Observatory, Alonso de C{\'o}rdova 3107, Vitacura 763 0355,
Santiago, Chile
\and
Centre for Star and Planet Formation, Niels Bohr Institute and Natural History Museum of Denmark, University of Copenhagen,
{\o}ster Voldgade 5-7, DK-1350 Copenhagen K, Denmark
\and
         	National Radio Astronomy Observatory, Charlottesville, VA 22903   	    
         	         	\and
         	European Southern Observatory, Alonso de C{\'o}rdova 3107, Vitacura, Santiago, Chile
         	\and
AIM, CEA, CNRS, Universit{\'e} Paris-Saclay, Universit{\'e} Paris Diderot, Sorbonne Paris Cit{\'e}, F-91191 Gif-sur-Yvette, France
         \and
         	Max-Planck-Institut f{\"u}r Extraterrestrische Physik, Giessenbachstrasse 1, D-85748 Garching, Germany       
\and
NAOJ Fellow
            \\
             }

  \abstract
  {Outflows are one of the first signposts of ongoing star formation. The fastest molecular component to the protostellar outflows -- extremely high-velocity (EHV) molecular jets -- are still puzzling since they are seen only rarely. As they originate deep inside the embedded protostar-disk system, they provide vital information about the outflow-launching process in the earliest stages.
  }
   {The first aim is to analyze the interaction between the EHV jet and the slow outflow by comparing their outflow force content. The second aim is to analyze the chemical composition of the different outflow velocity components and to reveal the spatial location of molecules.}
   {ALMA 3 mm (Band 3) and  1.3 mm (Band 6) observations of five outflow sources at $0\farcs3$  -- $0\farcs6$ (130 -- 260 au) resolution in the Serpens Main cloud are presented. Observations of CO, SiO, H$_2$CO and HCN reveal the kinematic and chemical structure of those flows.   Three velocity components are distinguished: the slow and the fast wing, and the EHV jet.}
   {Out of five sources, three have the EHV component. Comparison of outflow forces reveals that only the EHV jet in the youngest source Ser-emb~8~(N) has enough momentum to power the slow outflow. The SiO abundance is generally enhanced with velocity, while HCN is present in the slow and the fast wing, but disappears in the EHV jet. For Ser-emb~8~(N), HCN and SiO show a bow-shock shaped structure surrounding one of the EHV peaks suggesting sideways ejection creating secondary shocks upon interaction with the surroundings. Also, the SiO abundance in the EHV gas decreases with distance from this protostar, whereas that in the fast wing increases. H$_2$CO is mostly associated with low-velocity gas but also appears surprisingly in one of the bullets in the Ser-emb~8~(N) EHV jet. No complex organic molecules are found to be associated with the outflows.}
   {The high detection rate suggests that the presence of the EHV jet may be more common than previously expected. The EHV jet alone does not contain enough outflow force to explain the entirety of the outflowing gas. The origin and temporal evolution of the abundances of SiO, HCN and H$_2$CO through high-temperature chemistry are discussed. The data are consistent with a low C/O ratio in the EHV gas versus high C/O ratio in the fast and slow wings.}

%\date{Received 6th /
%Accepted <date>}

   \titlerunning{EHV jets in Serpens}
   \authorrunning{Tychoniec et al.}
   \maketitle

\section{Introduction}
Spectacular outflows are one of the crucial signposts of ongoing star formation. Outflows are invoked to release angular momentum, enabling a continuous flow of matter onto the disk and the young star \citep[e.g.,][]{Frank2014}. Their feedback from small to large scales can have a profound impact on the evolution of both the protostar and the entire parent star-forming region \citep[e.g.,][]{Arce2006, Plunkett2013}. Thus, probing the youngest and most powerful outflow sources is crucial for understanding the interactions between the outflows and their surroundings. 

\begin{figure*}[h]
\centering
  \includegraphics[width=0.7\linewidth,trim={2cm 27cm 1cm 0cm},clip]{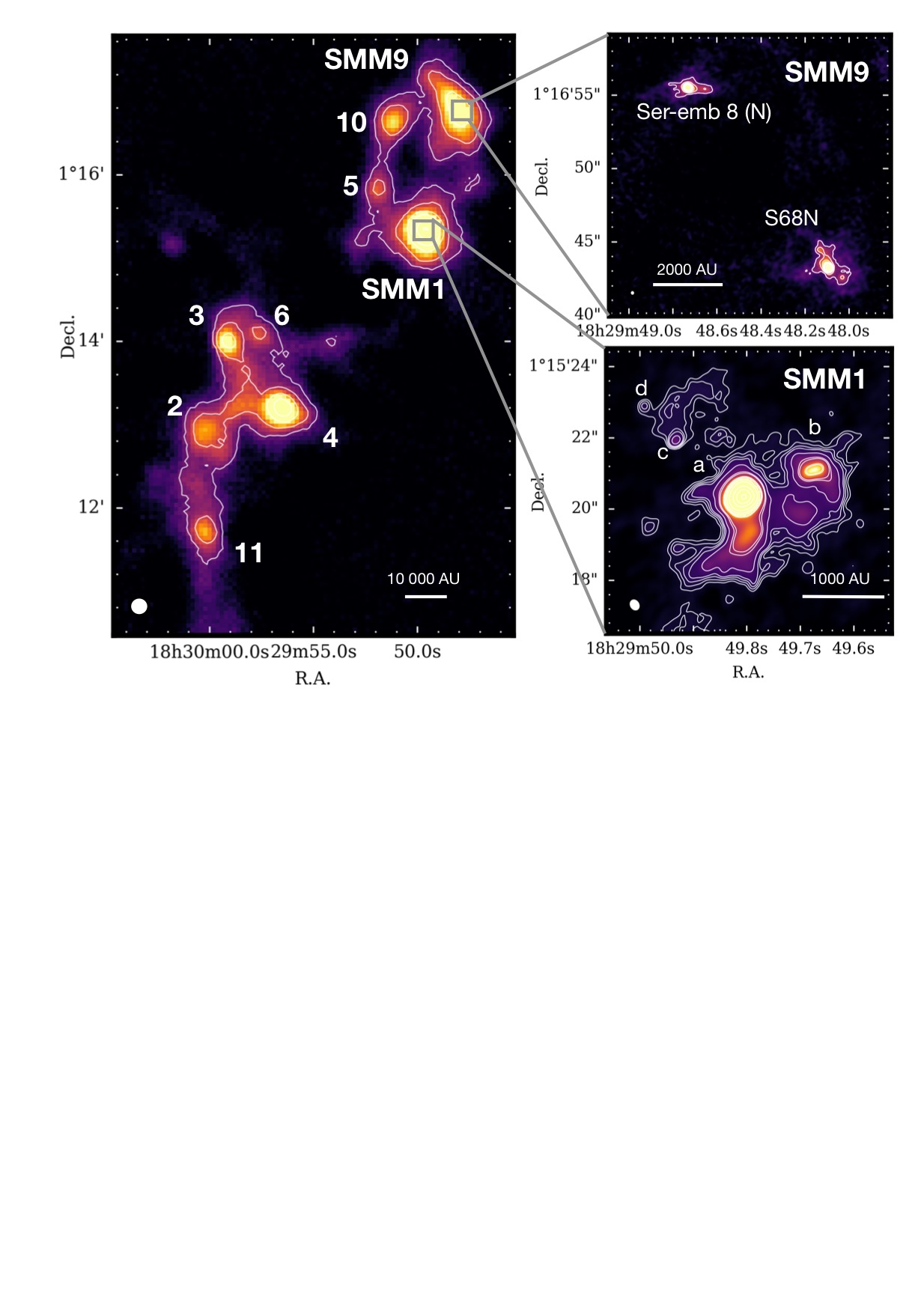}

  \caption{Left: JCMT/SCUBA 850-$\mu$m map of the Serpens Main region with numbers corresponding to SMM sources as classified by \cite{Davis1999}. Contours are [3, 6, 12, 20, 40] $\times$ 0.50 mJy arcsec$^{-2}$. Beam of the JCMT observations of 14 \arcsec\  is indicated in the bottom-left corner. Right: ALMA 1.3 mm continuum of the targeted protostars.  For SMM9 field contours are [3, 6, 9, 12] $\times$ 0.53 mJy beam$^{-1}$ and for SMM1 field contours are [3, 4, 5, 6, 9, 15, 40, 50]  $\times$ 0.62 mJy beam$^{-1}$. Synthesized beams of the ALMA observations are 0\farcs35 $\times$ 0\farcs33 for the SMM9 field and 0\farcs36 $\times$ 0\farcs30 for the SMM1 field.}
 \label{fig:introfig}
\end{figure*}

While the molecular emission from a typical protostellar outflow usually
 appears as slow and wide-angle entrained gas, there is a peculiar group of sources with high-velocity
collimated molecular emission. The extremely high-velocity (EHV)
molecular jets ($\varv>$ 30 km$\rm \ s^{-1}$) are found toward the youngest protostars
\citep[e.g.,][]{Bachiller1990, Bachiller1996} in the Class 0 stage \citep{Andre1993}.
 They were first detected as spectral features, as high-velocity peaks detached from the low-velocity outflow wings \citep{Bachiller1990}, and subsequently spatially resolved as discrete bullets embedded in a cocoon of low-velocity gas \citep[e.g.,][]{Santiago-Garcia2009, Hirano2010}.
%These molecular jetsoften show a clumpy structure in the form of high-velocity
These ‘bullets’ are
thought to arise from the variability of the outflow activity, possibly related to the variability of the accretion processes itself
\citep{Raga1993a}.  In the deeply embedded stage, EHV molecular jets
have been observed at submillimeter wavelengths
\citep[e.g.,][]{Bachiller1994, Tafalla2004}, as well as in far-IR
observations \citep{Kristensen2012, Mottram2014}.
They appear to be
quite rare. In a survey of 29 protostars with {\it Herschel Space Observatory}/HIFI,
water bullets were detected in only four sources, all of them being Class
0 \citep{Kristensen2012}. Thus, EHV jets are thought to be associated
exclusively with very young sources.

Apart from the spatial and spectral characteristics of the EHV jets relative to low-velocity outflows,
it appears that their chemical composition is significantly different
from that of the slow outflow. In observations with the IRAM-30m of
two young outflows with EHV jet components, \cite{Tafalla2010} show
that the molecular jets are more oxygen-rich compared with the slow and the fast wing component of the molecular outflow. The molecular jets are prominently seen in species like SiO
\citep[see also][]{Guilloteau1992}, SO, CH$_3$OH and H$_2$CO, whereas
emission from molecules like HCN and CS, which tend to be present in the slow and the fast wing, 
is missing at the highest velocities. These led  \cite{Tafalla2010}  to define three distinct velocity components: the slow and the fast wing, and the EHV jet (see Sect. \ref{velocity_regimes}). These studies presented spectrally resolved line profiles of different molecules, but their spatial location remains unclear. To date, only CO and SiO have been studied at high spatial resolution within the EHV jets \citep[e.g.,][]{Lee2008, Santiago-Garcia2009, Hirano2010, Codella2014, Hull2016}.
It is still not well understood what the spatial
distribution of other molecules is in the different kinematic
components of the outflow.

Additional important information on molecular jets and outflows comes
from observations with the HIFI instrument \citep{Graauw2010} on board
the {\it Herschel} \citep{Pilbratt2010} on scales of $12\arcsec$ --
$40\arcsec$.  Many water and high-$J$ CO transitions probing warm
shocked gas show complex line profiles that can be decomposed in two
main velocity components. The kinematic and chemical signatures of
those components are universal for all protostars, from low- to high
mass \citep{Kristensen2012, Mottram2014, SanJose-Garcia2016}:   a broad
component (FWHM $>20$ km s$^{-1}$), and an offset component ($20>$
FWHM $>5$ km s$^{-1}$), usually blue-shifted with respect to the
systemic velocity up to a few km s$^{-1}$.The CO excitation
  temperatures in the broad component are typically ~300\,K in the broad component and ~700\,K in the offset component. EHV bullets are also seen in HIFI
line profiles as discrete peaks detached from the main line profile;
 however, as noted above, these only appear in a few sources. The spatial origin
of those components can potentially be revealed with spectrally and
spatially resolved ALMA observations of low-$J$ CO and other
molecules. ALMA's high spatial resolution is needed since the water
analysis suggests that its emission originates from structures that
are only a few hundred au in size, much smaller than the region
encompassed by the HIFI beam at distances of nearby star-forming
regions \citep{Mottram2014}.

Here we target three protostars in the Serpens Main region at a distance of 436 pc \citep{Ortiz-Leon2017}, namely, the Serpens SMM1 (hereafter referred to as SMM1), S68N and Ser-emb~8~(N) protostellar systems.  
SMM1 is border-line between  a low and intermediate mass protostar \citep[100 L$_{\odot}$;][]{Kristensen2012}, and is known to host a massive disk-like structure \citep{Hogerheijde1999, Enoch2010}. The SMM1 source was discovered as a multiple system in the continuum observations \citep{Choi2009} and confirmed by the observations of the atomic jet \citep{Dionatos2014}. More recently, resolving the system with ALMA unveiled a total of 5 protostellar components \citep{Hull2017a} within a 2000 au radius, 3 of which show outflows (labelled a, b and d in Fig. \ref{fig:introfig}).
S68N and Ser-emb~8~(N) are deeply embedded protostars separated by 5000 au (Fig. \ref{fig:introfig}b). Both are powering outflows \citep{Hull2014}. The chemical structure of Serpens Main on cloud scale has been studied in detail by \citet{McMullin1994, McMullin2000,Kristensen2010}. A summary of the sources is provided  in Table \ref{table:1}.

\begin{table*}
\caption{Targeted protostars}             %
\label{table:1}      %
\centering                          %
\begin{tabular}{c c c c c c c c}        %
\hline\hline                 %
Name & Other names & R.A. & Decl. & $L_{\rm bol}$ &   $T_{\rm bol}$ & $M_{\rm env}$ &  Ref. \\
 & & (J2000) & (J2000) & (L$_{\odot}$) &   (K) & (M$_{\odot}$) &   \\

\hline                                   %
Serpens SMM1 & S68FIRS1 (1), Ser-emb 6 (5) & 18:29:49.765 & +1:15:20.506 & 109 &   39  & 58 &  (4)\\
S68N & Ser-emb 8 (5), SMM9 (2) & 18:29:48.087 & +1:16:43.260 & 6 &   58  & 10 &  (5) \\
Ser-emb 8 (N) & S68Nb (6), S68Nc (3) & 18:29:48.731 & +1:16:55.495 & --- &  ---  & --- &  ---\\

\hline                                   %

\end{tabular}
\begin{tablenotes}\footnotesize
\item {(1) \citealt{McMullin1994}, (2) \citealt{Davis1999}, (3) \citealt{Dionatos2010}, (4) \citealt{Kristensen2012}, (5) \citealt{Enoch2009}, (6) \citep{Maury2019}}.
\end{tablenotes}
\end{table*}

ALMA observations of CO\,$2-1$ and SiO\,$5-4$ reveal EHV jets toward the SMM1-a and SMM1-b sources in CO, both asymmetric, with only redshifted emission detected at high velocities. SMM1-b additionally shows EHV emission in SiO \citep{Hull2016, Hull2017a}.

In this paper we use ALMA to resolve both
spectrally and spatially the emission from different molecules,
allowing us not only to distinguish different kinematic components of the
outflows and jets from protostars but also to link them to the
specific physical components of the system, such as entrained gas, outflow cavity walls, or the protostellar jet.

\begin{figure*}[h]
\centering
  \includegraphics[width=0.95\linewidth,trim={1.2cm 17.3cm 2.1cm 0.4cm},clip]{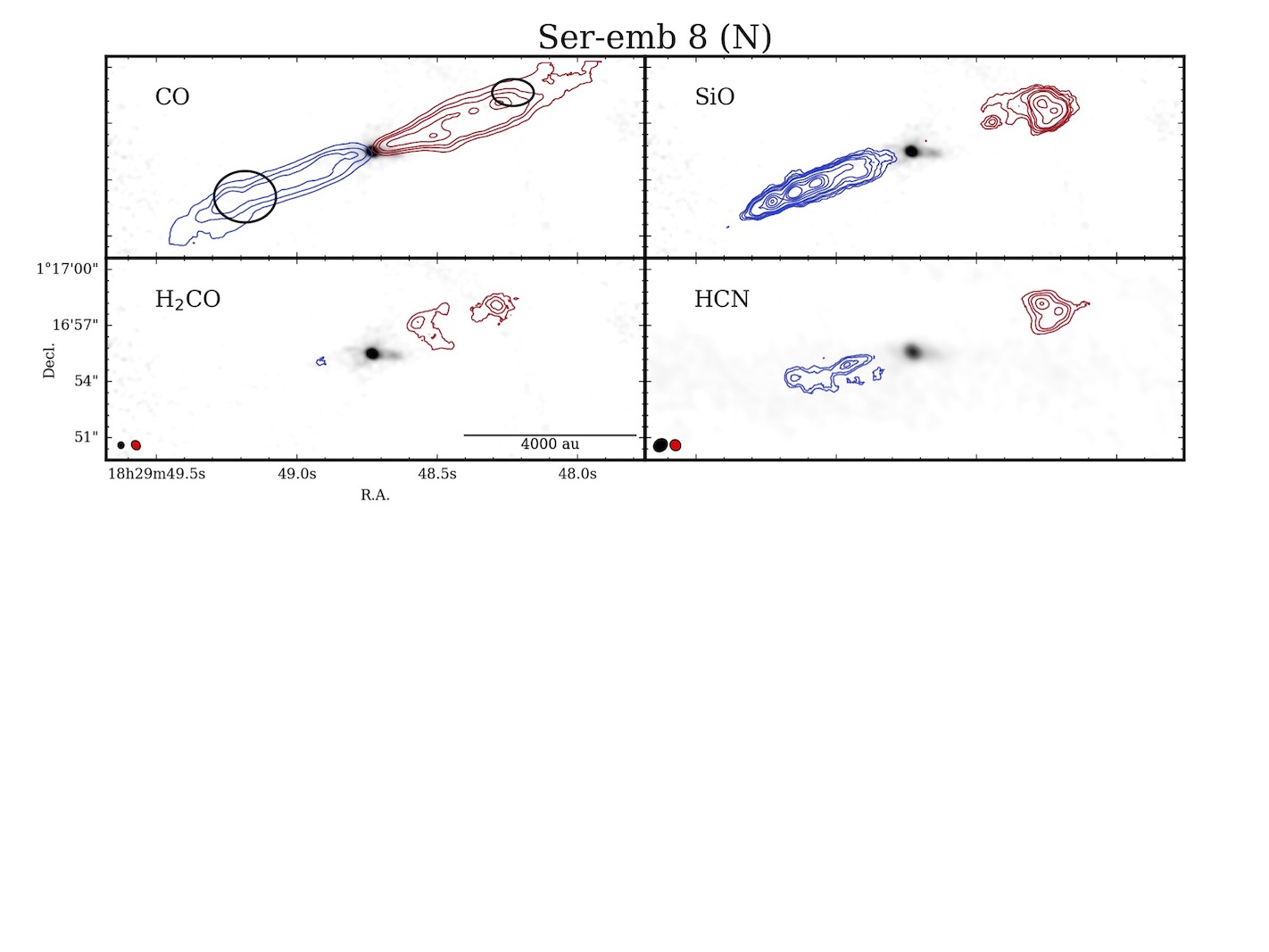}
  \caption{Integrated intensity maps of CO\,$2-1$,  SiO\, $5 - 4$, H$_2$CO\,$3_{03}-2_{02}$, and HCN\,$1-0$  overlaid on the Band 6 (Band 3 for HCN) continuum in grayscale for Ser-emb 8 (N). The emission is integrated from inner boundary of the slow wing component to the outer boundary of the EHV component as listed in Table \ref{tab:velocities} for the red and blueshifted emission. The exception are SiO and H$_2$CO maps where only the channels obtained at high spatial resolution are plotted (< 26 km s$^{-1}$ for H$_2$CO and < 40 km s$^{-1}$ for SiO). The synthesized beam size of the continuum images is 0\farcs35 $\times$ 0\farcs33  for Band 6 and 0\farcs79 $\times$ 0\farcs64 for Band 3; for spectral lines it is  0\farcs53 $\times$ 0\farcs45 (CO),  0\farcs55 $\times$ 0\farcs45 (SiO), 0\farcs53 $\times$ 0\farcs44 (H$_2$CO), and 0\farcs60 $\times$ 0\farcs56  (HCN). The beam size of the Band 6 spectral line  is presented in bottom-left corner of the H$_2$CO map and in HCN map for Band 3. Contour levels are [3, 6, 9, 15, 20, 40, 60, 80, 100] for CO, SiO, H$_2$CO, and redshifted HCN, and [2, 3, 5, 6, 12] for blueshifted HCN, multiplied by rms value of moment 0 maps. The rms values for blueshifted and redshifted, in K km s$^{-1}$:  CO [19.7, 14.4], SiO [2.2, 2.5], H$_2$CO [2.8, 2.1], and HCN [9.3, 12.2]. Black ellipses indicate regions from which spectra were extracted for Fig. \ref{fig:velregime_spectra} and  \ref{fig:velregimes_app}. }
 \label{fig:allmolecules_Emb8N}
\end{figure*}

\section{Observations}
ALMA observations of four molecular transitions, CO\,$2-1$, SiO\,$5-4$, H$_2$CO \,$3_{03}-2_{02}$ in Band 6 (ALMA project 2013.1.00726.S; PI: C. Hull) and HCN\,$1-0$ observed in Band 3 (ALMA project 2016.1.00710.S; PI: C. Hull) are presented. The synthesized beam of the observations is between $\sim$\,$0\farcs3$  and $\sim$\,$0\farcs6$, corresponding to 130 -- 260 au at the distance to Serpens Main. The largest recoverable scale in the data is  $\sim$\,$5\arcsec$ and $\sim 12\arcsec$ (2150 and 4960 au) for Band 3 and Band 6, respectively. The spectral resolution of the observations differs between the spectral windows, ranging from 0.04 to 0.3 km s$^{-1}$. For both bands, only 12-m array data were used. The Band 6 data were obtained in two configurations (C43-1 and C43-4 with resolutions of 1\farcs1 and 0\farcs3, respectively), and the final images are produced from the combined datasets. 

After obtaining the C43-4 configuration data, it became apparent that SiO and H$_2$CO emission is present at velocities extending further than the spectral setup. To capture the emission at high-velocities, the spectral configuration for SiO and H$_2$CO was changed for the compact C43-1 configuration. Thus the SiO and H$_2$CO emission at highest velocities  ($> 40 $ km s$^{-1}$ for SiO and $> 25 $ km s$^{-1}$ for H$_2$CO in both redshifted and blueshifted direction with respect to the systemic velocity of 8.5 km s$^{-1}$) are available only at lower spatial resolution. 

Continuum images were obtained from the dedicated broadband spectral windows and line-free channels. Self-calibration on continuum data was performed, and solutions were transferred to the emission line measurement sets. The line data were then continuum subtracted. The imaging was performed with the CASA 5.1.0 \citep{McMullin2007} \textit{tclean} task with masked regions selected by hand for each line. Data were imaged with Briggs weighting = 0.5 and re-binned to 0.5 km s$^{-1}$. Due to the large extent and complicated structure of the emission lines, the multiscale option in \textit{tclean}  was used for the lines, with scales manually adjusted for each line. Information about the observations is summarized in Table \ref{table:2}.

\section{Results}
\subsection{Images of outflows}

\begin{figure*}[h]
\centering
\includegraphics[width=0.95\linewidth,trim={0cm 15cm 0cm 11.2cm},clip]{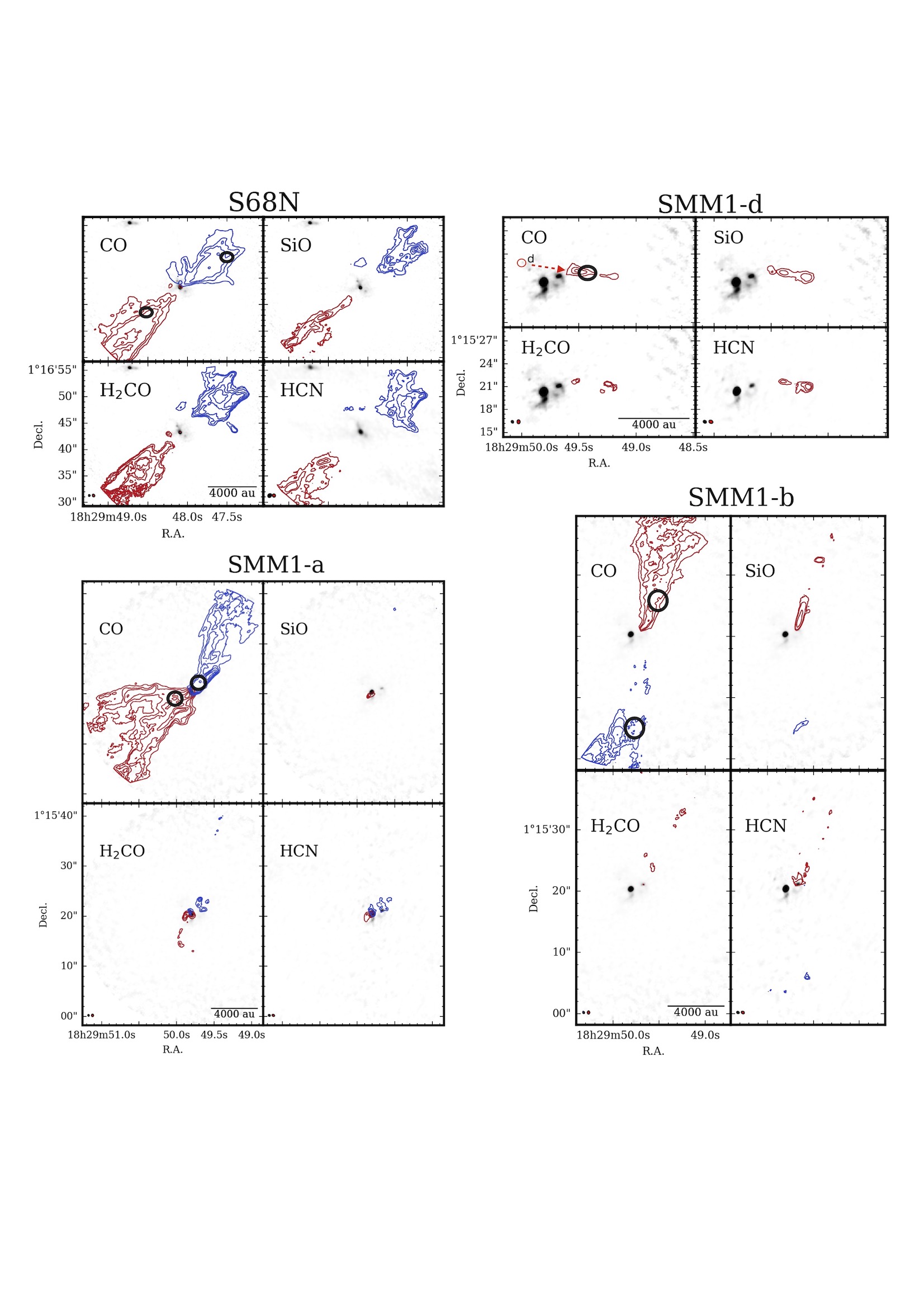}  
  \caption{Similar to  Fig.  \ref{fig:allmolecules_Emb8N} but for the remaining sources. S68N: Contour levels are [3, 6, 9, 15, 20, 40, 60, 80, 100] for CO and HCN; [3, 8, 15, 30, 45] for SiO and [3, 5, 9, 15, 20, 40] for H$_2$CO, multiplied by rms value of moment 0 maps. The rms values for blueshifted and redshifted, in K km s$^{-1}$:  CO [19.5, 14.1], SiO [1.6, 1.9], H$_2$CO [3.2, 2.0], and HCN [9.4, 12.7].  SMM1-a: Contour levels are [3, 6, 9, 15, 20, 40, 60, 80, 100] for all molecules, multiplied by rms value of moment 0 maps. The rms values for blueshifted and redshifted, in K km s$^{-1}$:  CO [20.2, 20.6], SiO [3.6, 4.0], H$_2$CO [2.0, 2.9], and HCN [7.5, 11.5]. SMM1-b: Contour levels are [3, 6, 9, 15, 20, 40, 60, 80, 100] for CO, [3, 9, 36] for SiO, and [3, 5] for H$_2$CO and HCN, multiplied by rms value of moment 0 maps. The rms values for blueshifted and redshifted, in K km s$^{-1}$:  CO [18.7, 20.3], SiO [3.6, 4.0], H$_2$CO [1.9, 2.9], and HCN [7.4, 11.5]. SMM1-d: Only redshifted moment 0 map is presented as no blueshifted component has been detected toward this source. Contour levels are [3, 6, 9, 15, 20, 40, 60, 80, 100] for CO and HCN,  [3, 12, 36] for SiO, and [2, 3] for H$_2$CO, multiplied by rms value of moment 0 maps. The rms values in K km s$^{-1}$:  CO [20.1], SiO [3.3], H$_2$CO [2.7], and HCN [9.1].  Black ellipses indicate regions from which spectra were extracted for Fig. \ref{fig:velregimes_app}.}
   \label{fig:allmolecules_S68N}
\end{figure*}

The highest resolution and sensitivity observations of the S68N and Ser-emb~8~(N) molecular outflows taken to date are presented here. For SMM1, H$_2$CO and HCN emission is shown in addition to the CO and SiO outflow presented in previous papers \citep{Hull2016, Hull2017a}. 
Figures \ref{fig:allmolecules_Emb8N} and\ref{fig:allmolecules_S68N} show the integrated emission maps of CO, SiO, H$_2$CO and HCN for all five sources. Various other molecules were detected as well in the ALMA observations \citep[e.g., DCO$^+$, C$^{18}$O, and complex organic molecules;][]{Tychoniec2018c}. Those molecules trace either the cold quiescent envelope or the warm inner envelope, but do not show the outflow components; thus, they are not further discussed here.

Ser-emb~8~(N) (Fig. \ref{fig:allmolecules_Emb8N}) shows a relatively symmetric outflow morphology in CO. It has a very small opening angle of 25$^\circ$, measured as an angle between the outflow cavity walls seen at the low-velocity CO. SiO emission toward this source traces both the central, most collimated part of the outflow, and the bow-shock structure at the redshifted part of the outflow, seen clearly also in the HCN. The structure is not so clear on the blueshifted side, although HCN is present mostly off the main axis of the outflow there, while there is no clear evidence for a blueshifted bow-shock from SiO emission. H$_2$CO is enhanced at the bow-shock position in the redshifted part of the outflow.

 S68N has an outflow with a wide opening angle of 50$^\circ$, although the cavity walls do not seem well defined for this source (Fig. \ref{fig:allmolecules_S68N}). The morphology of the outflow is similar in all molecules, but it can be noticed that peaks of the SiO emission generally appear in regions with weaker CO emission. There seems to be a narrow  on-axis ridge on the redshifted side of the S68N outflow where both SiO and HCN emission peaks, in contrast to H$_2$CO, which emits mostly off-axis.

The SMM1-a outflow has an asymmetric structure in CO, with blue- and redshifted lobes misaligned with respect to each other (30$^\circ$ difference in position angles) and having different opening angles: 65$^\circ$ and 35$^\circ$ for red- and blueshifted sides, respectively (Fig. \ref{fig:allmolecules_S68N}).
Other molecules are seen close to the protostar rather than throughout the full extent of the outflow, for example, SiO is found only very close to the protostar and only on the redshifted side and H$_2$CO and HCN are seen tracing the innermost regions of the outflow with irregular morphologies.

SMM1-b has an outflow with consistent position angles on both sides, but the redshifted part is much brighter in both CO and SiO (Fig. \ref{fig:allmolecules_S68N}). The CO outflow has a moderate opening angle of 45$^\circ$; the blueshifted part of the SiO emission is only detected several thousands of au away from the source as a clump of emission, very different from the bright, highly-collimated structures with several well-defined bullets on the redshifted side of the jet. HCN and H$_2$CO are only faintly detected toward SMM1-b at low-velocities.

The SMM1-d outflow has a peculiar morphology (Fig. \ref{fig:allmolecules_S68N}); the redshifted side is seen in three distinct clumps starting as far as 3000 au away from the SMM1-d protostar \citep{Hull2017a}, while no blueshifted side is observed. The CO emission peaks at the nearest clump while the SiO, HCN, and H$_2$CO are peaking in the most distant one.

\subsection{Velocity regimes}
\label{velocity_regimes}

\begin{figure}[h]
\centering
 \includegraphics[width=0.9\linewidth,trim={1.2cm 0cm 0.4cm 2.3cm},clip]{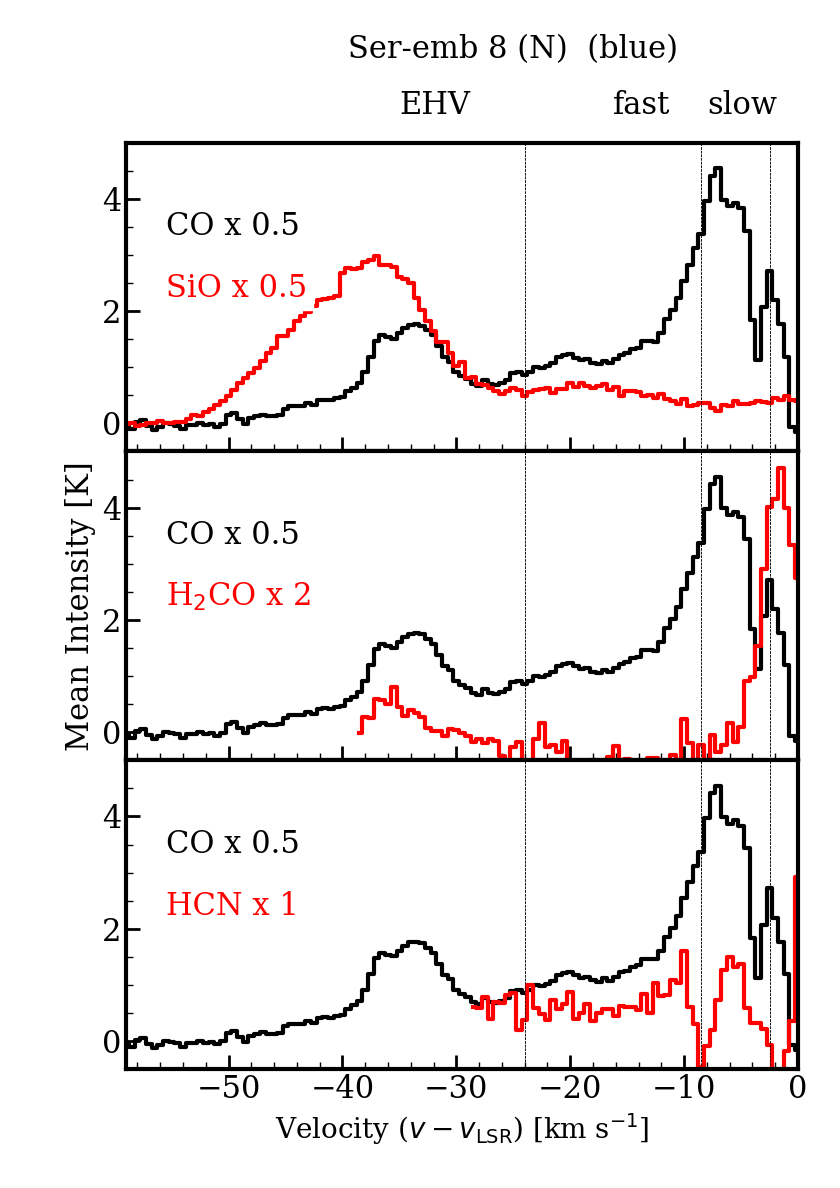}
 
\caption{Spectra of CO (black) and SiO, H$_2$CO and HCN (red) for the selected part of the blueshifted part of Ser-emb 8 (N) outflow, indicated in the Fig. \ref{fig:allmolecules_Emb8N}. The dashed lines show boundaries between different velocity components. Full set of spectra for the other sources is shown in the Appendix (Fig. \ref{fig:velregimes_app}).}
\label{fig:velregime_spectra}
\end{figure}

The high spectral resolution and high sensitivity observations of ALMA allow analysis of the different velocity components present in the outflows.  \cite{Tafalla2010} define three velocity components in molecular outflows; the slow wing is seen as a typical Gaussian profile
and the fast wing shows up as a broad component added to this profile; the transition between the two is smooth. The extremely high-velocity (EHV) component appears as a discrete peak at high velocities and is clearly separated from the wing profile.

\begin{figure}[h]
\centering
  \includegraphics[width=1.0\linewidth,trim={1cm 16.5cm 0 0cm},clip]{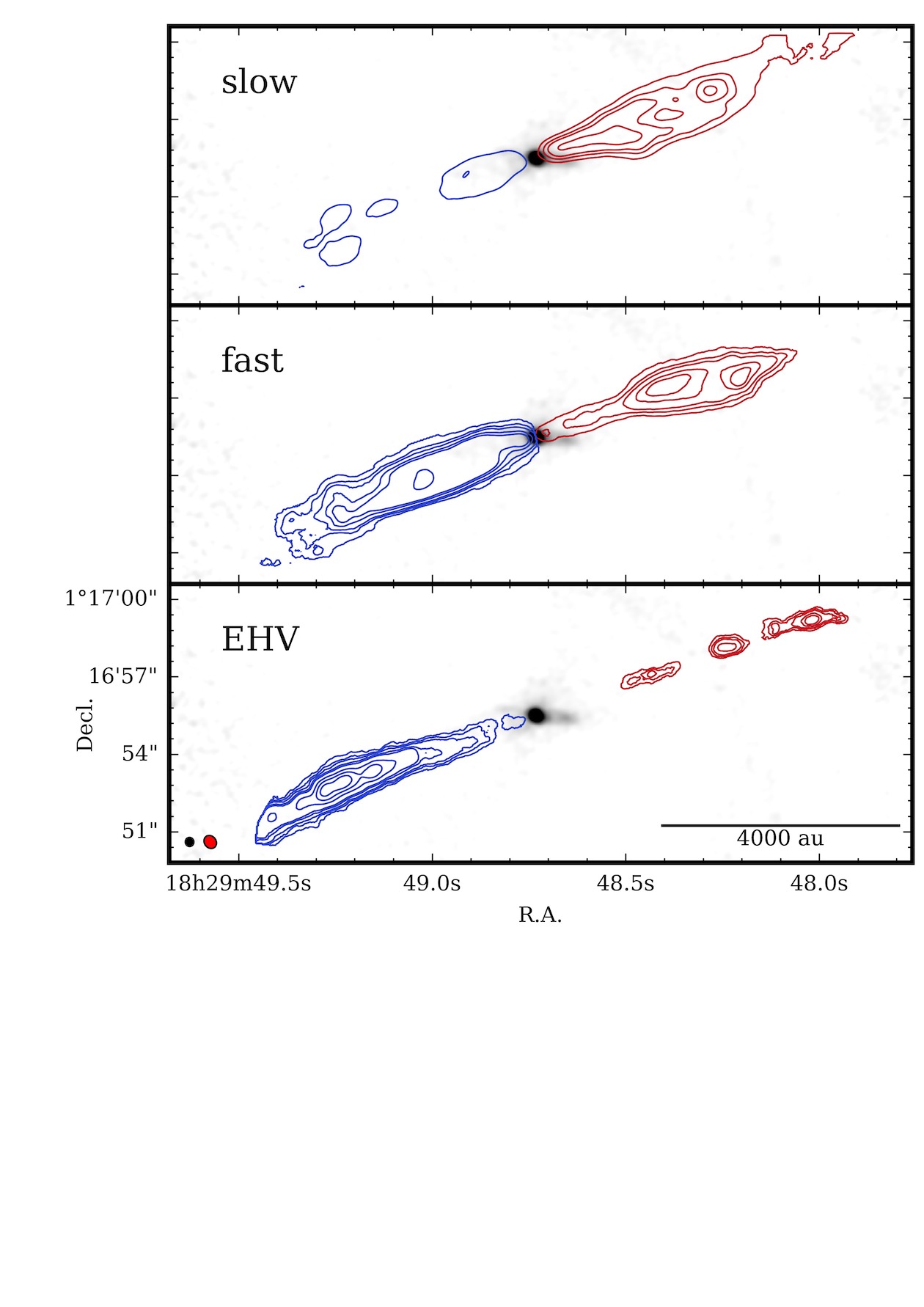}
  \caption{Integrated intensity maps of CO for different velocity regimes overlaid on the Band 6 continuum in grayscale for Ser-emb~8~(N). The emission is integrated over the velocities listed in Table \ref{tab:velocities}. The  synthesized beams of the CO (red) and continuum (black) are showed in bottom-left corner of EHV plot with sizes 0\farcs35 $\times$ 0\farcs33  and 0\farcs55 $\times$ 0\farcs45 for continuum and CO, respectively. The contours are [3, 6, 9, 15, 20, 40, 60, 80, 100] times the rms value. The rms values for each velocity channel, blueshifted and redshifted in K km s$^{-1}$, are slow [18.3, 13.7], fast [3.1, 4.5], EHV [1.7, 1.4].}
 \label{fig:velregimes_Emb8N}
\end{figure}

To define boundaries between the velocity regimes, especially to distinguish the slow from the fast wing, the examination of multiple molecules is needed. To avoid including the emission from the cold envelope in the measurement of the flux from the outflow, even though most of the envelope emission should be resolved out, C$^{18}$O spectra obtained within the Band 6 observations have been used to set constraints on possible contamination by the envelope emission in the outflow measurements. Spectra of C$^{18}$O of regions outside the outflow positions were used to assess by eye the velocity at which C$^{18}$O is still significant. Those values are set as the inner velocity limit for the slow wing.  

\cite{Tafalla2010} identify the transition between slow and fast wing by a decrease of intensity of H$_2$CO emission and enhancement of SiO and HCN, relative to CO; where possible, the same criteria are used here. Defining the EHV regime is more straightforward as it is the beginning of the increasing CO and SiO flux at high velocities. Figure \ref{fig:velregime_spectra} shows  spectra used to define the velocity regimes in Ser-emb~8~(N). Table \ref{tab:velocities} summarizes the velocity borders defined for each source. 
\begin{table*}
\centering
\caption{Boundary velocities of different components}
\label{tab:velocities}
\begin{tabular}{lcccccccc}
\hline \hline
 & &  blue & &  &  &  red  &  &  \\
 \cline{2-4}
 \cline{6-8}
 Source & EHV & fast & slow & & slow  & fast & EHV  \\
   & ($\mathrm{km\,s^{-1}}$)  & ($\mathrm{km\,s^{-1}}$) & ($\mathrm{km\,s^{-1}}$) &  & ($\mathrm{km\,s^{-1}}$)    & $\mathrm{km\,s^{-1}}$ & ($\mathrm{km\,s^{-1}}$)   \\

\hline
SMM1-a & --- & [-35,-8] & [-8, -1.5] & & [2, 12]  & [12, 50] & [50, 80] \\
SMM1-b &[-36, -29] & [-29, -8.5] & [-8.5, -2] & & [2, 9] & [9, 25] & [25, 56] \\
SMM1-d & --- & --- & --- & & [2, 7] & [7, 29] & --- \\
S68N & --- & [-22, -14] &  [-14, -2] & & [2,5, 12] & [12, 25] & --- \\
Ser-emb 8(N) &[-62,-24] & [-24, -8.5] & [-8.5, -2.5] &  & [2.5, 13.5] & [13.5, 35] & [35, 58] \\
\hline
\end{tabular}

\tablefoot{Velocities are given after subtracting the systemic velocity of the cloud $\varv_{\rm{lsr}} = 8.5$ km s$^{-1}$.
}

\end{table*}
 
Out of the five outflow sources observed, the EHV component is detected toward three sources. This is remarkable, as it is considered to be a rare phenomenon.
The new detection of the Ser-emb~8~(N) high-velocity molecular jet, along with further analysis of EHV jets toward SMM1-a and SMM1-b \citep{Hull2016,Hull2017a}, is presented here. 

Figure \ref{fig:velregimes_Emb8N} shows intensity maps of CO ($2-1$) integrated over velocity regimes defined in the previous section. %
Ser-emb~8~(N) has a high degree of symmetry between red and blueshifted emission at high velocities, with several peaks of emission, occurring at similar distances from the protostar on both sides. Three main clumps of EHV emission can be distinguished at 1500, 4000, and 6000 au away from the central protostar, although each of those clumps can be split into a more complex structure.
A similar bullet-like structure is observed toward the SMM1-b source in its redshifted jet, with bullets at roughly 1000, 3000, 5000, and 7000 au. The redshifted bullets seem to have only a single blueshifted counterpart - the furthermost EHV component at $\sim 7000$ au (Fig.  \ref{fig:velregimes_SMM1b}).

The EHV component from SMM1-a is very different from that of the first two jets described. It resembles a continuous stream emerging very close to the protostar, rather than forming discrete bullets. Hints of redshifted EHV emission further away are present as far as 7000 au from the protostar, although significantly off-axis compared with the stream close to the protostar; this may suggest precession, as discussed by \cite{Hull2016}. No corresponding blueshifted EHV emission is seen toward this source, in contrast to the slow and fast wing gas (Fig.  \ref{fig:velregimes_SMM1a}).

S68N shows no signs of the EHV component. (Fig.  \ref{fig:velregimes_S68N}). In the case of the SMM1-d outflow (Fig.  \ref{fig:velregimes_SMM1d})  it is difficult to assign the velocity components described above because almost all emission is confined to the low-velocity stream. SiO and HCN seem to follow CO in the spectral profile, and no enhancement is seen at higher velocities, but the CO profile appears broad and therefore slow and fast wing components are assigned. EHV emission is not present toward this source.

\subsection{Chemical abundances in velocity components}  \label{chemical_abundances}

Probing the composition of the wind at different velocities can shed light on physical conditions within the outflows, as a change in velocity also triggers a change in temperature and density. Moreover, a contrast between the chemical composition of wing and jet components can also point to a different physical origin of the outflowing gas \citep{Tafalla2010}, and thus help to understand the mechanism of the EHV jet formation and its interaction with entrained and quiescent gas. 

\subsubsection{Analysis method}

The emission from each pixel inside a region defined by hand was summed in order to measure the abundances in each flow.  The region was defined based on the extent of the low-velocity CO emission, for the red- and blueshifted parts of the outflow separately. These regions were then consistently used for all molecules and all velocity regimes.
We calculate an integrated intensity of every pixel within the region, with the integration going from fixed $\varv_{\rm in}$ to $\varv_{\rm out}$ specified for each velocity regime (see Table \ref{tab:velocities}).

Assuming that the emission is optically thin, the column density of the molecule in each pixel is computed as:
\begin{equation}
\frac{N_u}{g_u}=\frac{\beta \nu^2\int T(\varv) d\varv}{A_{ul}}   \,\,,
\label{eq:1}
\end{equation}
where $\beta = 8\pi k/hc^2$, $\nu$ is frequency, $A_{ul}$ is the Einstein coefficient of a transition, $g_u$ is the degeneracy of the transition, and $T(\varv)$ is an intensity of the emission in Kelvin in a single channel of velocity - $\varv$, with $d\varv$ being a width of a channel. For a given excitation temperature the column density of the molecule in a pixel is then:
\begin{equation}
N_{\rm tot} = N_u \times Q(T)\Big[g_{u}e^{-E_{u}/kT}\Big]  \,\,,
\end{equation}
where $Q(T)$ is the partition function at the assumed excitation temperature. Since only a single transition of each molecule was observed, it is not possible to derive an excitation temperature from these data. The CO excitation temperature is set to $~75 $ K, based on statistics of excitation temperatures for low-mass protostars \citep{Yildiz2015, Kempen2009} which show that the bulk of the low-$J$ CO emission can be fitted with this value.

Assessment of the excitation temperatures for other molecules is not straightforward. \cite{Tafalla2010} performed an LTE analysis of all  molecules included in this work for several transitions and obtained a very low values of $T_{\rm ex}$ of $\sim 7$~K. However, their  analysis was performed using low-energy transitions. \cite{Nisini2007} showed, based on SiO observations for a broader range of $E_{\rm up}$, that the conditions in the outflow may exhibit much higher kinetic temperatures. Their work showed an increase in temperature (up to 500 K) and density (up to $10^6$ cm$^{-3}$) for the high-velocity jet, consistent with the values derived from CO {\it Herschel} data \citep{Karska2018}. For SiO, H$_2$CO, and HCN we ran RADEX \citep{Tak2007}  calculations to constrain excitation temperatures under the conditions expected in the protostellar outflow (n$_{\rm H2}$ = $10^4$ -- $10^6$ cm$^{-3}$; T$_{kin}$ = 75 -- 700 K; $\Delta \varv = 10$ km s$^{-1}$). The extreme excitation temperatures found this way (low and high, see the column $T_{\rm ex}$ in Table \ref{table:4}) are used to calculate the column densities and associated uncertainties for those molecules. The excitation temperatures of the SiO, H$_2$CO and HCN are lower than the expected kinetic temperatures, as the critical density of the transitions are high, see column $n_{\rm crit}$ in Table \ref{table:4}. The low critical density of the CO transition justifies the assumption that its excitation temperature is equal to the kinetic temperature.

\begin{table*}
\caption{Outflow molecules}             % title of Table
\label{table:4}      % is used to refer this table in the text
\centering                          % used for centering table
\begin{tabular}{c c c c c  c c c c c  c c c c c c c c c }        % centered columns (4 columns)
\hline\hline                 % inserts double horizontal lines
& & & & & &  SMM1 &  & & Emb8 &  \\
\cline{7-8}
\cline{10-11}
Molecule & J$_{\rm U}$ - J$_{\rm L}$ & Frequency &   $n_{\rm crit}^{a}$& E$_{\rm up}$ & T{\rm$_{\rm ex}$} & Beam & RMS &  & Beam & RMS  \\    % table heading 
 &  &[GHz] & $[{\rm cm}^{3}$] &  [K] & [K] & & [mJy bm$^{-1}$] & & & [mJy bm$^{-1}$]  \\    % table heading 
\hline                        % inserts single horizontal line
CO &  2-1 & 230.538 &  2.7 x $10^{3}$  & 16.6 & 75 -- 700 & 0\farcs53 $\times$ 0\farcs43 & 3.2 & & 0\farcs54 $\times$ 0\farcs45 & 2.5\\
SiO &  5-4 & 217.104 &  1.7 x $10^{6}$ & 31.3 & 9 -- 47 & 0\farcs54 $\times$ 0\farcs43 & 4.8 & & 0\farcs55 $\times$ 0\farcs45 & 3.5 \\
H$_2$CO & 3(0,3)-2(0,2) & 218.222 &  4.7 x $10^{5}$ & 21.0 & 8 -- 46 & 0\farcs54 $\times$ 0\farcs42 & 4.1 &  & 0\farcs54 $\times$ 0\farcs45 & 3.4 \\ 
HCN & 1-0 & 88.631 & 2.3 x $10^{5}$ & 4.3 & 12 -- 41 & 0\farcs54 $\times$ 0\farcs41 & 2.3 & & 0\farcs60 $\times$ 0\farcs56 & 3.5  \\%Falgarone book
\hline                                   %inserts single line
\end{tabular}
\begin{tablenotes}\footnotesize
\item {  $^a$ Critical densities from \citep{Jansen1995} calculated in the optically thin limit for T{$_{kin}$} }
\end{tablenotes}
\end{table*}

Optically thin emission is assumed for all the molecules. SiO
emission has been suggested to be optically thick for the outflowing
gas \citep{Lee2008, Cabrit2012}.  Our calculations with RADEX show
  that within the conditions expected in the outflows, the SiO 5--4
  emission reaches $\tau \sim 0.1$ only for high gas densities n$_{\rm H2}$ = $10^6$ cm$^{-3}$ at low temperatures T$_{kin}$ = 75 K for the column densities inferred here (Section \ref{column_densities}; Tables \ref{tab:column_denisties_Emb8N}-\ref{tab:column_denisties_SMM1d}. High optical depths are found with our RADEX calculations only for much narrower linewidths,  but all the lines observed within our sample are broad. 
  
The H$_2$CO can become optically thick for high T$_{kin}$ = 700 K; regardless of gas density. Therefore if the emission is coming from the highest velocity material, the abundance of H$_2$CO may be underestimated. For the column densities we infer HCN 1--0 emission seems to be optically thick regardless of the conditions in the shock, and thus abundances of this molecule should be treated as lower limits.

For CO, our RADEX calculations show that $\tau \sim 0.3$  for the low-velocity gas with T$_{\rm kin}$ $\sim 75 $ K.   \cite{Dunham2014} suggest that CO lines can become optically thick at low velocities ($<2$km s$^{-1}$). By excluding channels at the lowest velocities using C$^{18}$O as a tracer of the dense gas, we probe mostly the optically thin gas, as the opacity rapidly decreases with velocity for CO wings \citep{Yildiz2015,Marel2013,Zhang2016}.

\subsubsection{Column densities and abundances}
\label{column_densities}
  
After calculating the column density in each pixel, the average of the column density within the pre-defined region is calculated from only those pixels with signal above 3 $\sigma$.
Calculated values for each molecule  are summarized in Tables \ref{tab:column_denisties_Emb8N}-\ref{tab:column_denisties_SMM1d},
where the boundary values calculated for the min and max T$_{\rm ex}$ are reported. 

\begin{figure}[h]
\centering
  \includegraphics[width=0.98\linewidth,trim={0.5cm 42cm 0 1.7cm},clip]{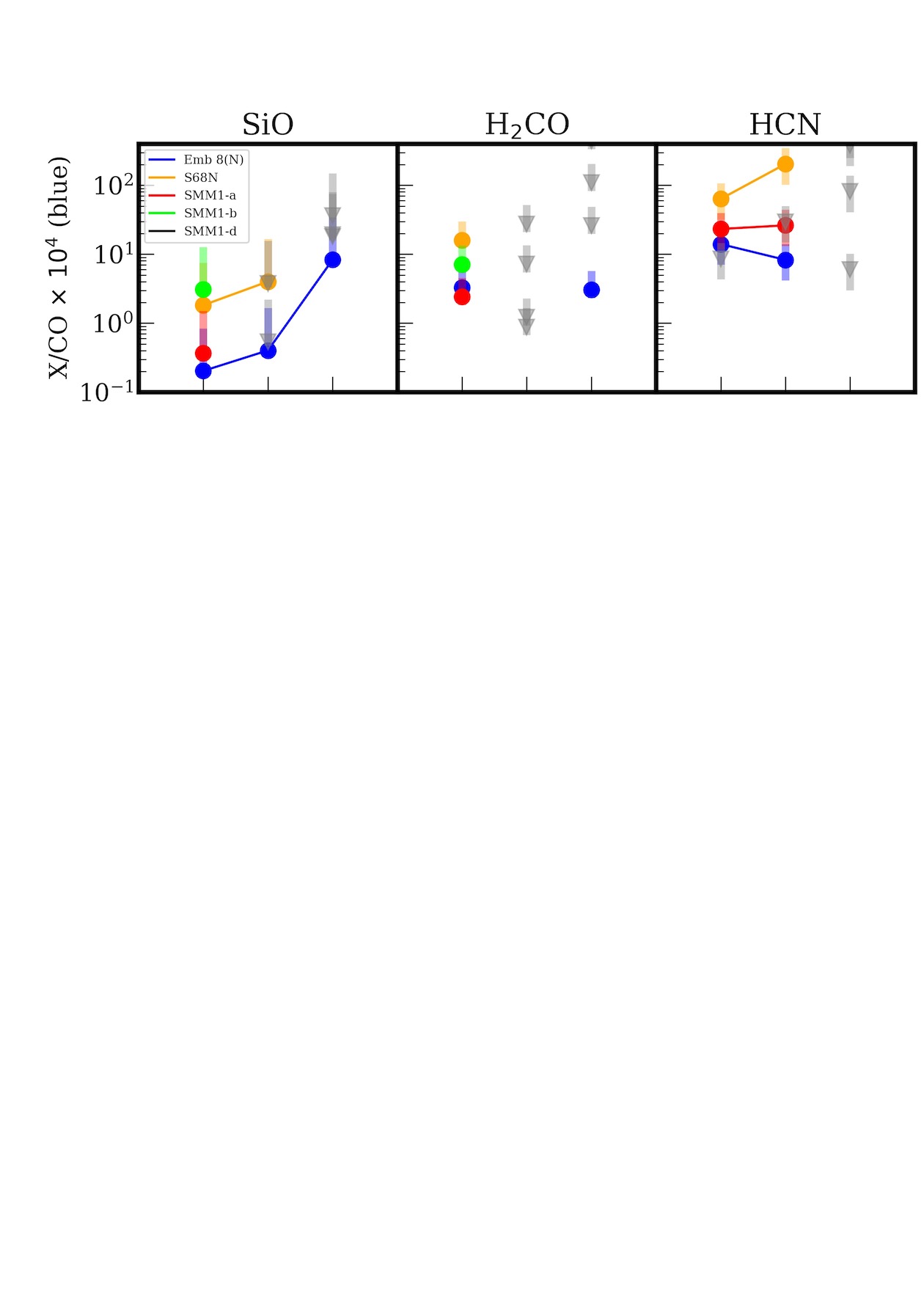}
   \includegraphics[width=0.98\linewidth,trim={0.5cm 42cm 0 5cm},clip]{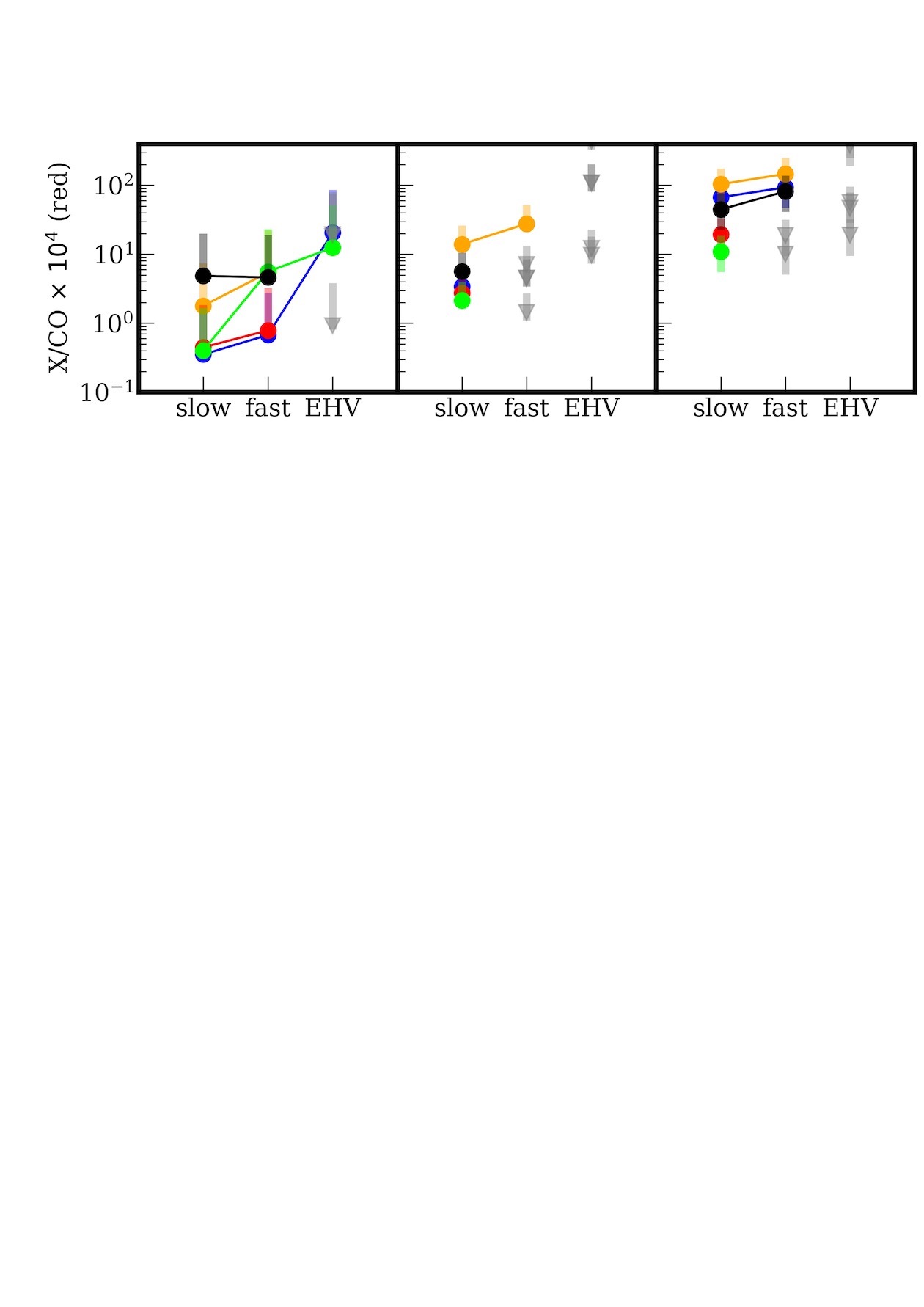}
   \caption{Molecular abundances with respect to CO scaled by $10^4$ for blueshifted (top) and redshifted (bottom) part of the outflow for all sources. Grey triangles represent upper limits.
Points on the plot show values calculated for the mean T$_{\rm ex}$ of the range defined for each molecule, see Table \ref{table:4}. Error bars represent the column densities calculated for min and max values of the excitation temperature. To obtain the abundance of the given molecule, the column density is divided by the CO column density (for T${\rm_{ex}}$ = 75 K.) measured in the region in which the emission from the molecule was above 3$\sigma$. The HCN emission is likely optically thick and therefore the abundance should be treated as a lower limit.}
 \label{fig:abundances_ratio_total}
\end{figure}

\begin{figure}[h]
\centering

  \includegraphics[width=0.95\linewidth,trim={ 1.0cm 44cm 0 1.9cm},clip]{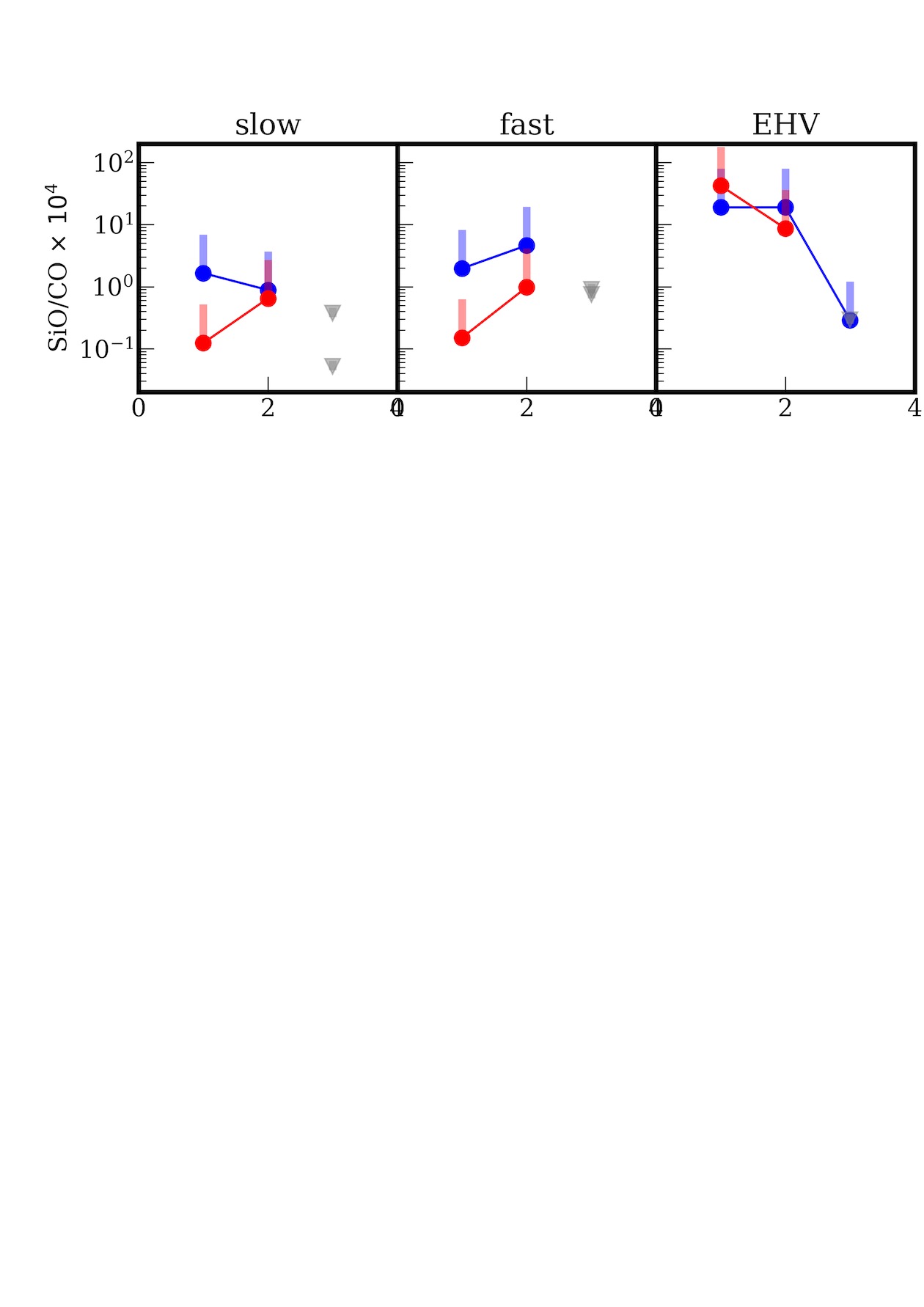}
    \includegraphics[width=0.95\linewidth,trim={ 1.0cm 44cm 0 3.0cm},clip]{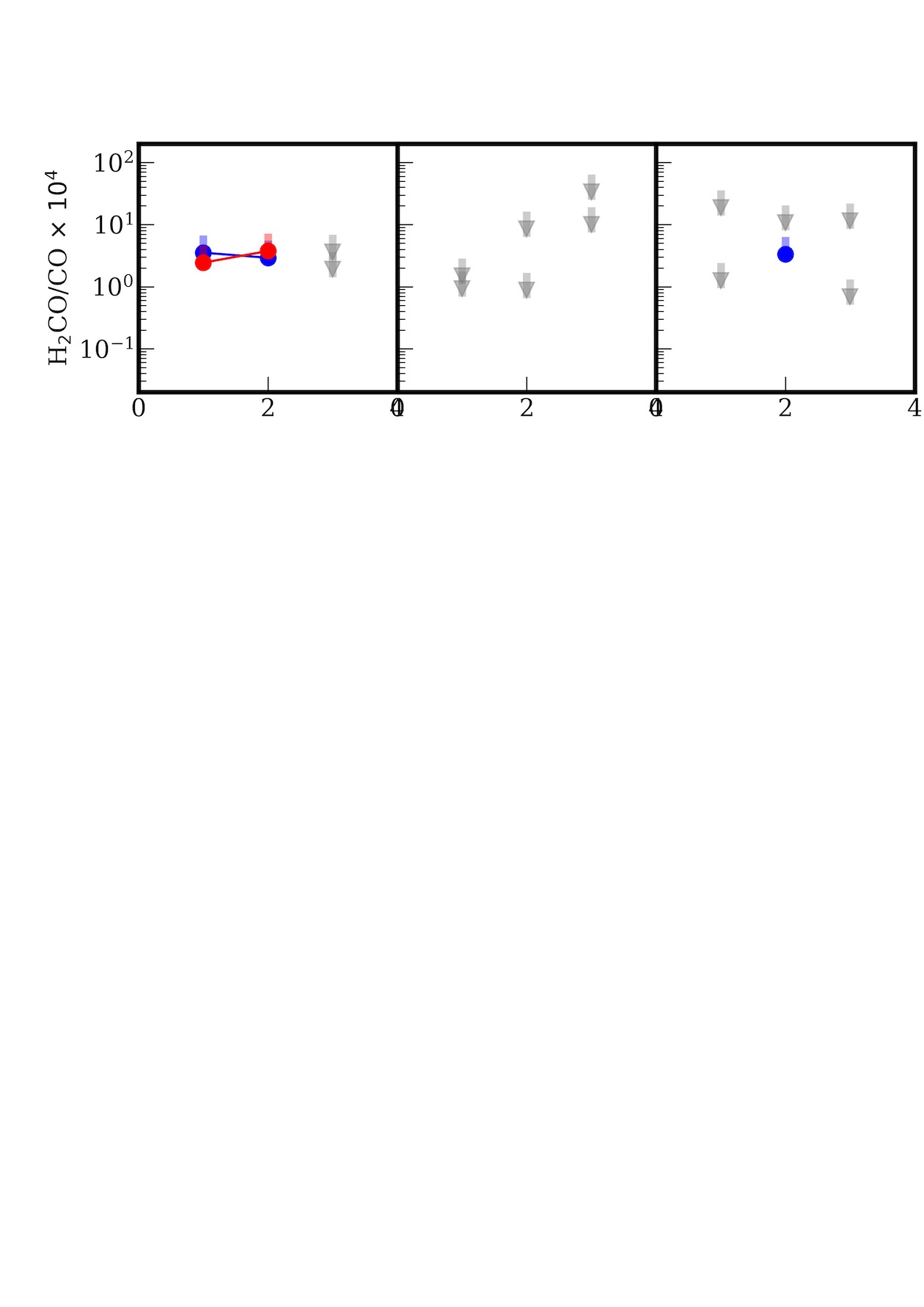}
      \includegraphics[width=0.95\linewidth,trim={ 1.0cm 41.5cm 0 3.0cm},clip]{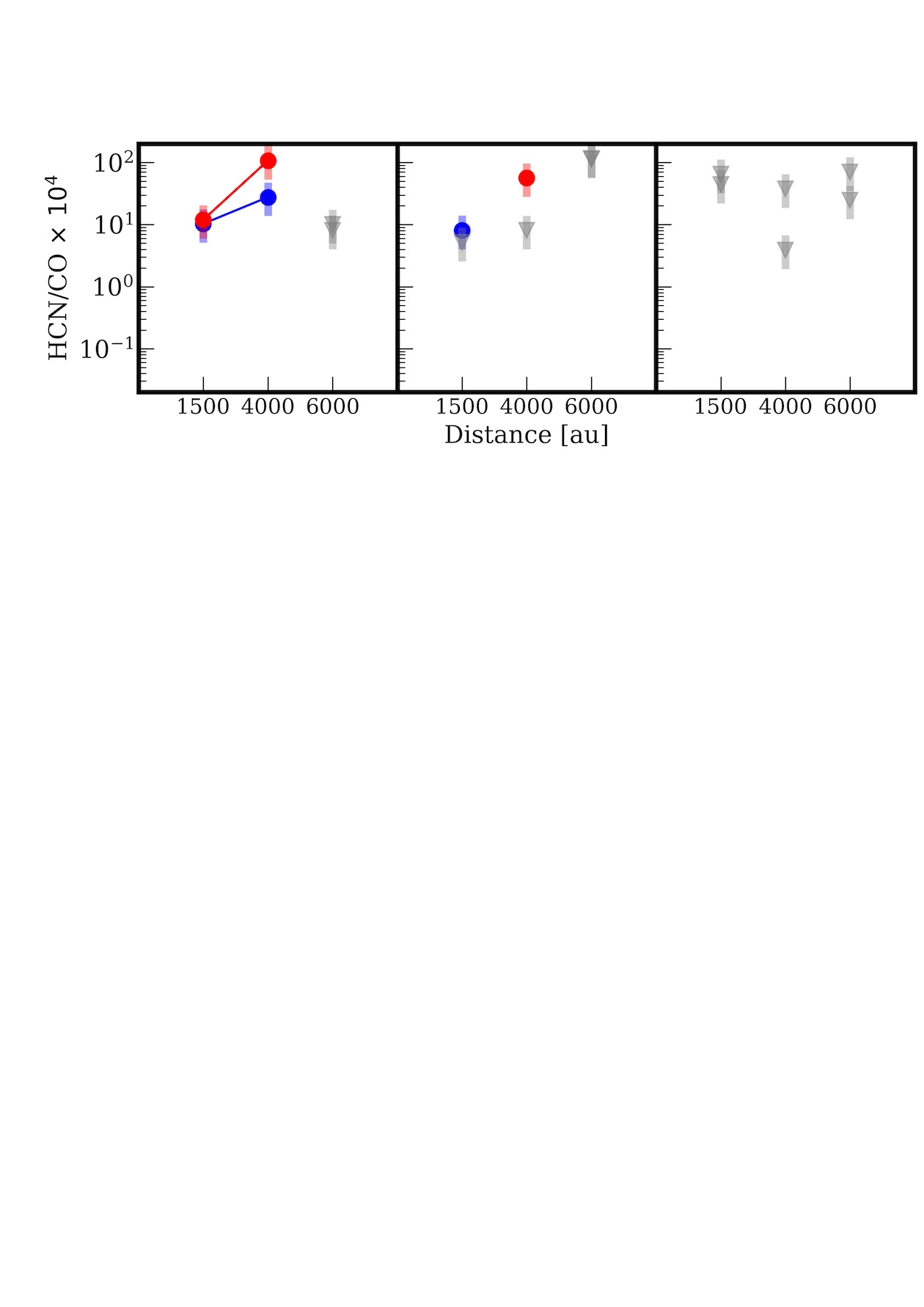}

   \caption{Molecular abundances with respect to CO scaled by $10^4$ for Ser-emb~8~(N). On the x-axis is the distance from the protostar. Panels from left to right are for the slow wing, the fast wing and the EHV component. The abundances measured for three different regions along the outflow are shown for blueshifted and redshifted part of the outflow separately. Abundances are measured in the same manner as in Fig. \ref{fig:abundances_ratio_total}. The HCN emission is likely optically thick and therefore the abundance should be treated as a lower limit.}
 \label{fig:sio_co}
\end{figure}

Abundances shown in Fig. \ref{fig:abundances_ratio_total}  and \ref{fig:sio_co} are obtained from the column density calculated for a mean temperature between the two extreme T$_{\rm ex}$ reported for each molecule in Table \ref{table:4}. To obtain the abundance with respect to CO, this column density is divided by the column density of CO calculated for T = 75 K. The CO column density is measured only in the region in which the emission from both molecules is above 3$\sigma$.

Figure \ref{fig:abundances_ratio_total} shows that the molecular abundances relative to CO change with velocity for each source. SiO increases in relative abundance from the slow to the fast wing for the redshifted SMM1-b outflow and both sides of the Ser-emb~8~(N) outflow. 
For the blueshifted Ser-emb~8~(N) flow, the abundance continues to rise toward the EHV regime, while it remains relatively constant for redshifted SMM1-b and Ser-emb~8~(N).
H$_2$CO is primarily associated with low-velocity gas, and it disappears in the fast wing for all sources. The only outflow to have the EHV H$_2$CO emission is the blueshifted part of the Ser-emb~8~(N) outflow, where H$_2$CO reappears in the EHV jet, with a relative abundance to CO around two times higher than in the slow gas. HCN is present in most of the outflows in both slow and fast wing, but it is never present in the EHV gas.

\begin{figure}[h]
\centering

  \includegraphics[width=0.95\linewidth,trim={ 0cm 0cm 0cm 3cm},clip]{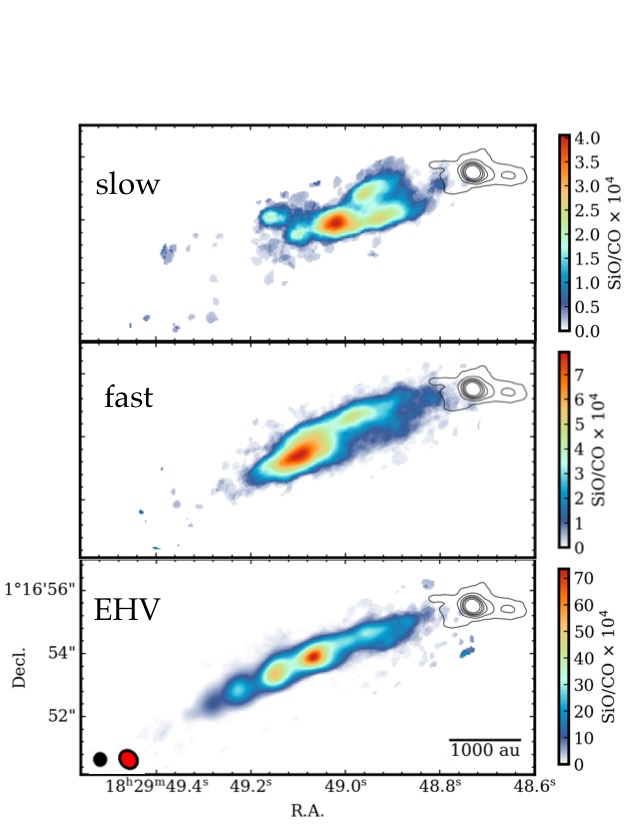}

   \caption{ Maps of the SiO/CO ratio for the blueshifted part of the Ser-emb~8(N) outflow for each velocity component. For the EHV component, only the channels  for which SiO emission was obtained at high spatial resolution are taken into account (< 40 km s$^{-1}$). The  synthesized beams of the CO (red) and continuum (black) are shown in the bottom-left corner of EHV plot with sizes 0\farcs35 $\times$ 0\farcs33  and 0\farcs55 $\times$ 0\farcs45 for continuum and CO, respectively. The black contours show 1.3 mm continuum emission.}
 \label{fig:sio_co_map}
\end{figure}

Even within the same velocity regime, the emission may be coming from different spatial regions, thus the analysis of the abundances over the entire outflow introduces additional uncertainties. Therefore, for the clearest case of the EHV jet --- Ser-emb~8~(N) --- we also measured the molecular abundances along the different positions of the outflow, in order to probe local abundances.

 Figure \ref{fig:sio_co} shows molecular abundances measured at three different positions on both sides of the Ser-emb~8~(N) outflow with regions defined appropriately to capture all of the lower-resolution SiO emission at the position. A remarkably similar behavior of SiO relative to CO can be noted on both sides of the outflow. The SiO abundance increases for the fast wing with distance  from the protostar, peaking at the second bullet at 4000 au and then disappears. In the EHV gas, the highest SiO abundance is observed close to the protostar, and then it drops with distance to the protostar by more than an order of magnitude. 
 
The furthermost region, associated with the CO bullet, is depleted in all the molecules except CO. The intermediate region at 4000 au appears as the most abundant in molecules, with HCN and SiO increasing for the slow and the fast wing. The H$_2$CO abundance is similar in the regions where it is detected.

To highlight the variations in the abundance ratios, maps of the SiO to CO ratio in the blueshifted part of the Ser-emb~8(N) are shown in Fig. \ref{fig:sio_co_map}. Only the blueshifted part is shown as an example, as the significant part of the redshifted EHV jet in SiO has been observed only at lower spatial resolution. It is clear that for the fast wing, the SiO/CO ratio peaks at a significant distance from protostar (3000 au; corresponding to dynamical age of ~500 years for a 30 km s$^{-1}$ outflow).  In the EHV jet, the SiO/CO ratio peaks at similar distance as in the fast wing and then decreases. 

\subsection{Outflow force}
Detection of the extremely high-velocity molecular jets provides a unique opportunity to probe the fastest and the most collimated part of the outflowing material. Quantifying the distribution of kinetic energy and mass among the different velocity components sheds light on their kinematic relationship, specifically determining if the jet is the driving force of the slow outflow. 

The mass of the gas must be derived from the number of molecules (see Sect. \ref{chemical_abundances}). The area of the pixel times the total number of molecules within pixel  N$_{\rm tot}$ times the ratio of H$_{2}/{\rm CO} = 1.2 \times 10^4$ \citep{Frerking1982}, with a molecular weight  $\mu = 2.8$ that takes helium into account  \citep{Kauffmann2008}, times the mass of the hydrogen atom $m_{\rm H}$ gives the amount of gas mass in a pixel  \citep{Yildiz2015}:

\begin{equation}
M = \mu m_{\rm H} A  \frac{\rm H_{2}}{\rm CO}  N_{\rm tot}   \,\,,
\end{equation}

The momentum of the outflowing material can then be defined accordingly:

\begin{equation}
P = M \times \varv_{\rm max} \,\,.
\end{equation}

We define the distance from the protostar to the edge of the integration region as  $R_{\rm lobe}$.  Note that the area of the ALMA observations  in all cases, except for SMM1-d and Ser-emb~8~(N), does not cover the full extent of the outflows, as evident in single dish observations \citep{Dionatos2010, Yildiz2015}. For that reason, parameters like outflow mass  or momentum do not provide information about the overall gas mass and kinetic energy content in the flow, but are rather local values or lower limits to those; the outflow force on the other hand, is dependent on  $R_{\rm lobe}$ and can be treated as a more general value, under the assumption that the outflow force content does not vary significantly at larger scales \citep{Marel2013}.

The contribution of the different velocity components to the overall outflow force is computed  for each side of the flow separately. In order to calculate the outflow mass loss rate -- $ \dot{M}$ it is convenient to make a velocity-weighted calculation per pixel since this is more sensitive to the velocity changes than using a single $\varv_{max}$ for the total outflow; this is method M7 as described in \cite{Marel2013}. According to this method, the Equation \ref{eq:1} is changed as follows:
\begin{equation}
\Bigg \langle \frac{N_u}{g_u}   \Bigg  \rangle_{\varv}=\frac{\beta \nu^2\int T(\varv)\varv d\varv}{A_{ul}}  \,\,,
\end{equation}
and the resulting velocity-weighted column density can be used to calculate the momentum in the same way as the column density is used to calculate the mass.

Finally, the outflow force in a pixel is given by:
\begin{equation}
F_{\rm out} = \frac{\dot{M}}{R_{\rm lobe}}\varv_{\rm max}  \,\,.
\end{equation}

 \begin{figure}[h]
\centering
\includegraphics[width=0.9\linewidth]{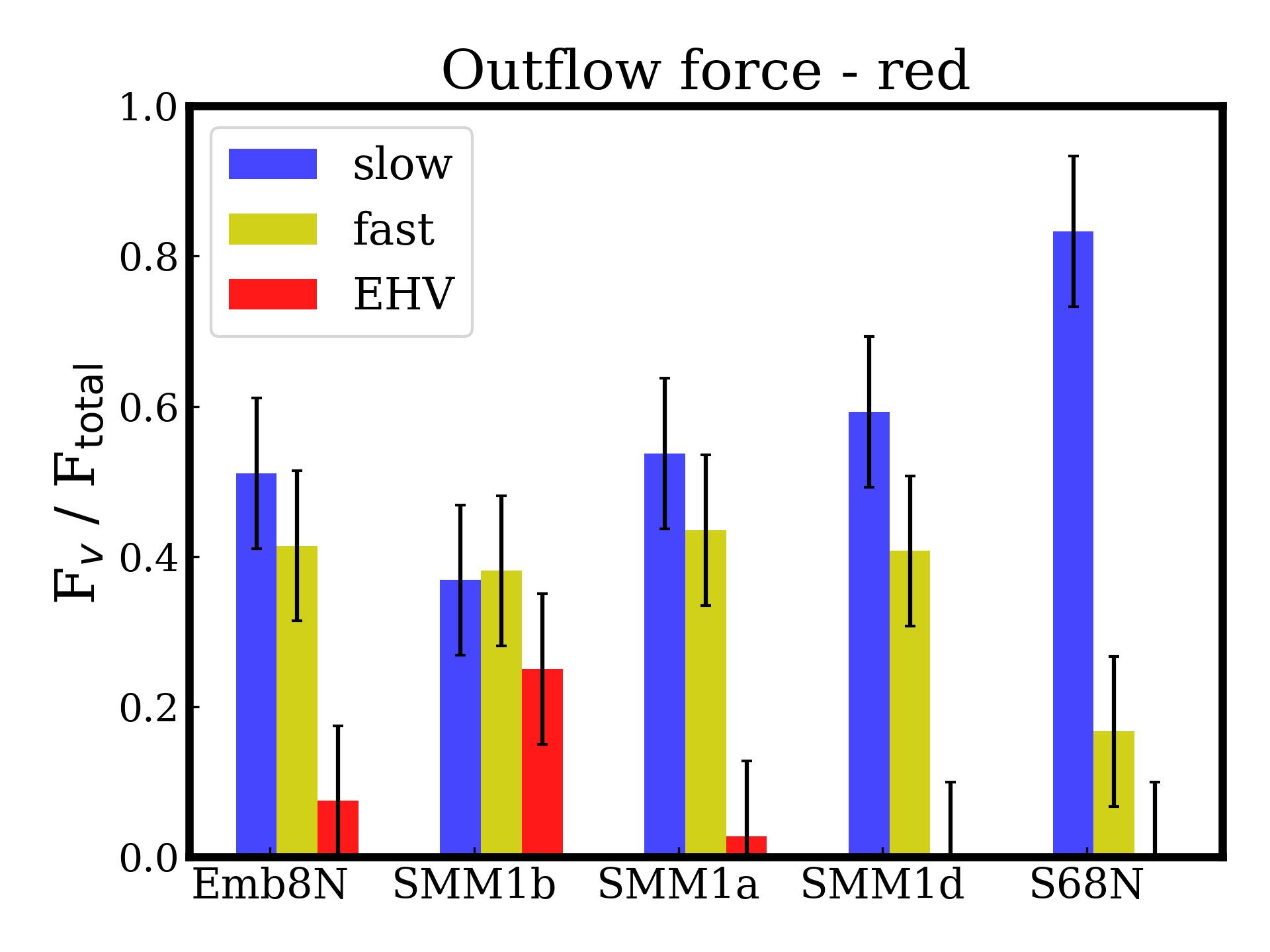}
  \includegraphics[width=0.9\linewidth]
  {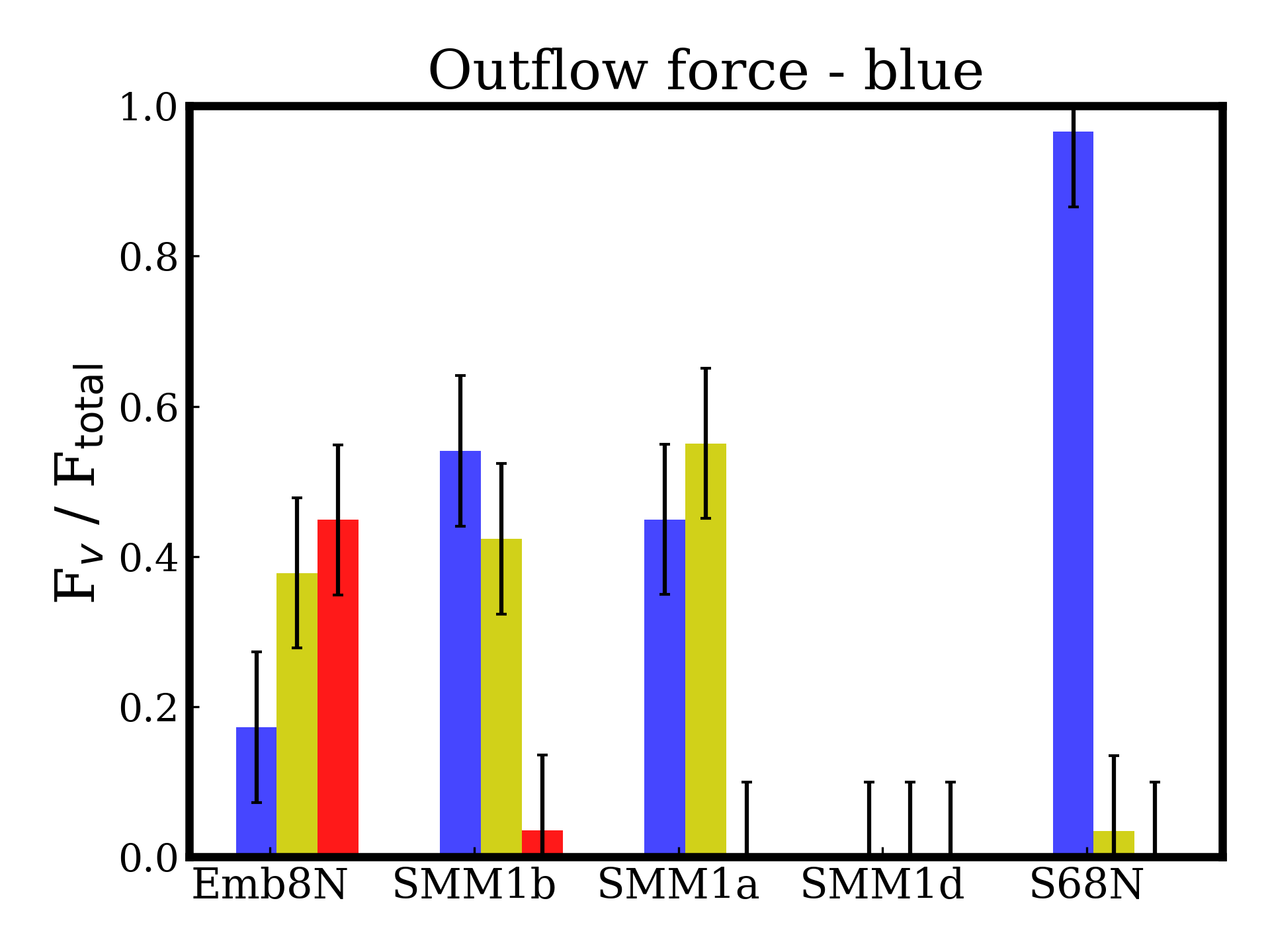}
   \caption{Fraction of the outflow force in each velocity regime, for the blueshifted (top) and redshifted (bottom) sides of the outflow for all sources. Approximate errors of 10\% are shown, resulting from uncertainty in the borders between the velocity regimes. }
 \label{fig:outflow_force_percent}
\end{figure}

Calculated values are presented in Tables C6-C10. As the choice of the velocity borders is done by eye, it introduces an uncertainty in the measurement of the outflow properties per velocity regime. Changing the velocity border by 5 km s$^{-1}$ between the fast wing and the EHV jet typically results in a change of $\sim$\,2--10\% in the outflow properties.

Figure \ref{fig:outflow_force_percent} shows the outflow force in each velocity regime relative to the total value. It shows that the contribution of the EHV jets to the total outflow force is between 5--40 \% of the total outflow force. The fraction of the fast wing component is similar for all outflows with a detected EHV jet (30--50 \%). The slow wing dominates the S68N outflow.

Inclination can introduce a significant uncertainty into the outflow parameters. For method M7, which has been adopted here to calculate the outflow force, \cite{Downes2007} provide a multiplication factor that should be used to account for inclination (Table 6 in their paper); values of the correction factor range between 1.2 -- 7.1. This correction largely affects the absolute values of the outflow forces; however, the relative ratios between the velocity components should not be affected \cite[Eq. 9 in ][]{Marel2013}

Although the outflows probed here often extend to much larger scales than those probed by ALMA,  the outflow force should be a conserved property. \cite{Yildiz2015} probed the outflow force of the SMM1 outflow in CO\,$3-2$ and CO\,$6-5$. They measured 1.5 and 8.7\,$\times$\,10$^{-4}$ M$_\odot$ yr$^{-1}$ km s$^{-1}$ for the blueshifted and redshifted emission, respectively, for CO\,$6-5$ and 6.7 and 23\,$\times$\,10$^{-4}$ M$_\odot$ yr$^{-1}$ km s$^{-1}$ for CO\,$3-2$ using the same M7 method, assuming the source inclination of 50$^\circ$.
From ALMA CO \,$2-1$ (slow\,+\,fast wing) we obtain 1.4 and 11 \,$\times$\,10$^{-4}$ M$_\odot$ yr$^{-1}$ km s$^{-1}$ for blueshifted and reshifted part of the outflow, respectively. Our results are thus consistent with single-dish data to within the typical uncertainties of a factor of few, even though no inclination correction is applied to ALMA observations.  The inclination correction applied by \cite{Yildiz2015} is based on Table 6 of \cite{Downes2007}, and it resulted in an increase of the outflow force by a factor of 4.4. Based on the similarity of the outflow force results between ALMA and single-dish data it appears that the observations obtained with the C43-1 configuration with a largest angular scale of $12\arcsec$ were sufficient to recover the bulk of the flux from those outflows. It is, however, plausible, that some of the emission has been resolved out, especially at low-velocities (see comparisons between the interferometric and single dish observations \citealt{Yildiz2015, Tafalla2017}). The similarity of the obtained outflow force values could be coincidental and related to the increased sensitivity of the ALMA observations.

\section{Discussion}
\subsection{Jet and wind kinematics. What is driving the outflows?}
The exact origin of the large-scale outflows from protostars is still unclear. It is suggested that the narrow, highly-collimated jet from the protostar or the inner disk could power the entirety of the outflow \citep{Raga1993b}.  However, models with jet bow-shocks powering the slow outflow fail to reproduce all of the observed kinematic features of the slow gas \citep{Lee2002}. Resolving the kinematic structure of the EHV bullets suggests, however, that significant fraction of the momentum of the jet is ejected sideways, impacting the surrounding envelope \citep{Santiago-Garcia2009, Tafalla2017}.

 Directly studying the relationship between the outflow and jet is difficult, as the atomic/ionized jet is invisible in the same wavelength regime as the colder molecular outflows. Thus, studying protostars in their earliest stages of formation, when the jet is still mostly molecular, gives a unique opportunity to study the relation between the outflow and the jet. Our ALMA observations allow us to study three remarkable outflows with EHV jet components within one cloud. Moreover, it is often difficult to study outflows at high resolution, since they are propagating to vast distances very rapidly. Only a few of them have been studied at their full extent with ALMA \citep[e.g.,][]{Arce2013}. While it appears that the SMM1-a,b, and S68N outflows have indeed already propagated to tens of thousands of au \citep{Dionatos2010, Yildiz2015}, it is plausible that Ser-emb~8~(N) outflow has not as apparent from the observations with a larger field of view \citep{Dionatos2010, Hull2014} . This source thus provides an opportunity to study the full extent of the outflow.

The relation between the different components here is quantified by measuring the outflow force in three velocity components: slow and fast wing, and in the EHV jet. From Fig. \ref{fig:outflow_force_percent} it is apparent that only for the blueshifted jet of Ser-emb~8~(N) the EHV contribution (45\%) to the total outflow force is higher than that of the slow and fast wing components. The contribution of the EHV components to the outflow force in the other two sources is smaller than the contribution from the wing. Based on these findings, it seems that the force contained in the jet is  generally not enough to power the total observed outflowing gas.

Not all of the jet can be probed with molecular emission alone. One of the explanations for the missing force is that the jet becomes atomic as the source evolves. Such a scenario is supported by the observations of atomic oxygen from {\it Herschel} \citep{vanKempen2010, Nisini2015}. For a small sample of protostars, \cite{Nisini2015} show that the atomic jet becomes an important dynamical agent in more evolved sources (late Class 0/ Class I), while younger outflows have a significant fraction of the jet in the form of molecular gas. Typical mass-loss rates in the jet derived from atomic oxygen for the Class 0 sources targeted by \cite{Nisini2015} are between 1--10~$\times$~10$^{-7 }$ M$_\odot$ yr$^{-1}$ whereas for the one Class I source HH46 they find ~2--8~$\times$~10$^{-6}$ M$_\odot$ yr$^{-1}$ which shows that the atomic jet becomes more important at the later stages of protostellar evolution.

The mass-loss rates of the molecular jets presented here are 7.0, 3.9, and 15.0 ~$\times$~10$^{-7 }$ M$_\odot$ yr$^{-1}$ for Ser-emb~8~(N), SMM1-a, and SMM1-b, respectively.
The atomic jet of SMM1-a has been probed in [O I] \citep{Mottram2017} and [Fe II]  \citep{Dionatos2014}. From these tracers, both authors find consistent mass flux of 2--4~$\times$~10$^{-7 }$ M$_\odot$ yr$^{-1}$, which is smaller than our molecular value by a factor of 2.
The total mass-loss of the slow and fast wing combined for SMM1-a is 1.4~$\times$~10$^{-5 }$~M$_\odot$~yr$^{-1}$. 
While these results are consistent with SMM1-a jet being mostly molecular, as is expected for a young Class 0 source, it appears that the jet cannot be solely responsible for driving the outflow, even when the atomic component is taken into account.

\begin{figure}[h]
\centering
  \includegraphics[width=0.9\linewidth,trim={0.6cm 40.5cm 0.6cm 1cm},clip]{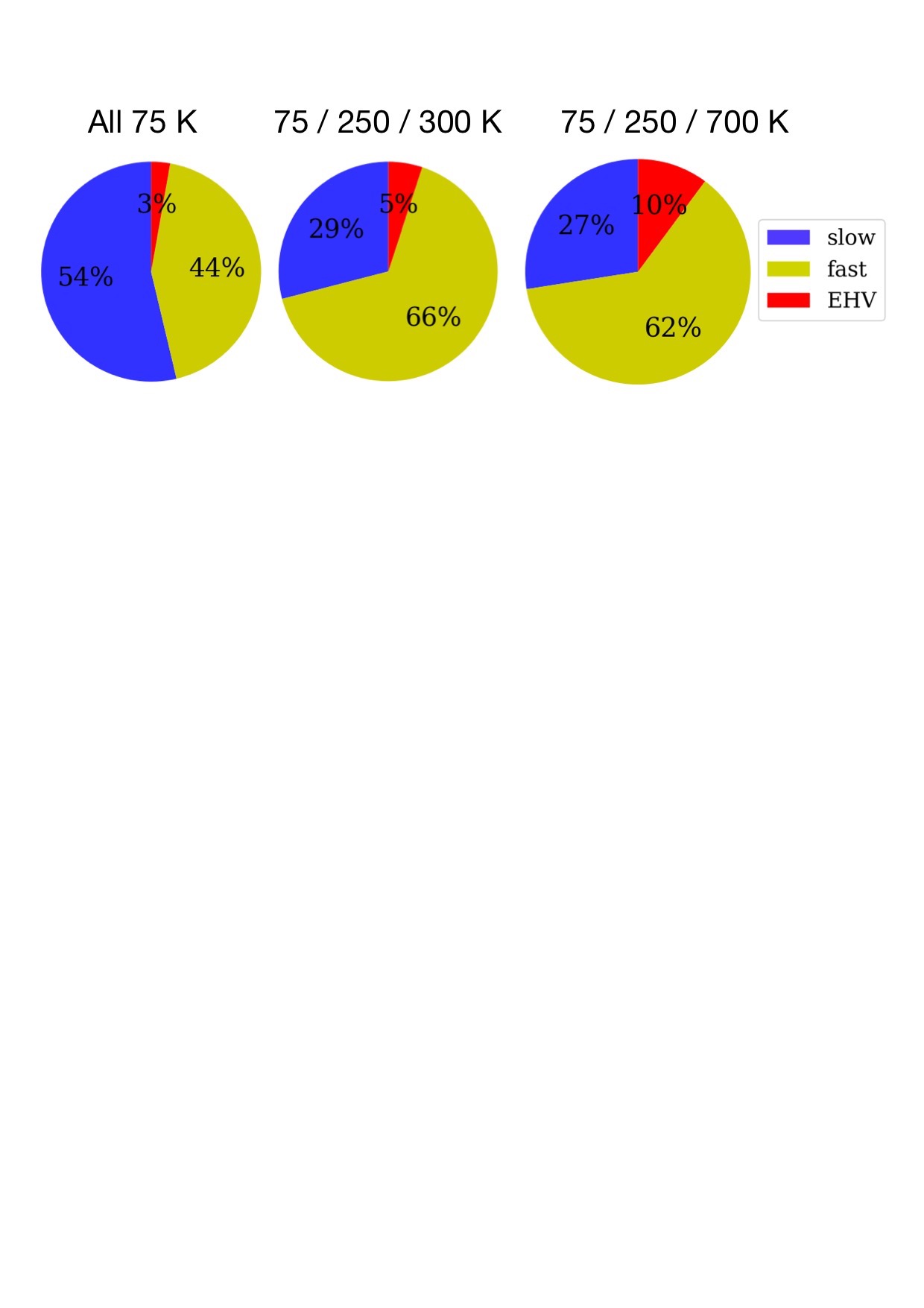}
 
   \caption{Fraction of the outflow force in the three different components (slow, fast, EHV) of the redshifted SMM1-a outflow for three different CO excitation temperatures used to calculate the outflow force. On the left plot all of the components have 75 K; in the middle plot, slow wing has  75 K, fast wing has 250 K, and EHV jet has 300 K;  on the right plot, slow wing has  75 K, fast wing has 250 K, and EHV jet has 700 K. The slow wing is yellow }
 \label{fig:fout_keynote1}
\end{figure}

Another explanation for the missing force in the molecular jet could be that the excitation temperature of the gas in the jet has been underestimated. Observations of high-$J$ CO and SiO suggest that excitation conditions change at higher velocities, with density and gas temperature rapidly rising \citep{Nisini2007, Lefloch2015, Kristensen2017}.
The assumed temperature here is 75 K, which is reasonable for a slow wing \citep{Yildiz2015, Kempen2016}. However, if the jet has different excitation conditions with higher temperatures, the CO mass of the gas will be underestimated. 
To test this possibility, we compare the change in relative contribution to the total outflow force for two other sets of temperatures. In one example we increase the temperature of the fast wing to 250 K, and the EHV temperature to 300 K --- this is the temperature of the warm component identified with PACS observations \citep{Karska2013, Karska2018, Kristensen2017, Dionatos2013}. In the second case we use 250 K for the fast wing again, and increase the temperature of the EHV component to 700 K --- fitted as the temperature of the hot component in PACS. 
   In Fig. \ref{fig:fout_keynote1} results of this comparison are presented for three cases for SMM1-a. The fraction of the EHV contribution to the total outflow force increases from the 3 to 10\%. A significant increase is seen in the fast wing with a change from 44 to 62 \% . For the case of SMM1-a, it does not change the general picture of the EHV jet contributing only a small fraction of the outflow force.
 
\begin{figure}[h]
\centering
  \includegraphics[width=0.9\linewidth,trim={0cm 0cm 0cm 0cm},clip]{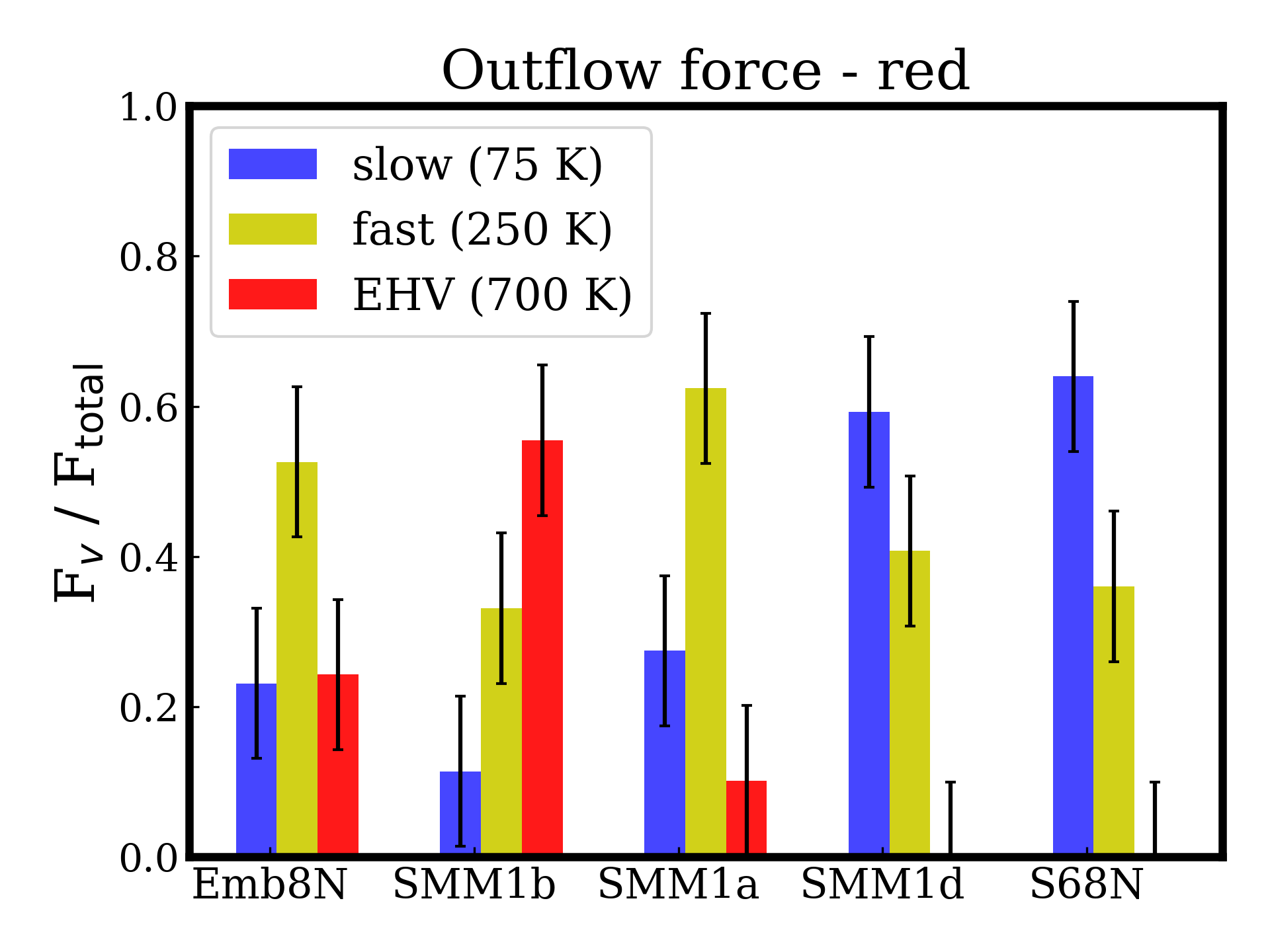}
 
   \caption{Fraction of the outflow force in each velocity regime, for the redshifted side of the outflow for all sources. Approximate errors of 10\% are shown, resulting from uncertainty in the borders between the velocity regimes. The excitation temperatures used to calculate the outflow force are: 75 K for the slow wing, 250 K for the fast wing, and 700 K for the EHV jet.}
 \label{fig:fout_keynote2}
\end{figure}

Fig. \ref{fig:fout_keynote2} shows how the outflow force contributions change for all of the sources in the redshifted outflow if the temperature is changed to 75 K, 250 K, and 700 K, for the slow wing, the fast wing, and the EHV jet, respectively. The SMM1-b EHV jet now contributes the majority of the outflow force, while for Ser-emb~8~(N) the fast wing becomes the primary component. This indicates that if the temperature of the gas in the jet is higher than assumed for the slow wing (75 K), the total mass of the gas and hence other properties derived from it can be significantly higher.

Nonetheless, the example of Ser-emb~8~(N) shows that young outflows that have not propagated to larger distances yet and, therefore, have a smaller number of shocks along the jet, can have a significant fraction of the outflow force in the EHV gas. Likely, older sources like SMM1-a that are more affected by precession have a more complicated jet-outflow relation and thus the interpretation is less straightforward. \\

While the SMM1-d outflow also lacks EHV emission, the contribution of the fast wing to the total outflow force is substantial ($\sim $40\%). Other characteristics of this source -- e.g., its bullet-like structure and lack of the well-defined cavity walls in CO -- suggest a peculiar nature of the outflow, and thus its lack of EHV emission cannot be attributed to the more evolved nature of the outflow.

For both SMM1-d and S68N, there is potentially another reason why the
EHV component is not detected: inclination. While for S68N we do not
see a clear bullet-like structure, for SMM1-d it might well be that
the bullets are seen moving at very high velocities but in the plane
of the sky. This is consistent with the fact that we see a significant
blueshifted component on the redshifted side of the flow, which is
consistent with the sideways expansion.  

We can see an evolution of the outflow force distribution among the different velocity components, that cannot be attributed only to the chemical changes in the jet. One way to explain this is that a significant amount of outflow force is deposited  in the fast and the slow wind very early in the protostellar evolution. Additional launching mechanisms like a wide-angle wind could also contribute to the bulk force released from the protostellar system.

\subsection{Relations with temperature/velocity components from HIFI}
Understanding the far-infrared (FIR) emission from outflows is crucial to quantify and describe cooling processes around young protostars, as the majority of cooling occurs in this regime \citep{Ceccarelli1996,Karska2013, Karska2018}. The {\it Herschel Space Observatory} provided new insight into the kinematics via FIR line profiles from the HIFI instrument \citep[e.g.,][]{Tafalla2013, Kristensen2013, Mottram2014}.
%Although high-$J$ CO and low-$J$ lines have different distributions in single-dish data, ALMA may pick up some of the excited emission at higher angular resolution even in CO\,$2-1$.

Specifically, observations with HIFI of large numbers of low-mass
  protostars have shown that the high-$J$ CO line profiles of
  shocked, outflowing gas can be decomposed universally into two
  velocity components. Subsequent radiative transfer modeling has
  linked these velocity components to the physical components of the protostellar system
  \citep{Kristensen2017}.
Unfortunately, the spatial information from {\it Herschel} is limited, and single-dish low-$J$ CO data show a different distribution from that of the high-$J$ lines, as the low-$J$ CO observations are sensitive to more extended emission \citep{Santangelo2012,Tafalla2013}.
ALMA data are sensitive to small scale emission, and thus offer the opportunity to relate the spatially unresolved components of the HIFI emission (estimated to arise on few hundred au scales, \citealt{Mottram2014}) with ALMA observations of low-$J$ lines, allowing us to unveil the physical origin of the emission observed with HIFI.

Here we compare the ALMA observations of CO\,$2-1$ toward Serpens SMM1 with {\it Herschel}/HIFI observations including CO\,$16-15$, CO\,$10-9$, and several water transitions \citep{Yildiz2013, Kristensen2012, Kristensen2013, Mottram2014}. Interferometric observations resolve the SMM1 system into at least five protostars, with three active outflows; this can help to disentangle the various components of the system blended into one HIFI beam of typically 20\arcsec. Fig. \ref{fig:HIFI2} shows three example comparisons between HIFI and ALMA spectral profiles.

There is some similarity between the HIFI velocity components for the SMM1 system and the ALMA low-$J$ CO spectra. The offset HIFI component is seen in the SMM1-a spectra and is spatially linked to the ridge of the blueshifted emission of the SMM1-a outflow. The broad component appears similar to the fast wing CO\,$2-1$ component and is present at both SMM1-a and SMM1-b outflows. The EHV bullet seen in water transitions from HIFI can be associated spatially with ALMA CO SMM1-b bullets, but peaks at higher velocities than the SMM1-b jet.
While it is impossible to spatially resolve the location of the water emission, this result suggests that water is formed in the higher velocity shock than CO or SiO.

  A detailed discussion of the comparison of ALMA observations with {\it Herschel} data is presented in the Appendix \ref{hifi_appendix}.

\subsection{The case of Ser-emb 8 (N): a pristine outflow-jet system}

\begin{figure}[h]
\centering
  \includegraphics[width=0.95\linewidth,trim={6.3cm 9.2cm 8cm 5.5cm},clip]{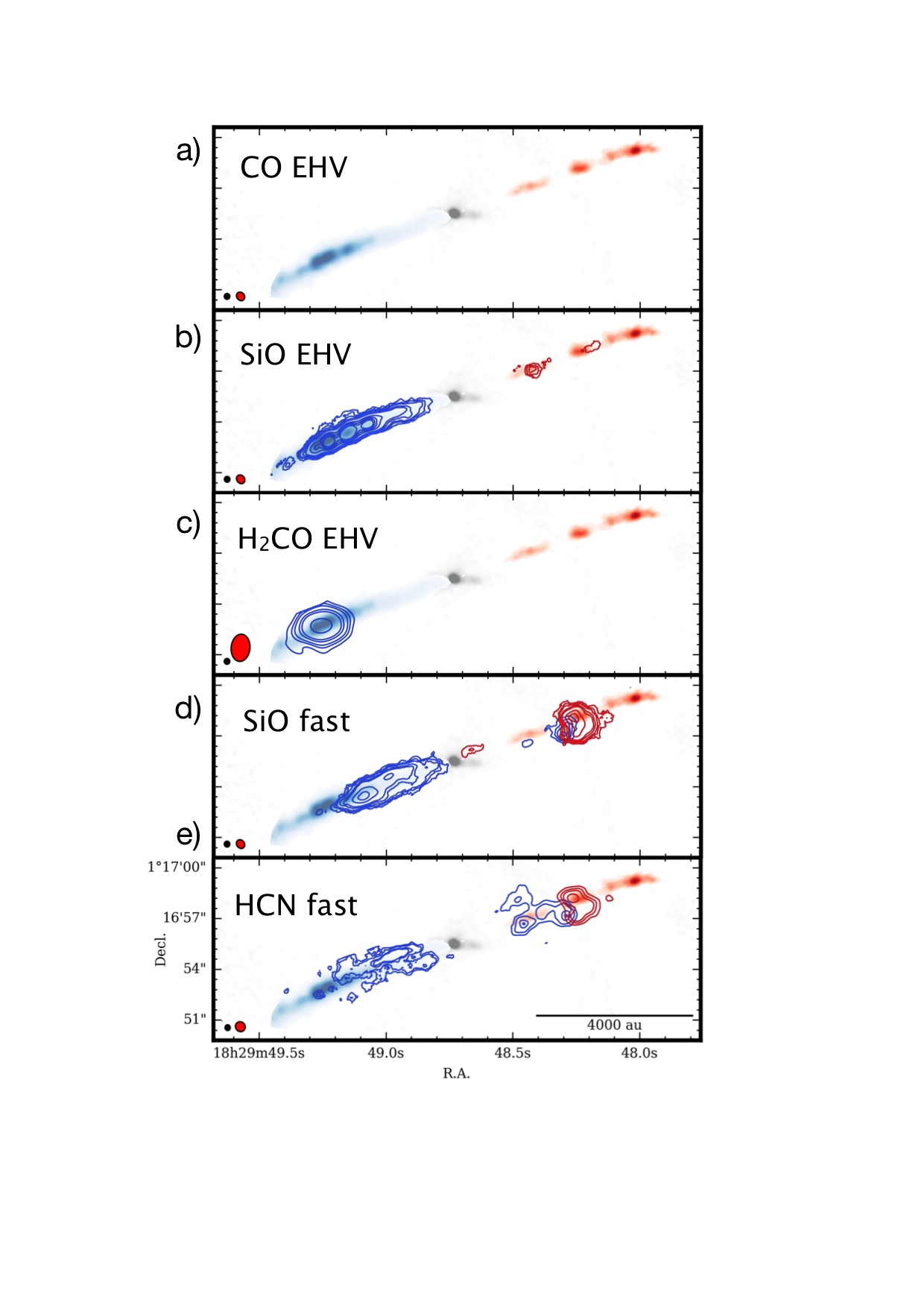}
  \caption{Schematic view of spatial distribution of different molecules and their relation with different velocity components in the Ser-emb 8 (N) outflow: a) in colorscale the CO moment 0 map is shown  integrated over the EHV velocities, also overlaid on the following plots; b) contours are SiO EHV emission captured at high spatial resolution i.e. below 40 km s$^{-1}$; c) H$_2$CO EHV emission (available only at low spatial resolution - synthesized beam is 1\farcs65 $\times$ 1\farcs13); d) SiO fast wing emission; e) HCN fast wing emission.
    %On d and e plots with emission from -5 to -2 km s$^-1$ is showed on the redshifted side of the outflow. EvD: put in text, confusing here in caption
    The synthesized beams of continuum (black) and contour map (red) is shown in bottom-left corner. }
 \label{fig:fig5}
\end{figure}

Many characteristics of Ser-emb~8~(N), such as its narrow opening angle of 25$^\circ$ and the high contribution of the molecular jet to the total force of the outflow, show that it is likely the youngest of the sources in the sample and therefore the best example of a pristine molecular jet and outflow system. It is also likely that we see most of the outflow within the ALMA field of view, in contrast to SMM1-a and b, which are known to extend to much larger scales \citep{Davis1999, Dionatos2010}.  If so, the most distant bullet at 4500 au would have a dynamical age of only 350 years for a velocity of 60 km s$^{-1}$.
In this section, we explore the spatial distribution of the analyzed velocity components of other molecules of the Ser-emb~8~(N) outflow. Figure \ref{fig:fig5} shows the spatial distribution of the fast and EHV velocity components for the CO, SiO, H$_2$CO, and HCN.

One thing that is immediately apparent is the very similar shape of the SiO and HCN emission with both forming a redshifted bow-shock in the fast velocity component. 
On the blueshifted side, the shape of emission does not resemble a bow-shock, but both HCN and SiO appears mostly off the jet axis.  The SiO and HCN bow-shock on the red side (Figs. \ref{fig:fig5}d,e) is 
surrounding one of the EHV bullets seen in CO (Figs. \ref{fig:fig5}a). The weak blueshifted emission on the redshifted side of the outflow seen in SiO and HCN (velocities from --5 to --2 km s$^{-1}$ with respect to the source velocity) is consistent with the sideways expansion of the gas due to interaction with the internal shock in the EHV bullet \citep{Tafalla2017}. This suggests a relation between EHV jets with the fast wing. Sideways ejections of the EHV gas can create slow shocks along the cavity walls.  {\it Herschel} line profiles show that when the source exhibits EHV emission, the broad component is always present \citep{Kristensen2012}.  The nearly identical shape of the SiO and HCN emission in the fast wing can be related to the same physical process that is responsible for the production of the SiO and HCN gas, as both species are enhanced in shocks \citep{Schilke1997,Pineau1990}.

The most distant EHV bullet at$~6000$ au -- corresponding to the dynamical age of $~500$ yrs --  is seen mostly in CO with SiO emission much fainter compared with the 'younger' bullets. It is possible that grains have started to reform, causing the SiO depletion from the gas. The decrease in the SiO emission can however also be caused by the change in the excitation conditions along the jet: the density and the temperature of the gas is likely decreasing in the more distant bullets \citep{Nisini2007}.

H$_2$CO is seen in only one bullet on the blueshifted side of Ser-emb~8~(N). This H$_2$CO bullet is coincident with CO peak of intensity along the jet at $\sim 4000$ au. Thus, the presence of H$_2$CO  can be related to the total density of the gas at that position - CO formation in the EHV jet is enhanced with density \citep{Glassgold1991}.

\subsection{Chemistry of the velocity components}

\begin{figure*}[h]
\centering
  \includegraphics[width=0.95\linewidth,trim={ 1.5cm 14cm 1cm 10.5cm},clip]{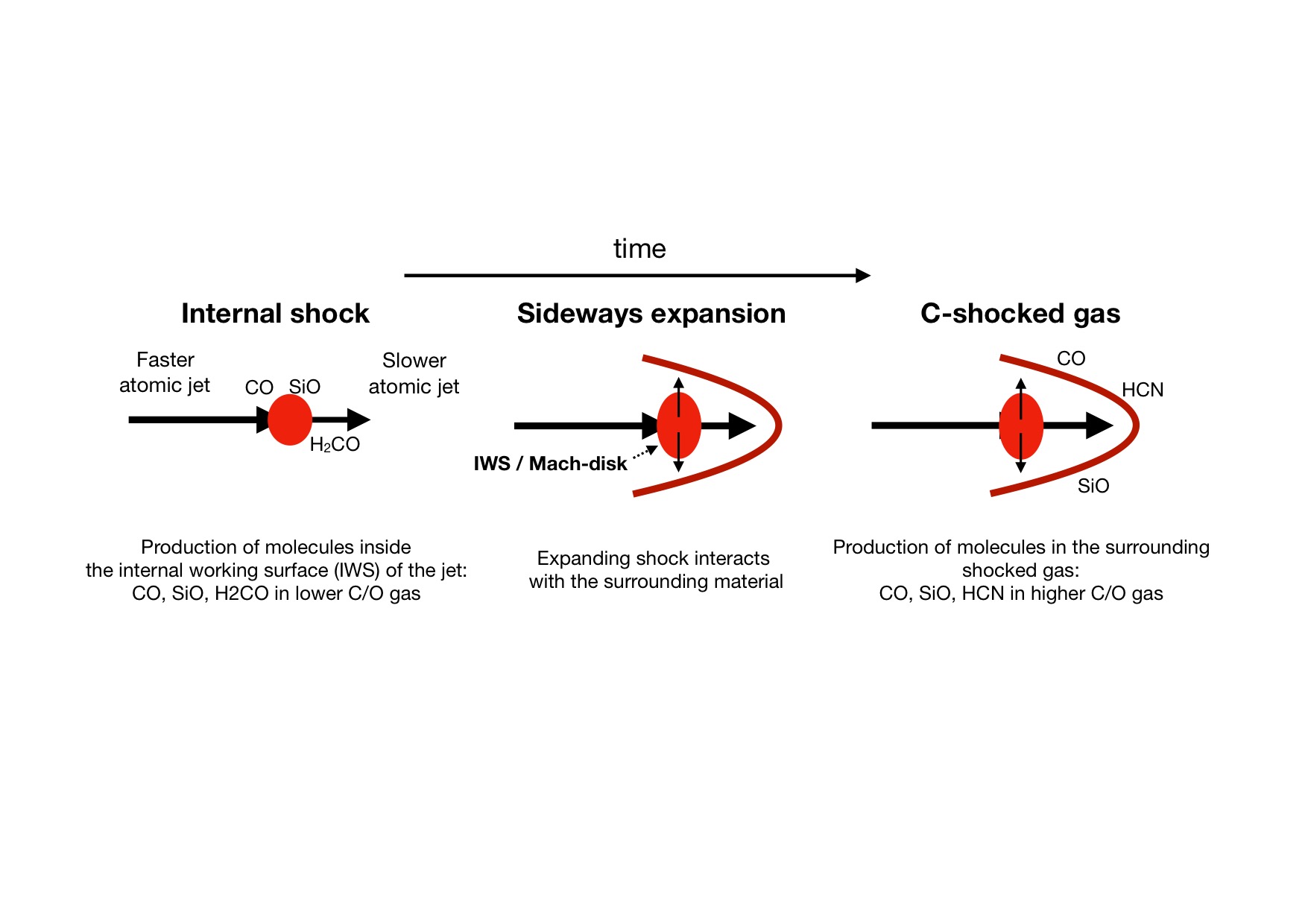}

  \caption{Cartoon presenting the interaction between a molecular bullet and the surrounding material. From the left to right a time evolution is shown starting with an internal shock within the molecular jet where atomic gas produce molecules inside a high-density internal working surface. As the bullet expands both forwards and sideways it creates a shock with the surrounding ambient material; in the shocked gas molecules are formed.
 The molecules observed in the EHV bullet are produced in lower C/O gas originating from the inner Mach disk, while the molecules from the shocked gas are formed from ambient gas with higher C/O ratio.}
 \label{fig:cartoon1}
\end{figure*}

%EvD: moved to introduction, out of place here.
%Extremely high-velocity jets were first detected as spectral features, as peaks detached from the low-velocity outflow wings \citep{Bachiller1990}, and further spatially resolved as discrete bullets embedded in the cocoon of low-velocity gas \citep[e.g.,][]{Santiago-Garcia2009, Hirano2010}.

The first extensive chemical survey of the molecular jets revealed differences in chemical composition of the slow and fast components and the EHV jet \citep{Tafalla2010}, the main conclusion being that the EHV component has more oxygen-containing molecules than the slow and the fast wing gas which are carbon-rich (abbreviated as a higher C/O ratio).
The high-resolution interferometric observations presented here are consistent with these single-dish studies: SiO abundances are enhanced with velocities up to those of the EHV jet for Ser-emb~8~(N) and redshifted SMM1-b. H$_2$CO appears in one EHV bullet of Ser-emb~8(N). The HCN is present in the slow and the fast wing, but it does not appear in the EHV jet.  Unique to our analysis is the ability to not only study the spectra but also relate the abundances with different spatial and velocity components of the outflow.

The spatial distribution of molecules can indeed provide essential clues about the relation between different velocity components. The bow-shock structure in the redshifted part of the Ser-emb~8(N) outflow (fast wing, Fig. \ref{fig:fig5}) is co-spatial with a gas bullet moving at much higher velocities. The interaction between the EHV jet and the ambient gas, and the origin of the chemical composition of the fast wing component and the jet, is described in Fig. \ref{fig:cartoon1}. If the jet indeed has a low C/O ratio \citep{Tafalla2010}, the production of oxygen-bearing molecules will take place in the internal working surface of the jet.  Then, the (sideways) expanding internal shock interacts with the surrounding ambient material (with higher C/O ratio), where production of other species like HCN can take place.
  %Production of SiO occurs not only in the jet but also (with some time delay) in the cooled down shocked gas (fast wing), therefore explaining the presence of the abundance enhancement at significant distance from the protostar.

Our results can also be compared with interferometric studies of the prototypical chemically rich outflow L1157 \citep[e.g.,][]{Gueth1996, Arce2008, Codella2009, Codella2017}, which is also known to have a molecular jet \citep{Podio2016}. The L1157 data show a chemical evolution with time along the outflow, with the jet impacting already shocked gas.
%Below we show how interactions between the high-velocity jet and ambient material can explain some of the images and abundance trends seen toward Ser-emb~8~(N).

\subsubsection{SiO}
SiO is enhanced consistently for Ser-emb~8~(N) from the slow to the fast wing and then to EHV jet, where it peaks in abundance. The enhancement of SiO in supersonic gas is commonly explained by sputtering and grain destruction, and subsequent formation of the SiO in the gas phase through reactions of Si with OH in the shocked gas \citep{Schilke1997, Gusdorf2008a, Gusdorf2008}. If the high-velocity jet is ejected in an atomic state thus containing ample atomic Si), SiO molecules can also be efficiently formed in the internal shocks in the jet that trigger the density enhancement \citep{Glassgold1991,Tafalla2010}.

There are differences among the SiO velocity profiles of the various sources. Ser-emb~8~(N) and SMM1-b --- the two sources with the EHV emission --- show weak emission at low velocities, with SiO emission peaking at high velocities. Such offsets in the peak of the emission can be caused by shock enhancement of the SiO abundance, consistent with models described above. S68N and SMM1-d, on the other hand, have SiO profiles that peak  close to the systemic velocity and then decrease with velocity.

\cite{Nisini2007} see a similar dichotomy of the profiles for two protostellar outflows -- L1448-mm, the prototypical EHV source and L1157-mm, a classic example of the chemically rich outflow, with EHV bullets detected by \citet{Tafalla2015} and \citet{Podio2016}. These authors attributed this difference to the temporal evolution of the outflow, where young shocks show offset peak profiles, while wing profiles peaking at low velocities correspond to the gas after the passage of a shock, where gas slowed down but retained its enhanced SiO abundance  \citep{Jimenez-Serra2009}. It is possible that this temporal evolution can be observed within one outflow. The SiO abundance along the Ser-emb~8~(N) outflow decreases with the distance from the source for the EHV jet. On the other hand, the fast wing abundance increase with the distance from the source up to $\sim$ 4000 au and then decreases toward the most distant CO bullet. This can be interpreted as the SiO being produced in the EHV gas and then consistently slowing down as the shell of the internal shock is expanding.
%*** do we want to add a comment on water bullet velocity: produced first?***

The similarity of the HCN and SiO emission in the bow shock of the Ser-emb~8(N) poses a challenge to this scheme. Their similar spatial and kinematic structure in the fast wing would suggest a similar origin; however, HCN is not seen in the EHV gas, and therefore its formation in the jet is unlikely. An alternative explanation for the SiO emission in the fast wing is a $C-$shock along the cavity walls. Fig. \ref{fig:cartoon1} presents a schematic of this scenario. The formation of the SiO in the $C-$shocked gas is a process with a timescale of $>$ 100 yr \citep{Gusdorf2008a}, which would explain an enhancement at some distance from the protostar. If the EHV SiO emission arises from the production in the dense atomic jet gas \citep{Glassgold1991}, this process would occur much faster, explaining the high EHV SiO abundance close to the protostar \citep{Hirano2010, Podio2016}. 
The observed H$_2$O line with HIFI, which appears faster than EHV jet toward SMM1-b, can thus be interpreted as having been formed even earlier, i.e., in the fastest component of the internal working surface of the jet.

\subsubsection{H$_2$CO}

\begin{figure}[h]
\centering
  \includegraphics[width=0.95\linewidth,trim={ 3.5cm 30.5cm 3.5cm 4cm},clip]{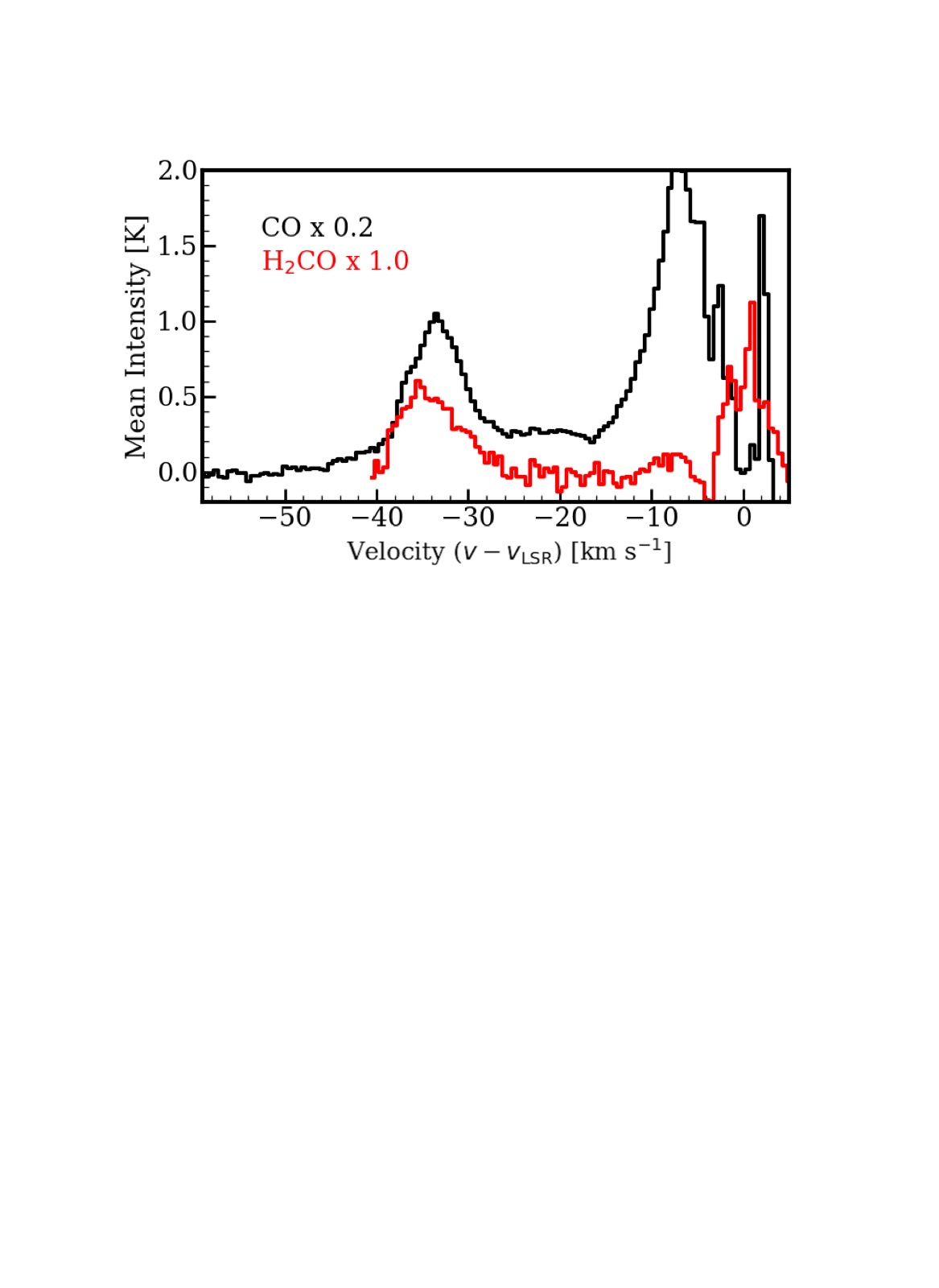}

  \caption{Spectra of CO\,(black) and H$_2$CO\,(red) of Ser-emb~8~(N) integrated on the region where H$_2$CO high-velocity emission is present.}
 \label{fig:h2co_bullet_spectra}
\end{figure}

 \cite{Tafalla2010} detected H$_2$CO in EHV gas for the first time in only one source in their study of two EHV jets. In the case of L1448-mm, H$_2$CO is also accompanied by CH$_3$OH emission. In the slow wing, the H$_2$CO abundance swiftly decreases with increasing velocity, likely being easily destroyed in shocks, similar to CH$_3$OH  \citep{Suutarinen2014}. It is then remarkable that we see the H$_2$CO in the high-velocity bullet of Ser-emb~8(N)~(see Fig. \ref{fig:fig5}c).
More recently several transitions of H$_2$CO have been detected in the high-velocity component of the IRAS 2A outflow, while CH$_3$OH has  only been seen at low velocities \citep{Santangelo2015}. Figure \ref{fig:h2co_bullet_spectra} compares CO and H$_2$CO spectra integrated toward the H$_2$CO bullet for Ser-emb~8~(N).

Surprisingly H$_2$CO is seen in only one particular EHV bullet in the blueshifed jet of Ser-emb~8~(N). It is located at the peak of CO EHV emission but ahead of the SiO peak emission. The presence of the EHV H$_2$CO emission for one of the three sources, and exclusively in one side of the emission, is puzzling. Two main explanations can be considered. One is the hydrogenation of CO on the reformed post-shock grains and then subsequent release from the grains \citep[e.g.,][]{Watanabe2004, Chuang2016}. 
Within the bandwidth of the ALMA observations, many complex organic molecules are detected toward the protostars SMM1-a and S68N, with their location indicating an origin in the warm inner envelopes of the protostars \citep{Tychoniec2018c}. Those molecules are not detected toward the position of H$_2$CO EHV bullet. If the release from the ices were a mechanism that is responsible for the H$_2$CO  emission at high velocities, one would expect the presence of other ice mantle components in the gas-phase. This is not seen in the case of this high-velocity bullet. Releasing H$_2$CO from the ices is usually associated with lower outflow velocities -- toward the L1157 outflow H$_2$CO is present in the shell of low-to-intermediate velocity gas. It is argued that the release of H$_2$CO from the ices can trigger formation of the complex organic molecules in the gas-phase \citep{Codella2017}. Again, this is not seen here.

An alternative explanation for the H$_2$CO emission in the
high-velocity jet is gas-phase formation, mainly through the
CH$_3$\,+\,O reaction \citep{Dalgarno1973, Millar1975} with CH$_3$
abundance enhanced due to the high temperature.  In particular,
  the C + H$_2$ $\to$ CH + H reaction has a barrier of $\sim$12000 K,
  with subsequent reactions of CH and CH$_2$ with H$_2$ leading to
  CH$_3$ having only somewhat smaller barriers
  \citep{Agundez2008,Bast2013}.  In this case, the abundance of the
H$_2$CO increases from the slow wing to the EHV component by least a factor of two; therefore, the mechanisms responsible for the
production and excitation of H$_2$CO can be more efficient at higher
velocities where temperatures are higher. A high abundance of
atomic oxygen in the jet can further facilitate the reaction.
  This scenario would require the presence of some free atomic C in
  the jet, which would form H$_2$CO but not HCN before  all of the carbon is  locked up in CO.

\subsubsection{HCN}
HCN traces the most energetic outflows associated with young, Class 0 sources \citep[][]{Jorgensen2004,Walker-Smith2014}. High temperatures and densities of the shocked gas are responsible for HCN production. The enhancement of the HCN emission in shocks arises due to the  H$_2$ + CN $\rightarrow$ HCN + H reaction \citep{Bruderer2009,Visser2018}, which has an activation barrier of 960 K \citep{Baulch2005}. 
Both models and observations suggest orders of magnitude increase in HCN abundance for gas temperatures above 200 K \citep{Boonman2001,Lahuis2007}.

We see HCN present in the slow and the fast wing, but it is depleted in the EHV jet. However, it appears that the presence of the fast HCN and SiO strongly depends on the presence of the EHV jet, as both HCN and SiO are observed in the bow shock in which the EHV bullet is embedded.
It appears that, as an EHV bullet is present and as it ejects gas sideways at locations where it can interact with the cavity wall, both HCN and SiO are produced in these lower velocity $C-$type shocks. This interpretation is straightforward only for Ser-emb~8~(N); it is much harder to interpret the HCN in SMM1, as no HCN emission is observed toward SMM1-b and very little in SMM1-a.  %

\cite{Tafalla2010} argue that HCN enhancement in the fast wing and depletion in the EHV jet is related to the atomic carbon abundance in the gas phase, specifically to a much lower C/O ratio in the EHV gas which leads to the efficient formation of CO and SiO, but not HCN. It is unlikely that the gas in the EHV jet is colder than in the fast wing, so temperature difference can not explain the lack of HCN in the EHV gas. Therefore our results support different chemical compositions of the EHV gas compared with the slow and the fast wings.

\section{Summary}
In this work, we use ALMA to study extremely high-velocity molecular jets in the Serpens Main region. The relationship between the fast jet and slow outflow is studied, in an attempt to unveil the chemical composition of the different velocity components. The conclusions are as follows:
\begin{enumerate}
\item{Out of five observed outflows, three show the extremely high-velocity jet component. The high-sensitivity ALMA observations reveal that the EHV component in outflows from protostars is more frequent than previously thought.}

\item {Comparison of outflow forces between the slow outflow and EHV jet reveals that the observed force in the molecular jet is not sufficient to power the slow outflow in 3/5 sources. The most narrow and compact outflow (i.e., likely very young) in Ser-emb~8~(N) -- drives the jet with the highest EHV contribution of outflow force relative to the total energetic content of the flow. These results suggest an evolutionary sequence of the molecular emission from protostellar outflows where the EHV component is present in the youngest sources. The EHV and the fast wing components then subsequently disappear as the protostellar system evolves. Even accounting for the atomic component, we conclude that the outflow force in the jet component is not sufficient to carry the entirety of the flow for all observed sources. This shows that a large fraction of the outflow force could already have been deposited in the fast and the slow wind, or that another launching mechanism(i.e., a wide-angle wind) is also at play; however, the latter option cannot explain the bow-shock structures we observe in the fast wing component of Ser-emb~8(N).}

\item {The spatial distribution of the different molecular species is revealed in 0\farcs4 ALMA observations; we focus in particular on the newly reported EHV jet from Ser-emb~8~(N). The fast wing SiO and HCN emission on the redshifted side of this outflow resembles bow-shocks, surrounding the EHV bullet, which indicates a relationship between the fast wing and the sideways ejections of the EHV jet. 
}

\item  {The chemical composition of the velocity components of the outflow has been probed: the SiO abundance is enhanced from the slow to fast gas; the HCN is present from slow to fast wing but disappears in the EHV jet; H$_2$CO is seen only in the slow gas and in one EHV bullet exclusively in the blueshifted part of the Ser-emb~8~(N) jet. These results are in agreement with the single-dish results from \cite{Tafalla2010} where the EHV jet has a lower C/O ratio than the entrained slow and fast gas. Consistent velocity profiles of both molecules suggest that gas-phase formation  is a plausible explanation for H$_2$CO emission in the EHV jet. The HCN presence at the bow-shock (fast wing) is consistent with an increased temperature in the $C-$shocked region compared with the lower velocity gas. HCN depletion in the EHV gas can be associated with the lower C/O ratio in that gas.}
  
\item {The decrease in the SiO abundance in the EHV gas with distance from the protostar, combined with increase in the fast wing, suggests that SiO produced in the EHV gas is slowed down, but remains abundant at lower velocities. Production of SiO and HCN in C-shocks (fast wing) after some time from the passage of the shock front, as expected by models, provides an alternative explanation to an apparent temporal evolution of the abundances.}

\item{We compare ALMA observations with the {\it Herschel}/HIFI velocity profiles of high-$J$ CO and water, specifically comparing the offset and broad components seen universally in the HIFI observations \citep{Mottram2014, Kristensen2017} with the slow wing, the fast wing and the EHV jets explored with ALMA CO\,$2-1$ line profiles.  The spatial location of the HIFI profiles is revealed; the fast wing has a similar profile to the HIFI broad component and EHV features are seen in both HIFI water emission and in ALMA spectra. However, the water EHV bullet peaks at higher velocities  and is therefore formed first in the internal working surface of the jet.}

\end{enumerate}

\begin{acknowledgements} The authors are grateful to the referee for comments that helped to improve the manuscript. {\L}T would like to thank Beno\^{i}t Tabone for stimulating discussions. 
This paper makes use of the following ALMA data: ADS/JAO.ALMA\#2013.1.00726.S  and ADS/JAO.ALMA\#2016.1.00710.S. ALMA is a partnership of ESO (representing its member states), NSF (USA) and NINS (Japan), together with NRC (Canada), MOST and ASIAA (Taiwan), and KASI (Republic of Korea), in cooperation with the Republic of Chile. The Joint ALMA Observatory is operated by ESO, AUI/NRAO and NAOJ. Astrochemistry in Leiden is supported by the Netherlands Research School for Astronomy (NOVA), by a Royal Netherlands Academy of Arts and Sciences (KNAW) professor prize, and by the European Union A-ERC grant 291141 CHEMPLAN. The research of L.E.K. is supported by a research grant (19127) from VILLUM FONDEN.  C.L.H.H. acknowledges the support of both the NAOJ Fellowship as well as JSPS KAKENHI grant 18K13586. This research made use of Astropy, a community-developed core Python package for Astronomy \citep{Astropy2013},
http://astropy.org); Matplotlib library \citep{Hunter2007}; NASA's Astrophysics Data System.
\end{acknowledgements}

\interlinepenalty=10000
\bibliography{mybib}

\appendix

\raggedbottom

%\begin{minipage}{\textwidth}
%\includegraphics{}
%\label{}
%\end{minipage}

\section{Relations with temperature/velocity components from HIFI}
\label{hifi_appendix}

\begin{figure*}[h]
\centering
  \includegraphics[width=0.85\linewidth,trim={3cm 4cm 1.5cm 2cm},clip]{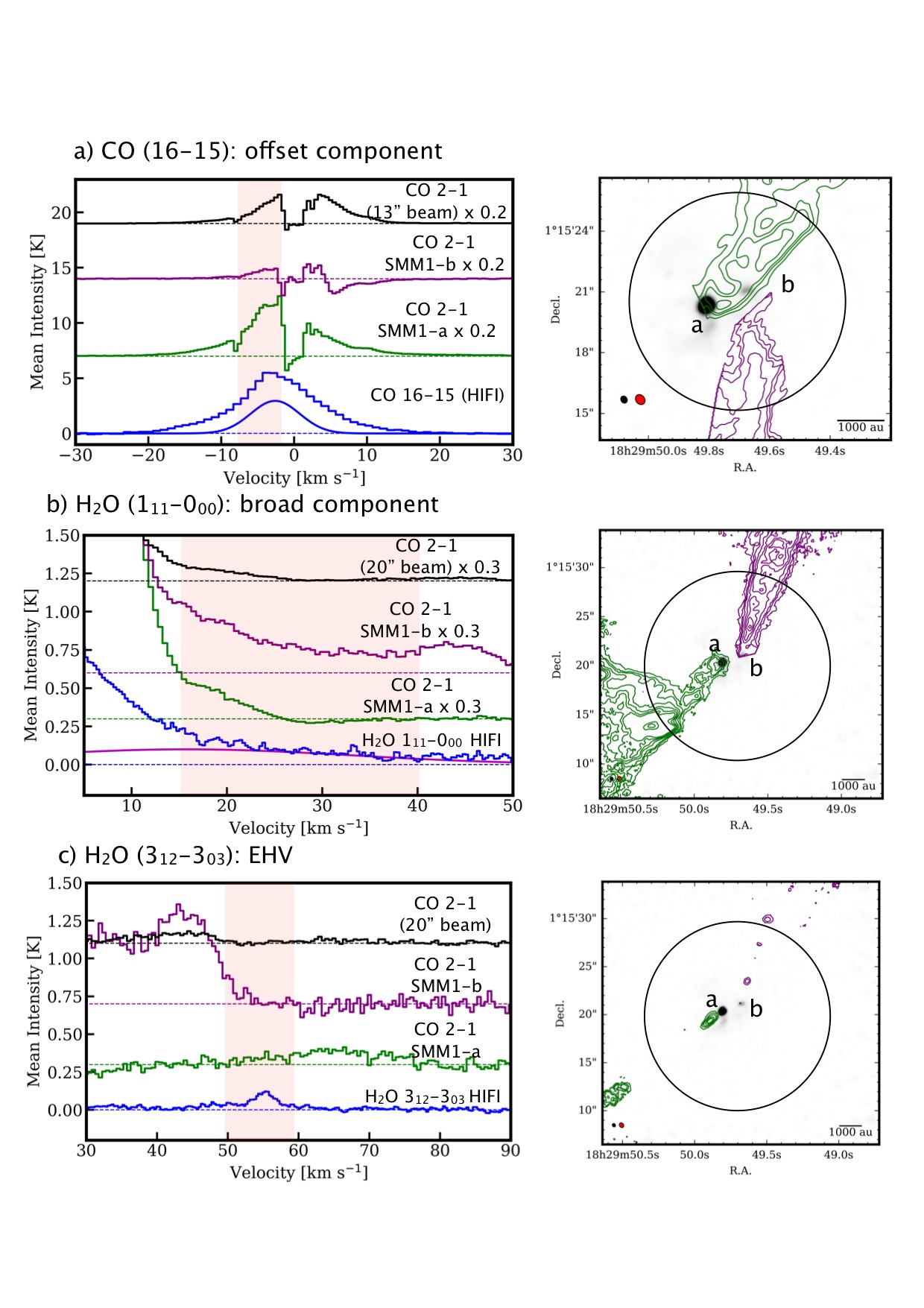}

  \caption{Comparison of CO \,$2-1$  emission from SMM1 system with HIFI observations. Left: Spectra from HIFI (blue), ALMA integrated with HIFI beam (black), ALMA spectra integrated with regions drawn to capture all emission from SMM1-a (purple) and SMM1-b (green) within HIFI beam. The velocity range from which the moment 0 map on the right was produced is indicated with the red shade. The Gaussian profiles for the relevant velocity components that were fitted to the HIFI profiles for CO 16-15 \citep{Kristensen2012} and H$_2$O \citep{Mottram2014} are showed.  Right: moment 0 maps made by integrating the emission from range indicated by the red, shaded box on the left. Colors are corresponding to the spectra with SMM1-a outflow in green and SMM1-b outflow in purple. HIFI beam is plotted as a black circle. The beam size of the ALMA Band 6 spectral line (red) and continuum (black) is presented in the bottom-left corner of images.}
 \label{fig:HIFI2}
\end{figure*}

The comparison of ALMA CO\,$2-1$ observations with HIFI high$-J$ CO and H$_2$O  line profiles is presented in Fig.  \ref{fig:HIFI2}.

In Fig. \ref{fig:HIFI2}a we plot the HIFI CO\,$16-15$ spectrum is plotted alongside three ALMA spectra, one of which is averaged over the HIFI beam (13\arcsec) for CO\,$16-15$, and two of which are averaged over the region dedicated to SMM1-a and SMM1-b but limited to the borders of the HIFI beam. The HIFI spectra are shown with the Gaussian offset component overlaid based on the fit from \cite{Kristensen2013}.

The offset component for SMM1 is seen in all water transitions targeted by \cite{Mottram2014}, and in CO\,$16-15$ \citep{Kristensen2013}. At the same time CO\,$10-9$ does not show a clear offset component \citep{Yildiz2013}. This suggests high temperatures in the offset component, and indeed the offset profile has been linked to the hot gas component (700 K) seen in the rotational diagrams from PACS observations \citep{Karska2013,Green2013, Kristensen2017}. Radiative transfer modeling of the physical conditions for this component \citep{Kristensen2013} suggests that this emission comes from a small emitting area ($\sim$100 AU; 0\farcs25) with high densities (10$^6$-10$^7$ cm$^{-3}$).

Hints for the spatial origin of the offset component can be seen with ALMA: the offset component is likely associated with the prominent blueshifted emission from SMM1-a close to the source (Fig. \ref{fig:HIFI2}a). This particular position has also been associated with a series of blueshifted water maser emissions \citep{Kempen2009} and a bright spot of ion emission in the near-IR \citep{Dionatos2014}. 
In Fig. \ref{fig:HIFI2}b the H$_2$O $1_{11}-0_{00}$ spectrum is plotted with ALMA spectra: one averaged over the HIFI beam (20\arcsec) for  H$_2$O H$_2$O $1_{11}-0_{00}$, and two averaged over the region dedicated to SMM1-a and SMM1-b but limited to the borders of the HIFI beam. The HIFI spectrum is shown with the broad Gaussian component overlaid based on \cite{Mottram2014} fit.
The redshifted broad component is seen mostly in the lower energy levels of water in the HIFI data \citep{Mottram2014}. It is explained by emission tracing warm (300 K) gas from the outflow cavity shocks, where the protostellar wind interacts with the outflow cavity walls \citep{Kristensen2013, Mottram2014}. It is also proposed that this component can trace the protostellar wind itself \citep{Panoglou2012, Yvart2016}.
Figure \ref{fig:HIFI2}b shows that ALMA CO $2-1$ for SMM1-a and SMM1-b has a very similar line profile to the H$_2$O\,$1_{11}-1_{00}$  in the fast wing component. Spatially, the emission is widespread, coming from both sources.

 It is worth noting that the broad component is much more prominent in the redshifted part of the HIFI spectra; similarly, the EHV jets are associated only with the redshifted jets for both SMM1-a and b. This shows that the presence of the jet could be linked to the presence of the broad component, and possibly part of the component arises as the high-velocity jet material ejected sideways interacts with the outflow cavity walls.
 
The broad component from the HIFI water emission seems coincident spectrally with the fast wing component in CO $2-1$. The chemical signatures of the fast wing such as abundance enhancement of the SiO and presence of HCN can then be linked to the outflow cavity shocks.
In Fig. \ref{fig:HIFI2}c the H$_2$O\,$3_{12}-3_{03}$ spectrum is plotted with ALMA spectra: one averaged over the HIFI beam  (20\arcsec) for  H$_2$O\,$3_{12}-3_{03}$ , and two averaged over the region dedicated to SMM1-a and SMM1-b but limited to the borders of the HIFI beam. The EHV bullets seen prominently in CO\,$2-1$ are not bright in the HIFI spectra. Although no EHV detection for this source with HIFI has been claimed, it seems that there is a faint emission in the two most energetic transitions observed by \cite{Mottram2014}: 3$_{12}$-$3_{03}$ and $3_{12}-2_{21}$, which suggests that the water bullets might be associated with high temperatures, although due to higher frequency of the transitions and thus a smaller beam, less dilution can also play a role.  H$_2$O (3$_{12}$-$3_{03}$) shows a peak at 55 km s$^{-1}$ while the $3_{12}-2_{21}$ shows a tentative detection at 72 km  s$^{-1}$.

 At the same time CO 16-15 and 10-9 show no EHV emission, which suggests that the temperatures are not high enough to populate those levels, or even if the temperatures reach 700 K, the bullets are very compact and filling factor is too small. %
 
   Fig. \ref{fig:HIFI2}c shows that  H$_2$O 3$_{12}$-$3_{03}$  spectral EHV feature peaks outside the EHV peak for both SMM1-a and SMM1-b. The CO $2-1$ emission at the peak velocities of the water bullet shows that the emission could be associated with both jets. Spectra show that the water EHV feature peaks just as the SMM1-b CO\,$2-1$ feature decreases. It is possible that the water feature is associated with the jet at higher temperatures, where CO\,$2-1$ emission is weak.
   
\clearpage

\raggedbottom

%\begin{minipage}{\textwidth}
%\includegraphics{}
%\label{}
%\end{minipage}

\section{Additional figures}
\begin{figure*}[h]
\centering
    \includegraphics[width=0.22\linewidth]{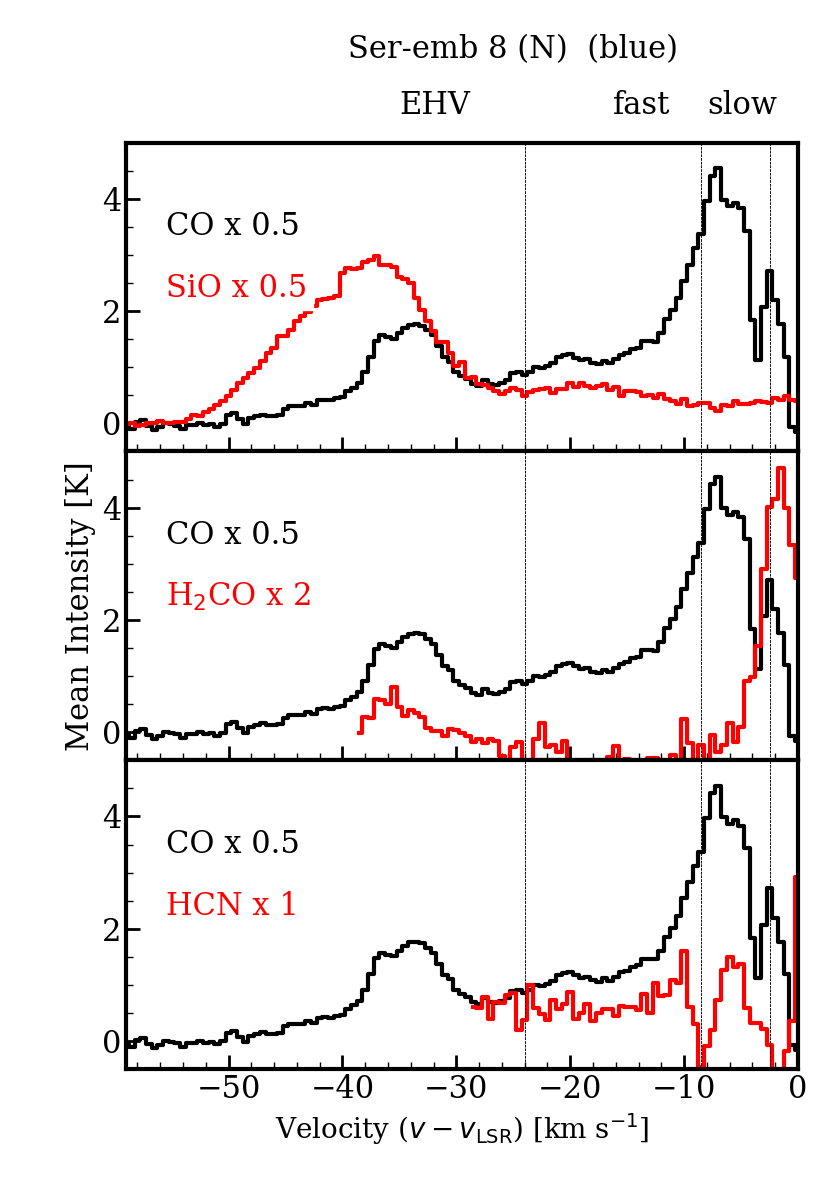}  
  \includegraphics[width=0.22\linewidth]{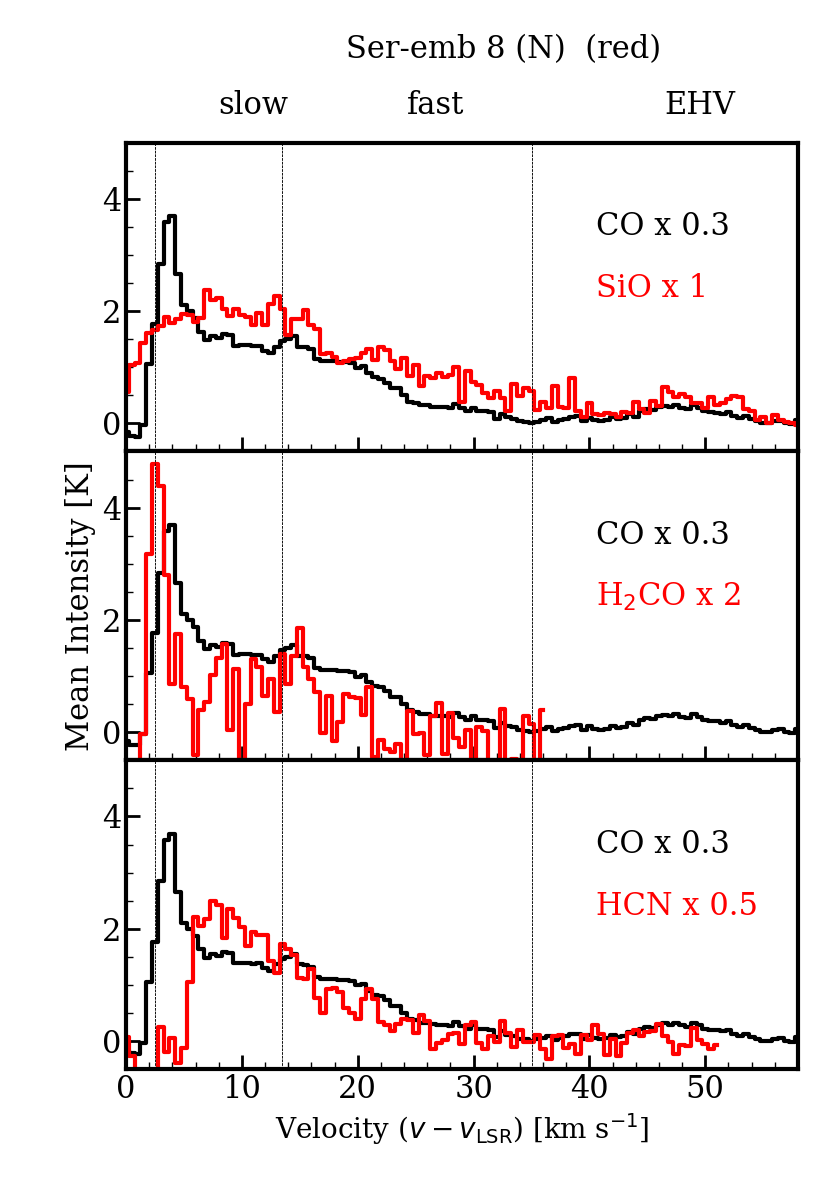}  
    \includegraphics[width=0.22\linewidth]{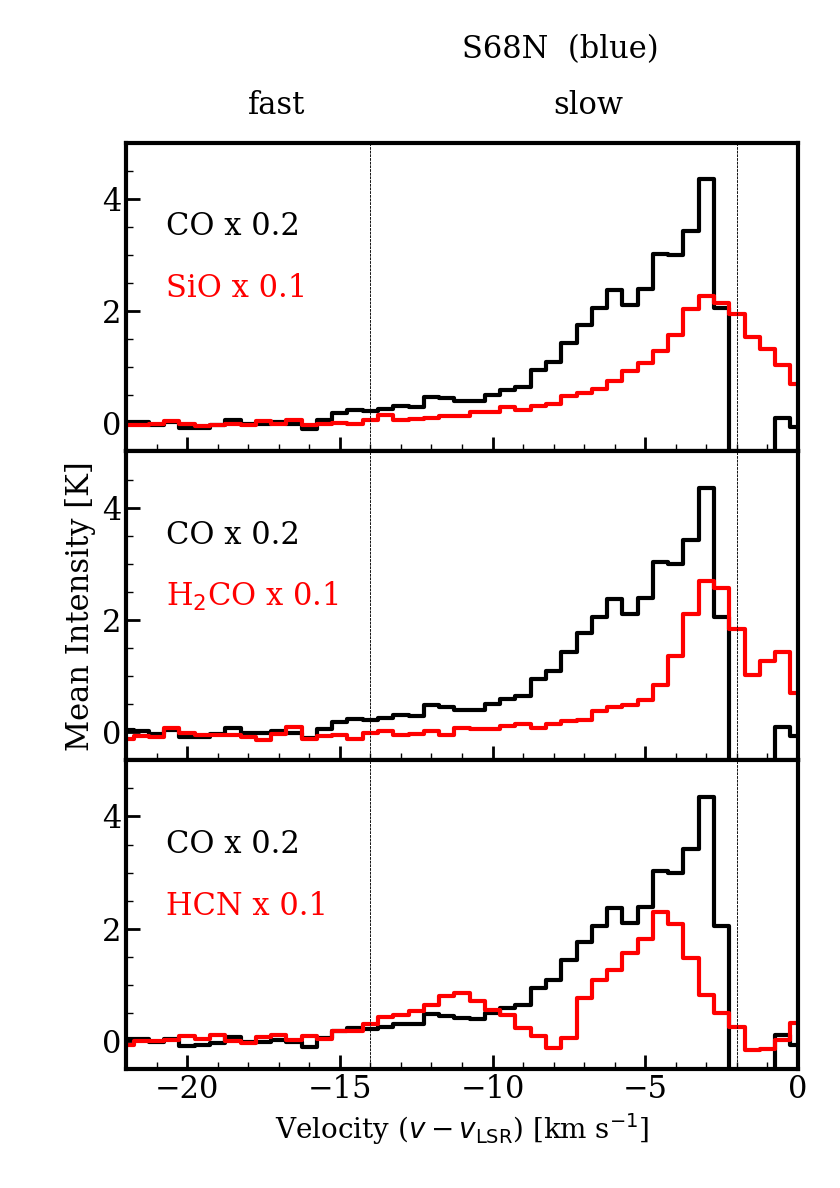}
  \includegraphics[width=0.22\linewidth]{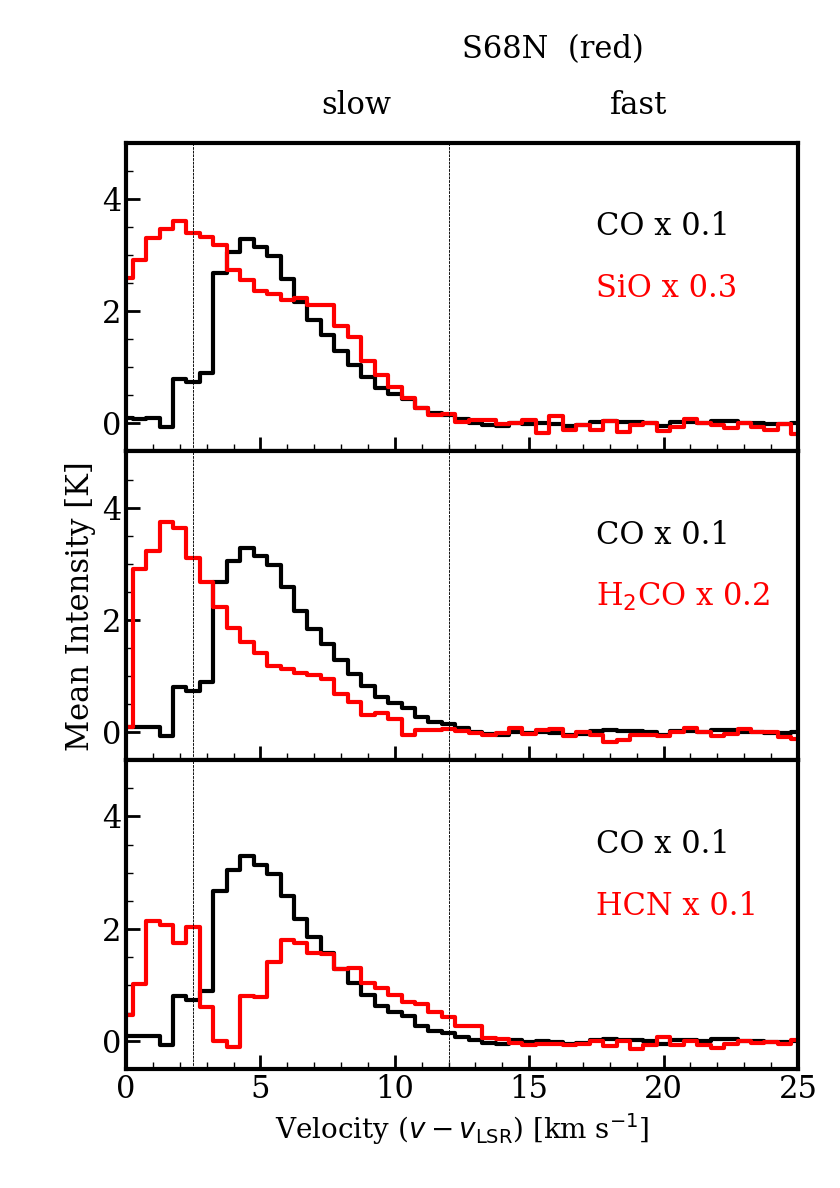}
    \includegraphics[width=0.22\linewidth]{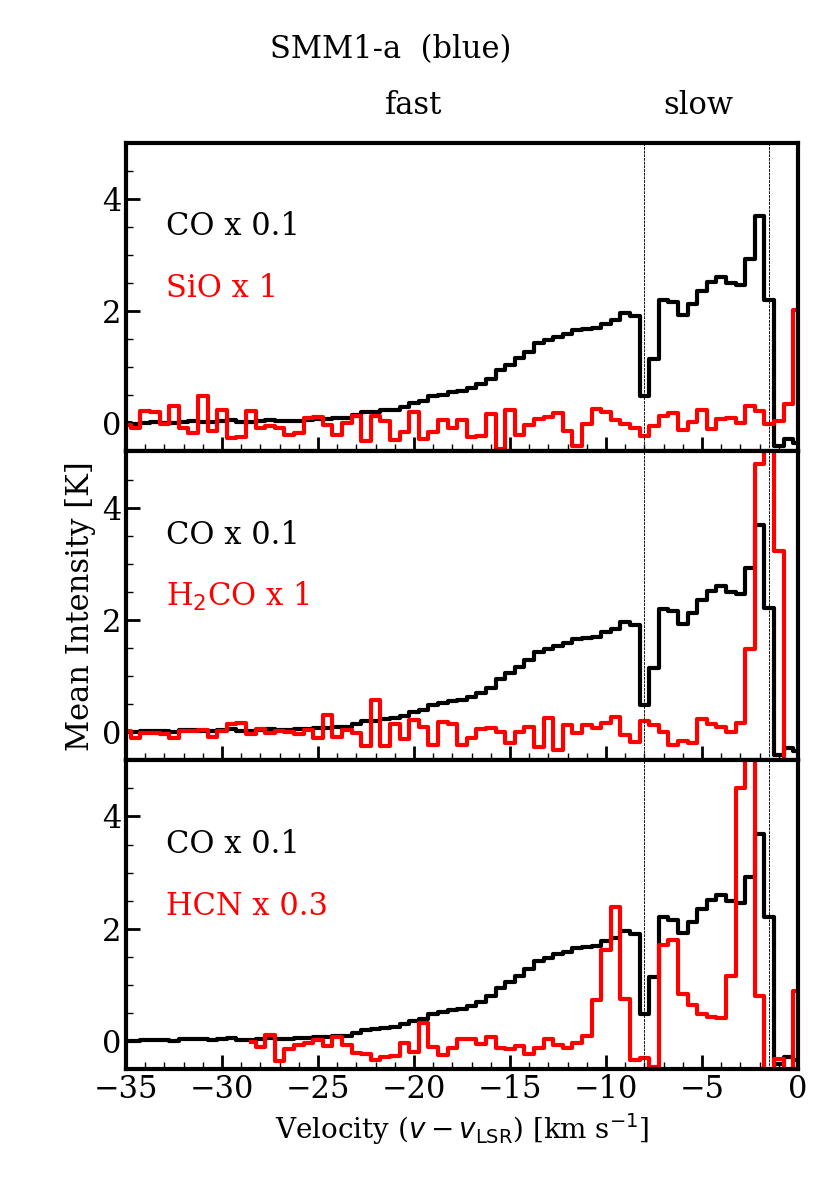}
  \includegraphics[width=0.22\linewidth]{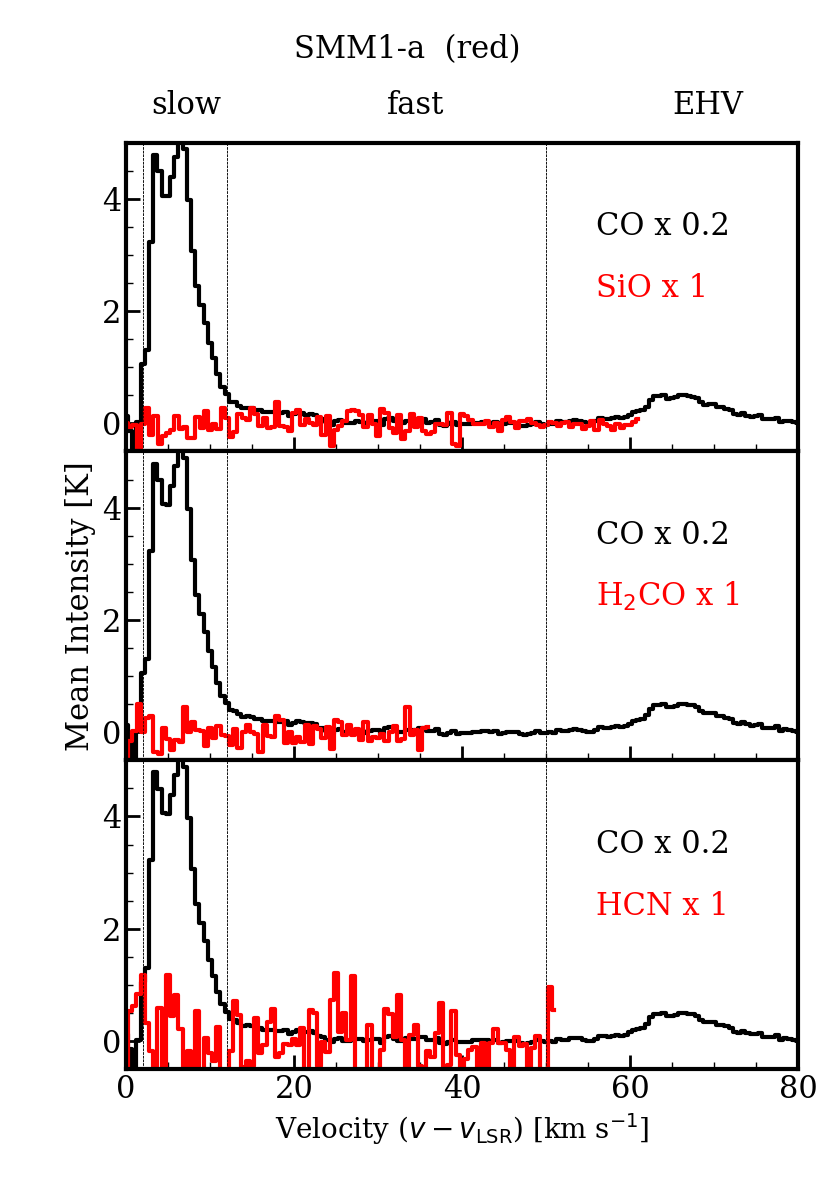}
    \includegraphics[width=0.22\linewidth]{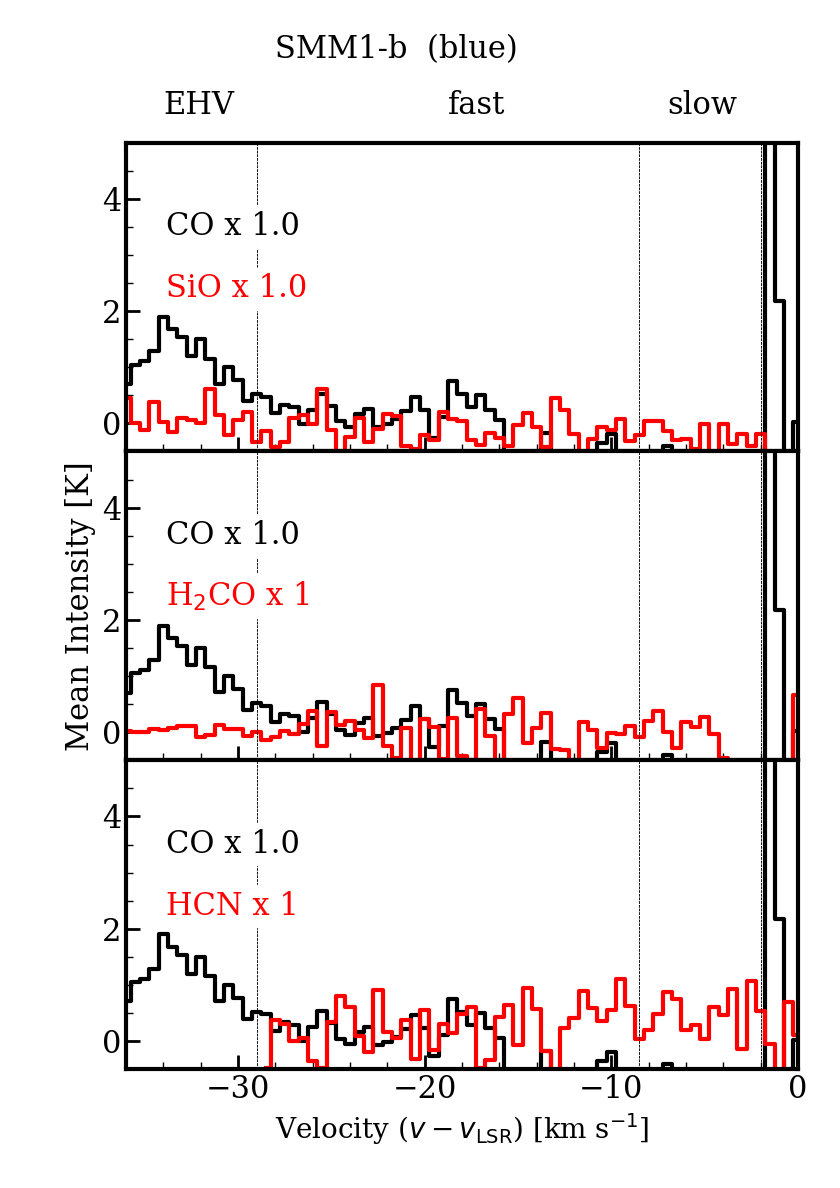}
  \includegraphics[width=0.22\linewidth]{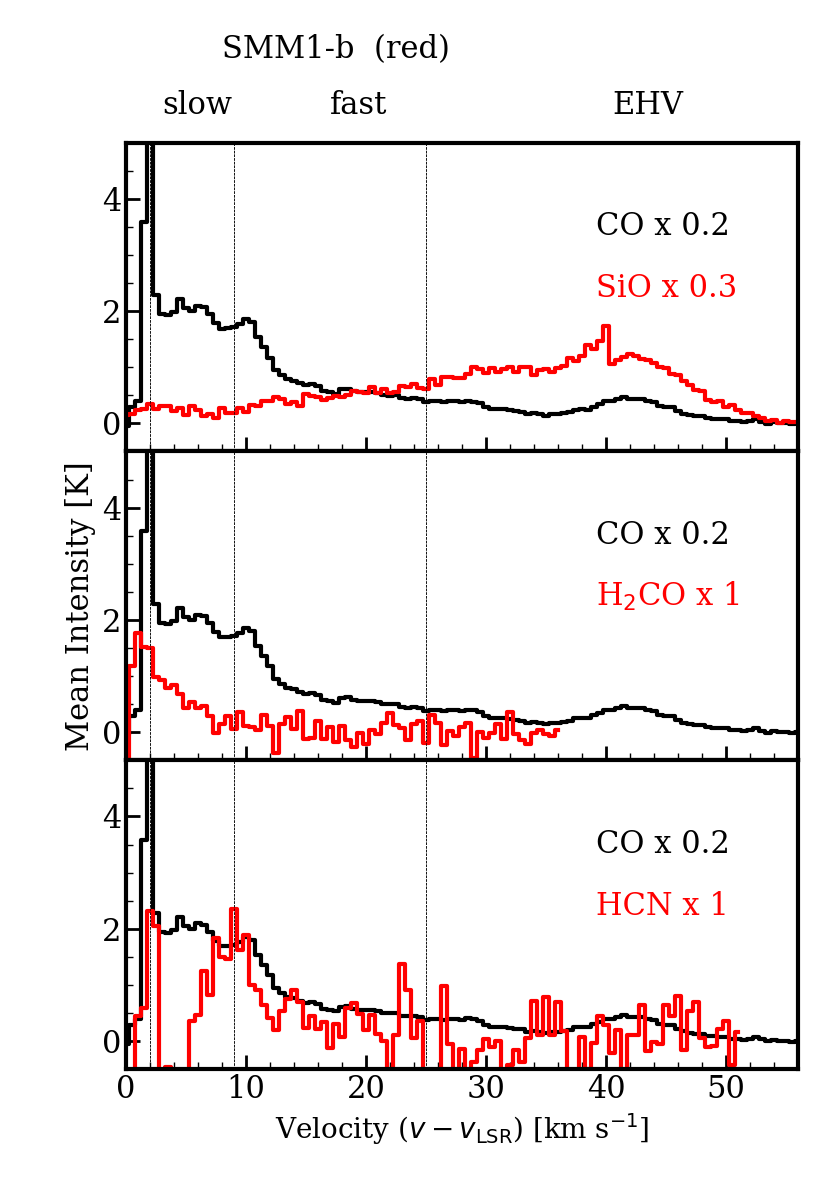}
    \includegraphics[width=0.22\linewidth]{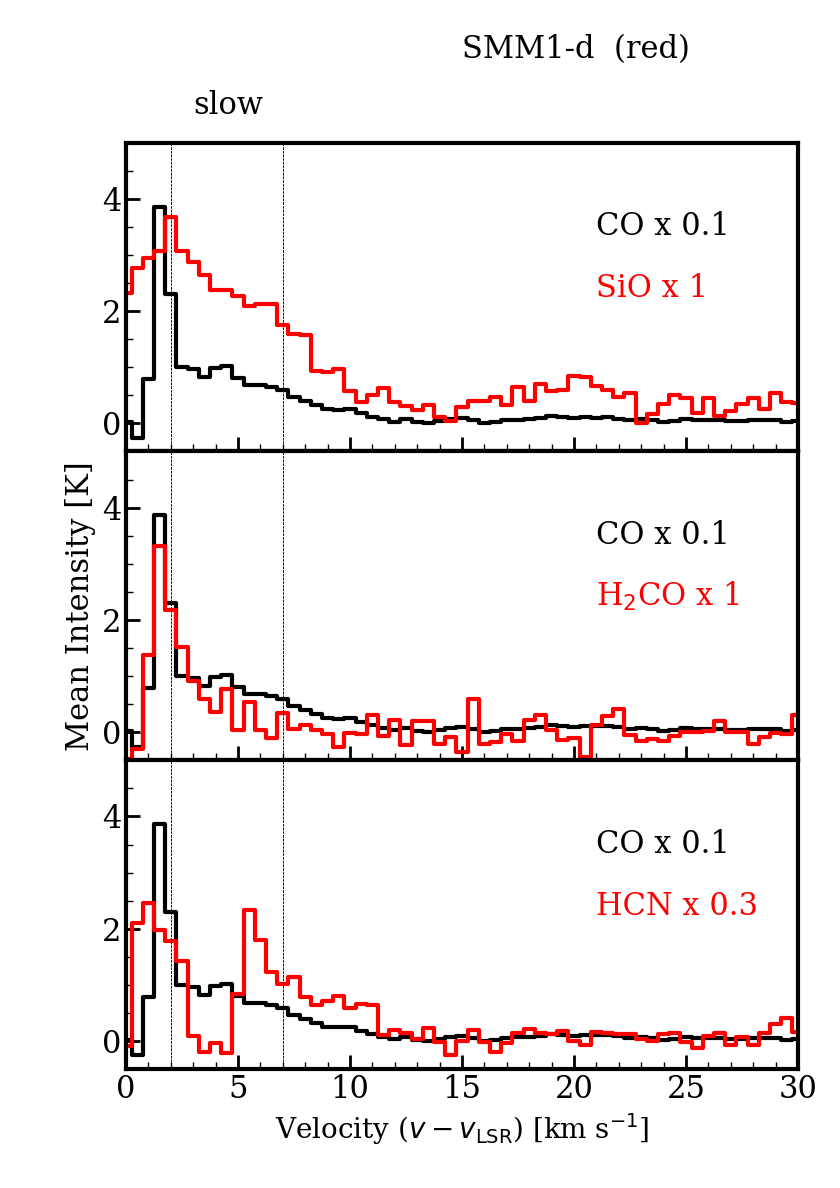}
  \caption{Spectra of CO in black, and different molecules as annotated in top-right corner in red. Spectra extracted from selected part of the outflow to highlight the key velocity components, indicated in the Fig. \ref{fig:allmolecules_Emb8N} and \ref{fig:allmolecules_S68N}. Vertical dashed lines represent boundaries between different velocity regimes: slow wing, fast wing and extremely high-velocity.}
 \label{fig:velregimes_app}
\end{figure*}

\begin{figure*}[h]
\centering
 \includegraphics[width=0.96\linewidth, trim={3cm 25cm 0cm  20cm},clip]{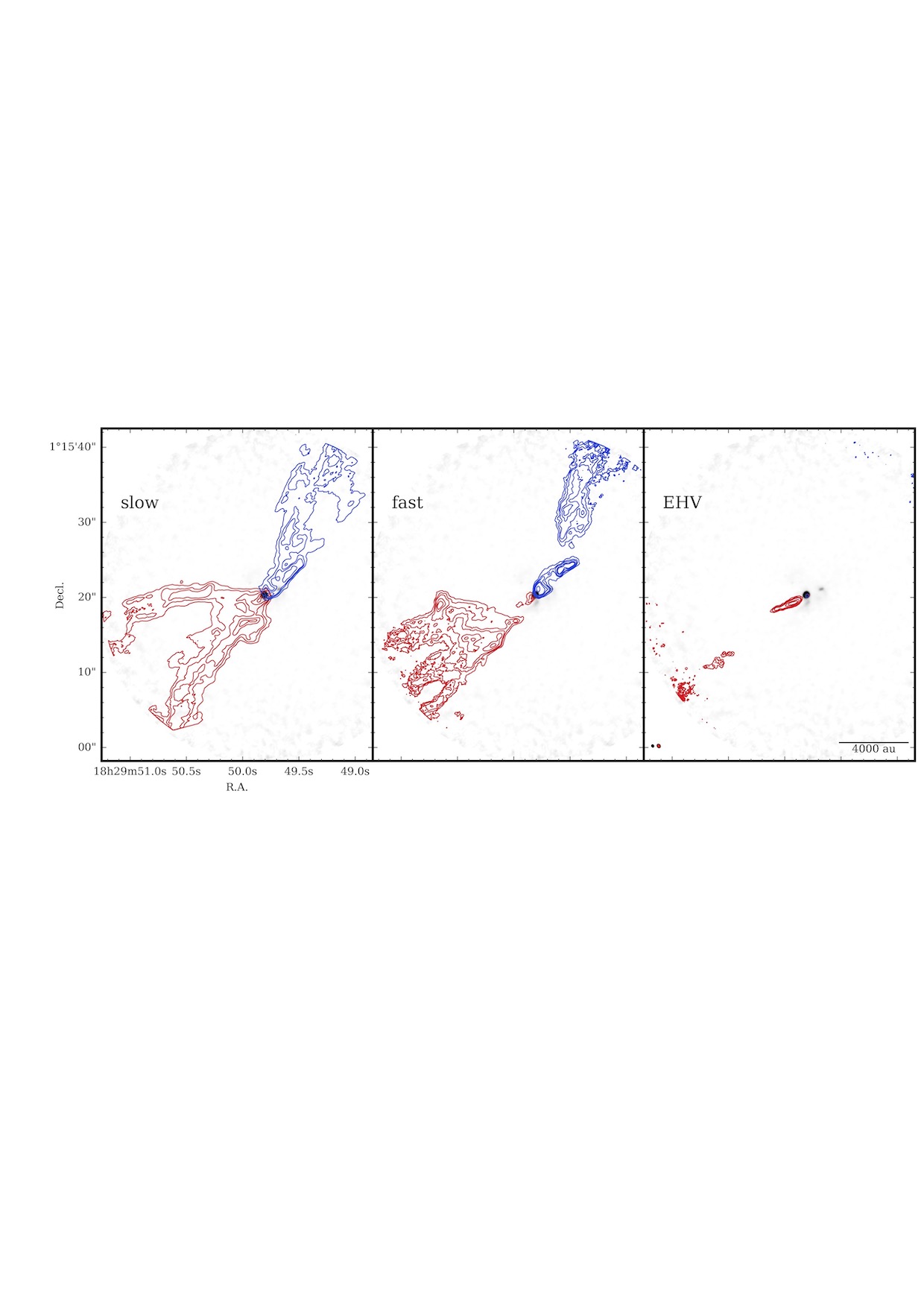}
  \caption{Integrated intensity maps of CO for different velocity regimes overlaid on the Band 6 continuum in grayscale for SMM1-a. The emission is integrated over the velocities listed in Table \ref{tab:velocities}. The  synthesized beam of the CO (red) and continuum (black)  is presented in bottom-left corner of EHV plot. The synthesized beam size of the continuum images is 0\farcs35 $\times$ 0\farcs33 and 0\farcs55 $\times$ 0\farcs45 for CO map. The contours are [3, 6, 9, 15, 20, 40] times the rms value. The rms values for each velocity channel, blueshifted and redshifted in K km s$^{-1}$, are slow [18.8, 20.5], fast[5.8, 7.2], EHV [2.0, 2.8].}
  
 \label{fig:velregimes_SMM1a}
\end{figure*}

\begin{figure*}[h]
\centering
 \includegraphics[width=0.96\linewidth,trim={4cm 28.5cm 0 20.5cm},clip]{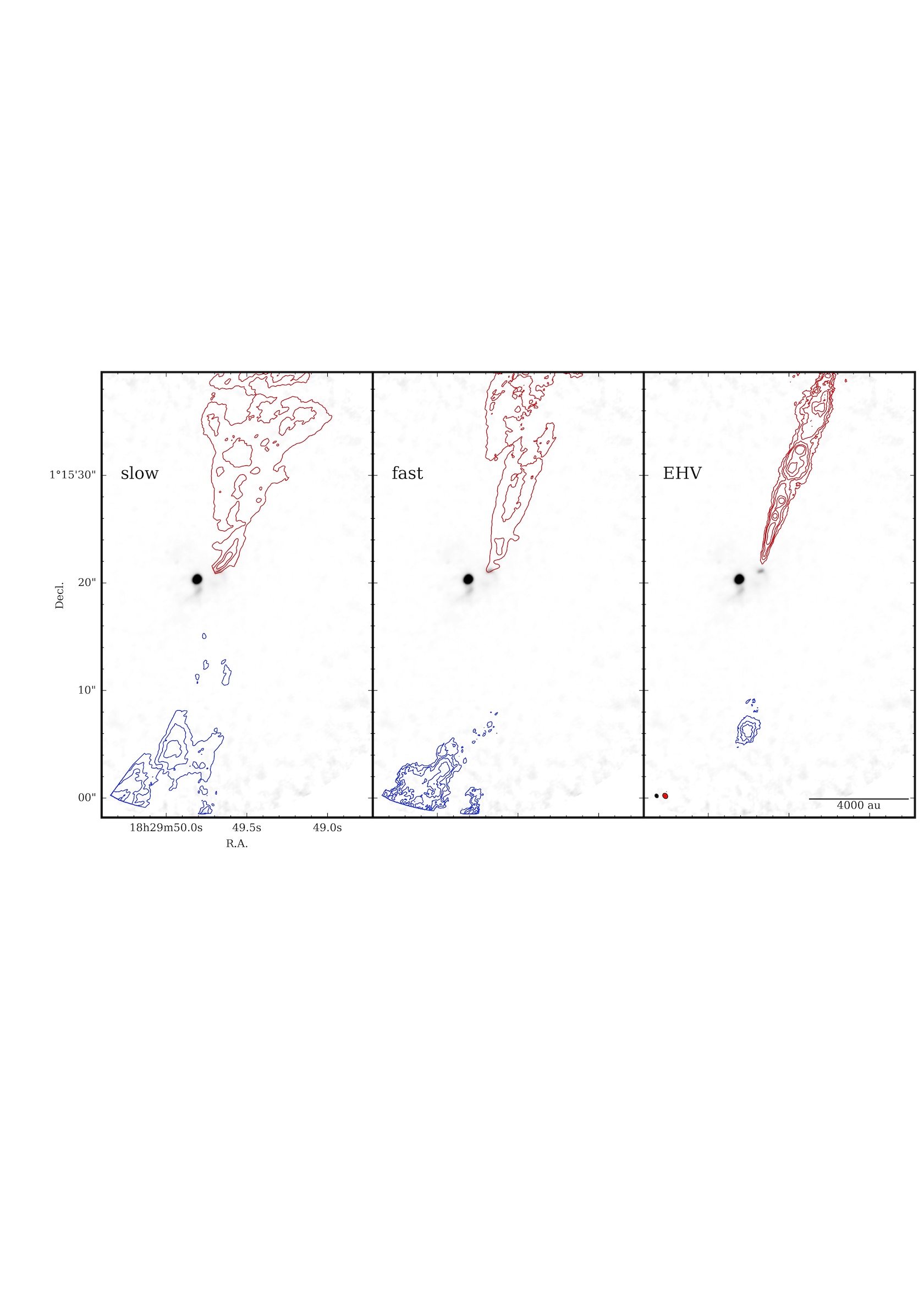}
  \caption{Same as Fig. \ref{fig:velregimes_SMM1a} for SMM1-b. The rms values for each velocity channel, blueshifted and redshifted in K km s$^{-1}$, are slow [17.5, 21.0], fast [5.3, 12.6], EHV  [2.1, 3.7].}
 \label{fig:velregimes_SMM1b}
\end{figure*}

\begin{figure*}[h]
\centering
 \includegraphics[width=0.96\linewidth, trim={1cm 28cm 0 22cm},clip]{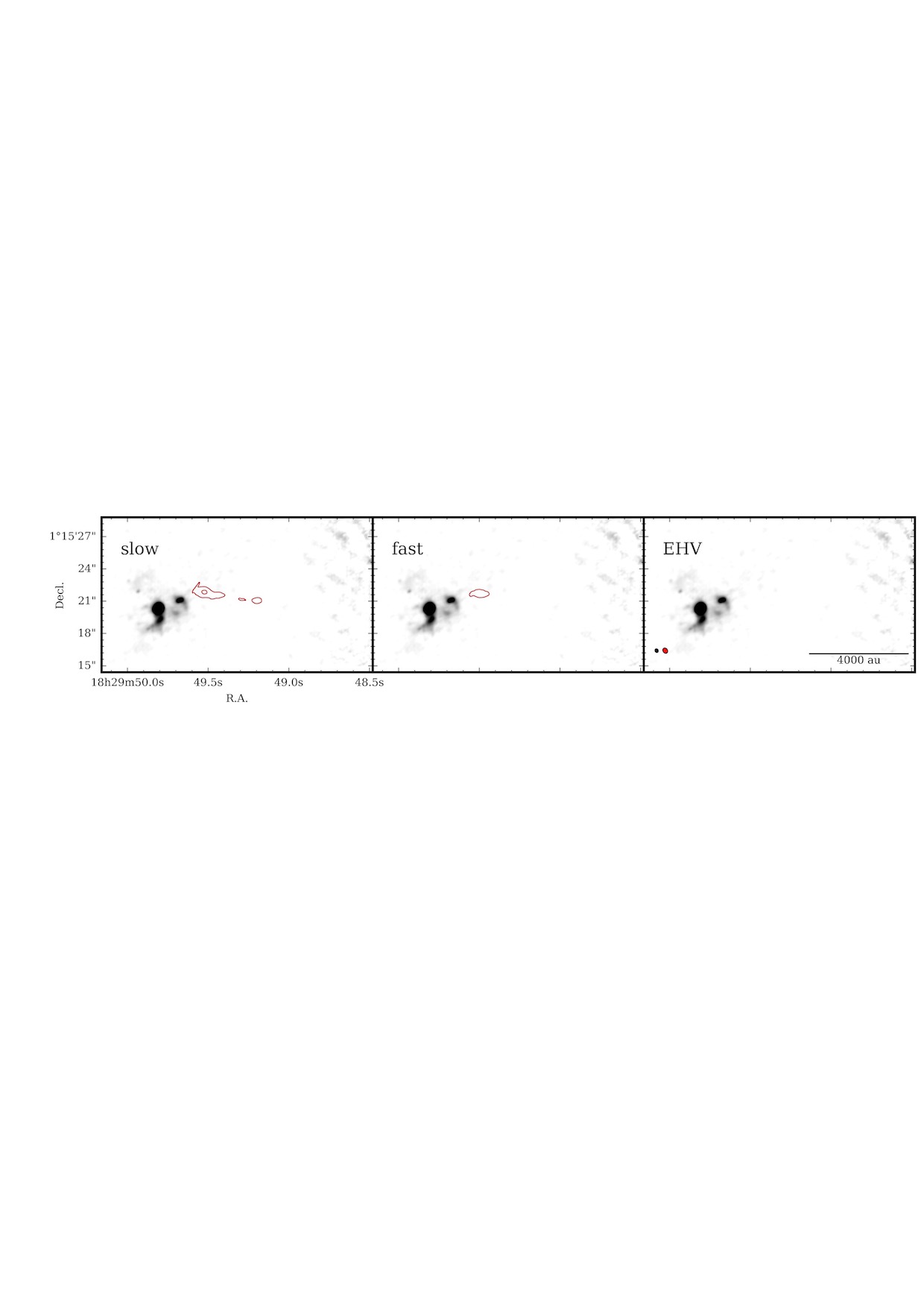}
  \caption{Same as Fig. \ref{fig:velregimes_SMM1a} for SMM1-d. The rms values for each velocity channel,  redshifted in K km s$^{-1}$ [20.7],[15.72]. No blueshifted and no EHV emission is detected toward this source.}
 \label{fig:velregimes_SMM1d}
\end{figure*}

\begin{figure*}[h]
\centering
  \includegraphics[width=0.95\linewidth, trim={2cm 25cm 0cm 20cm},clip]{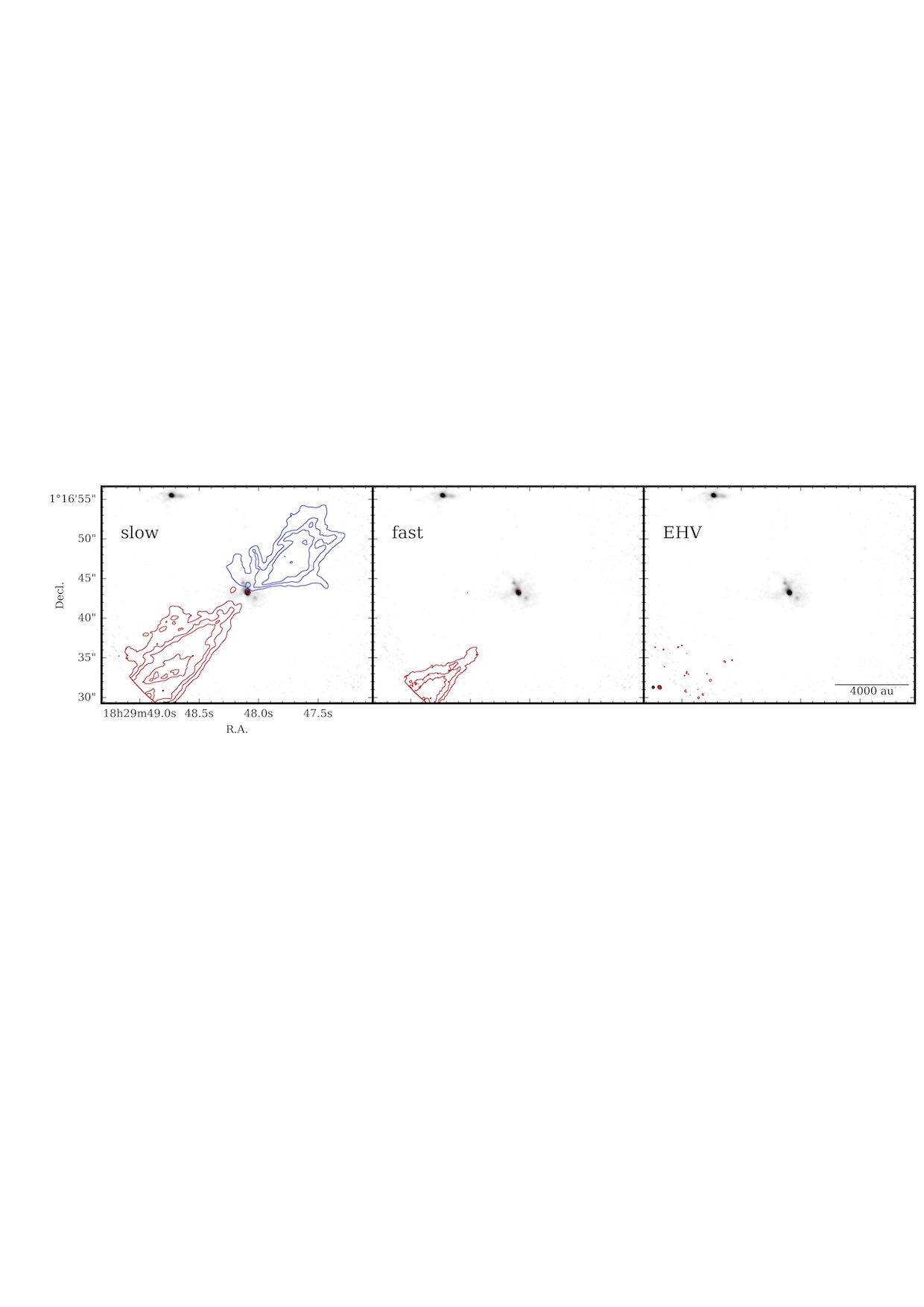}
  \caption{Same as Fig. \ref{fig:velregimes_SMM1a} for S68N. The rms values for each velocity channel, blueshifted and redshifted in K km s$^{-1}$, are slow [19.2, 13.8], fast [-, 4.7]. No EHV emission is detected toward this source.}
 \label{fig:velregimes_S68N}
\end{figure*}

\clearpage
\section{Tables}
Details of the observations used in this paper are listed in Table \ref{table:2}

Calculated abundances of each molecule are shown in Tables \ref{tab:column_denisties_Emb8N}-\ref{tab:column_denisties_SMM1d}. For each molecule we calculate column density for minimum and maximum expected T$_{\rm ex}$  as listed in Table \ref{table:4}. We assume that values have 20 \% uncertainty arising from calibration, arbitrary defining velocity regimes and other factors. 

Tables C.7-C.10 present the outflow forces not corrected for
inclination (see main text). Their absolute values are therefore
uncertain by factors of a few.

\begin{table*}
\caption{Specifications of observations}             %
\label{table:2}      %
\centering                          %
\begin{tabular}{c c c c c c c c c c c}        %
\hline\hline                 %
Configuration & $\lambda$ & Max. Baseline & Date & Calibration$^{\rm a}$  & Bandpass & Phase & Flux  \\
\hline                                   %
Band 6 (C43-1) & 1.3 mm& 378 m & 06/04/2015   &4.2.2 - m  & J1733-1304 &  J1751+0939 & Titan \\
Band 6 (C43-4) & 1.3 mm&1250 m &  18/08/2014 & 4.3.1 - p &  J1751+0939 &  J1751+0939 &  J1751+096  \\ 
Band 3  (C43-5) &3 mm& 2500 m &  04/10/2016  &  4.7.38335 - p   & J1751+0939 & J1838+0404 & J1838+0404 \\

\hline                                   %

\end{tabular}
\tablefoot{
\tablefoottext{a}{Version of CASA used for calibration (m - manual calibration, p - pipeline calibration)}
}

\end{table*}

\begin{table*}
\centering
\caption{Column densities of targeted molecules per velocity regime for Ser-emb 8 (N)}
\label{tab:column_denisties_Emb8N}
\begin{tabular}{lcccccccc}
\hline \hline
Ser-emb 8 (N) & CO & SiO & H$_2$CO & HCN \\
& cm$ ^{-2}$ & cm$ ^ {-2}$ & cm$ ^{-2}$ & cm$ ^{-2}$ \\
\hline
red & & & &\\
\hline
slow &  6e+16  --   4e+17 &  1e+13  --   3e+12 &  3e+13  --   5e+13 & > 2e+14  --   6e+14 \\
fast &  3e+16  --   2e+17 &  1e+13  --   4e+12 & < 7e+12  --   1e+13 & > 2e+14  --   4e+14 \\
EHV &  6e+15  --   4e+16 &  2e+13  --   6e+12 & < 7e+12  --   1e+13 & < 2e+13  --   4e+13 \\
\hline
blue & & & &\\
\hline
slow &  4e+16  --   3e+17 &  4e+12  --   1e+12 &  1e+13  --   3e+13 & > 4e+13  --   1e+14 \\
fast &  4e+16  --   3e+17 &  1e+13  --   3e+12 & < 7e+12  --   1e+13 & > 2e+13  --   5e+13 \\
EHV &  3e+16  --   2e+17 &  8e+13  --   2e+13 &  1e+13  --   2e+13 & < 2e+13  --   4e+13 \\
\hline
\end{tabular}
\end{table*}

\begin{table*}
\centering
\caption{Column densities of targeted molecules per velocity regime for S68N}
\label{tab:column_denisties_S68N}
\begin{tabular}{lcccccccc}
\hline \hline
S68N & CO & SiO & H$_2$CO & HCN \\
& cm$ ^{-2}$ & cm$ ^ {-2}$ & cm$ ^{-2}$ & cm$ ^{-2}$ \\
\hline
red & & & &\\
\hline
slow &  7e+16  --   5e+17 &  8e+13  --   2e+13 &  1e+14  --   2e+14 & > 1e+14  --   3e+14 \\
fast &  1e+16  --   1e+17 &  3e+13  --   9e+12 &  7e+12  --   1e+13 & > 5e+13  --   1e+14 \\
EHV & < 8e+14  --   5e+15 & < 5e+12  --   1e+12 & < 7e+12  --   1e+13 & < 2e+13  --   4e+13 \\
\hline
blue & & & &\\
\hline
slow &  8e+16  --   6e+17 &  8e+13  --   2e+13 &  1e+14  --   2e+14 & > 2e+14  --   4e+14 \\
fast &  5e+15  --   3e+16 &  1e+13  --   4e+12 & < 7e+12  --   1e+13 & > 2e+13  --   6e+13 \\
EHV & < 8e+14  --   5e+15 & < 5e+12  --   1e+12 & < 7e+12  --   1e+13 & < 2e+13  --   4e+13 \\
\hline
\end{tabular}
\end{table*}

\begin{table*}
\centering
\caption{Column densities of targeted molecules per velocity regime for SMM1-a}
\label{tab:column_denisties_SMM1a}
\begin{tabular}{lcccccccc}
\hline \hline
SMM1-a & CO & SiO & H$_2$CO & HCN \\
& cm$ ^{-2}$ & cm$ ^ {-2}$ & cm$ ^{-2}$ & cm$ ^{-2}$ \\
\hline
red & & & &\\
\hline
slow &  9e+16  --   6e+17 &  4e+13  --   1e+13 &  4e+13  --   8e+13 & > 1e+14  --   4e+14 \\
fast &  4e+16  --   3e+17 &  6e+12  --   2e+12 & < 8e+12  --   2e+13 & < 2e+13  --   6e+13 \\
EHV &  8e+15  --   5e+16 & < 2e+12  --   4e+11 & < 8e+12  --   2e+13 & < 2e+13  --   6e+13 \\
\hline
blue & & & &\\
\hline
slow &  8e+16  --   5e+17 &  1e+13  --   4e+12 &  3e+13  --   5e+13 & > 7e+13  --   2e+14 \\
fast &  4e+16  --   3e+17 & < 6e+12  --   2e+12 & < 2e+12  --   4e+12 & > 3e+13  --   7e+13 \\
EHV & < 9e+14  --   6e+15 & < 6e+12  --   2e+12 & < 2e+12  --   4e+12 & < 2e+13  --   6e+13 \\
\hline
\end{tabular}
\end{table*}

\begin{table*}
\centering
\caption{Column densities of targeted molecules per velocity regime for SMM1-b}
\label{tab:column_denisties_SMM1b}
\begin{tabular}{lcccccccc}
\hline \hline
SMM1-b & CO & SiO & H$_2$CO & HCN \\
& cm$ ^{-2}$ & cm$ ^ {-2}$ & cm$ ^{-2}$ & cm$ ^{-2}$ \\
\hline
red & & & &\\
\hline
slow &  7e+16  --   5e+17 &  1e+13  --   4e+12 &  2e+13  --   3e+13 & > 4e+13  --   1e+14 \\
fast &  3e+16  --   2e+17 &  1e+14  --   3e+13 & < 8e+12  --   2e+13 & < 2e+13  --   6e+13 \\
EHV &  2e+16  --   2e+17 &  1e+14  --   3e+13 & < 8e+12  --   2e+13 & < 2e+13  --   6e+13 \\
\hline
blue & & & &\\
\hline
slow &  6e+16  --   4e+17 &  4e+13  --   1e+13 &  1e+13  --   2e+13 & < 2e+13  --   6e+13 \\
fast &  2e+16  --   2e+17 & < 6e+12  --   2e+12 & < 2e+12  --   4e+12 & < 2e+13  --   6e+13 \\
EHV &  5e+15  --   4e+16 & < 6e+12  --   2e+12 & < 2e+12  --   4e+12 & < 2e+13  --   6e+13 \\
\hline
\end{tabular}
\end{table*}

\begin{table*}
\centering
\caption{Column densities of targeted molecules per velocity regime for SMM1-d}
\label{tab:column_denisties_SMM1d}
\begin{tabular}{lcccccccc}
\hline \hline
SMM1-d & CO & SiO & H$_2$CO & HCN \\
& cm$ ^{-2}$ & cm$ ^ {-2}$ & cm$ ^{-2}$ & cm$ ^{-2}$ \\
\hline
red & & & &\\
\hline
slow &  2e+16  --   2e+17 &  7e+13  --   2e+13 &  2e+13  --   4e+13 & > 8e+13  --   2e+14 \\
fast &  1e+16  --   9e+16 &  4e+13  --   1e+13 & < 8e+12  --   2e+13 & > 8e+13  --   2e+14 \\
EHV & < 9e+14  --   6e+15 & < 6e+12  --   2e+12 & < 8e+12  --   2e+13 & < 2e+13  --   6e+13 \\
\hline
blue & & & &\\
\hline
slow & < 9e+14  --   6e+15 & < 6e+12  --   2e+12 & < 8e+12  --   2e+13 & < 2e+13  --   6e+13 \\
fast & < 9e+14  --   6e+15 & < 6e+12  --   2e+12 & < 8e+12  --   2e+13 & < 2e+13  --   6e+13 \\
EHV & < 9e+14  --   6e+15 & < 6e+12  --   2e+12 & < 8e+12  --   2e+13 & < 2e+13  --   6e+13 \\
\hline
\end{tabular}
\end{table*}

\begin{table*}
\centering
\caption{Outflow properties per velocity regime for Ser-emb 8 (N)}
\label{tab:ourflow_propertiess_Emb8N}
\begin{tabular}{lcccccccc}
\hline \hline
Ser-emb 8 (N) & M & $\dot {\rm M}$ & P & F$_{\rm out}$ \\
& ﻿$\mathrm{M_{\odot}}$ &﻿$\mathrm{M_{\odot}\,yr^{-1}}$ & ﻿$\mathrm{M_{\odot}\,km\,\,s^{-1}}$  & ﻿$\mathrm{M_{\odot}\,km\,s^{-1}\,yr^{-1}}$ \\
\hline
red & & & &\\
\hline
slow & 2.4e-06  & 5.1e-07 & 2.1e-02 & 2.9e-05  \\
fast & 7.3e-07  & 4.1e-07 & 1.7e-02 & 2.4e-05  \\
EHV & 5.2e-08  & 7.4e-08 & 3.1e-03 & 4.3e-06  \\
\hline
blue & & & &\\
\hline
slow & 1.3e-06  & 2.4e-07 & 1.0e-02 & 1.4e-05  \\
fast & 1.2e-06  & 5.3e-07 & 2.2e-02 & 3.1e-05  \\
EHV & 6.2e-07  & 6.3e-07 & 2.6e-02 & 3.7e-05  \\
\hline
\end{tabular}
\end{table*}
\begin{table*}
\centering
\caption{Outflow properties per velocity regime for S68N}
\label{tab:ourflow_propertiess_S68N}
\begin{tabular}{lcccccccc}
\hline \hline
S68N & M & $\dot {\rm M}$ & P & F$_{\rm out}$ \\
& ﻿$\mathrm{M_{\odot}}$ &﻿$\mathrm{M_{\odot}\,yr^{-1}}$ & ﻿$\mathrm{M_{\odot}\,km\,\,s^{-1}}$  & ﻿$\mathrm{M_{\odot}\,km\,s^{-1}\,yr^{-1}}$ \\
\hline
red & & & &\\
\hline
slow & 1.1e-05  & 2.4e-06 & 1.0e-01 & 5.9e-05  \\
fast & 1.1e-06  & 4.8e-07 & 2.0e-02 & 1.2e-05  \\
EHV & ---  & --- & --- & ---  \\
\hline
blue & & & &\\
\hline
slow & 1.2e-05  & 2.6e-06 & 1.1e-01 & 5.8e-05  \\
fast & 1.7e-07  & 9.4e-08 & 3.9e-03 & 2.1e-06  \\
EHV & ---  & --- & --- & ---  \\
\hline
\end{tabular}
\end{table*}
\begin{table*}
\centering
\caption{Outflow properties per velocity regime for SMM1-a}
\label{tab:ourflow_propertiess_SMM1a}
\begin{tabular}{lcccccccc}
\hline \hline
SMM1-a & M & $\dot {\rm M}$ & P & F$_{\rm out}$ \\
& ﻿$\mathrm{M_{\odot}}$ &﻿$\mathrm{M_{\odot}\,yr^{-1}}$ & ﻿$\mathrm{M_{\odot}\,km\,\,s^{-1}}$  & ﻿$\mathrm{M_{\odot}\,km\,s^{-1}\,yr^{-1}}$ \\
\hline
red & & & &\\
\hline
slow & 2.6e-05  & 7.6e-06 & 3.5e-01 & 6.1e-04  \\
fast & 7.2e-06  & 6.2e-06 & 2.9e-01 & 4.9e-04  \\
EHV & 2.2e-07  & 3.9e-07 & 1.8e-02 & 3.1e-05  \\
\hline
blue & & & &\\
\hline
slow & 1.4e-05  & 1.8e-06 & 8.7e-02 & 6.4e-05  \\
fast & 4.5e-06  & 2.2e-06 & 1.1e-01 & 7.8e-05  \\
EHV & ---  & --- & --- & ---  \\
\hline
\end{tabular}
\end{table*}
\begin{table*}
\centering
\caption{Outflow properties per velocity regime for SMM1-b}
\label{tab:ourflow_propertiess_SMM1b}
\begin{tabular}{lcccccccc}
\hline \hline
SMM1-b & M & $\dot {\rm M}$ & P & F$_{\rm out}$ \\
& ﻿$\mathrm{M_{\odot}}$ &﻿$\mathrm{M_{\odot}\,yr^{-1}}$ & ﻿$\mathrm{M_{\odot}\,km\,\,s^{-1}}$  & ﻿$\mathrm{M_{\odot}\,km\,s^{-1}\,yr^{-1}}$ \\
\hline
red & & & &\\
\hline
slow & 1.2e-05  & 2.2e-06 & 9.1e-02 & 1.2e-04  \\
fast & 4.4e-06  & 2.2e-06 & 9.4e-02 & 1.2e-04  \\
EHV & 1.5e-06  & 1.5e-06 & 6.2e-02 & 8.2e-05  \\
\hline
blue & & & &\\
\hline
slow & 5.9e-06  & 1.1e-06 & 4.4e-02 & 3.8e-05  \\
fast & 1.6e-06  & 8.3e-07 & 3.5e-02 & 3.0e-05  \\
EHV & 7.0e-08  & 7.0e-08 & 2.9e-03 & 2.5e-06  \\
\hline
\end{tabular}
\end{table*}
\begin{table*}
\centering
\caption{Outflow properties per velocity regime for SMM1-d}
\label{tab:ourflow_propertiess_SMM1d}
\begin{tabular}{lcccccccc}
\hline \hline
SMM1-d & M & $\dot {\rm M}$ & P & F$_{\rm out}$ \\
& ﻿$\mathrm{M_{\odot}}$ &﻿$\mathrm{M_{\odot}\,yr^{-1}}$ & ﻿$\mathrm{M_{\odot}\,km\,\,s^{-1}}$  & ﻿$\mathrm{M_{\odot}\,km\,s^{-1}\,yr^{-1}}$ \\
\hline
red & & & &\\
\hline
slow & 5.6e-07  & 1.5e-07 & 6.3e-03 & 4.5e-06  \\
fast & 2.5e-07  & 1.0e-07 & 4.3e-03 & 3.1e-06  \\
EHV & ---  & --- & --- & ---  \\
\hline
blue & & & &\\
\hline
slow & ---  & --- & --- & ---  \\
fast & ---  & --- & --- & ---  \\
EHV & ---  & --- & --- & ---  \\
\hline
\end{tabular}
\end{table*}
\end{document}